\pdfoutput=1
\documentclass[11pt]{article}
\usepackage{threeparttable}

\usepackage[top=0.8in, bottom=0.78in, left=0.87in, right=0.87in]{geometry}
\usepackage{setspace}
\usepackage[T1]{fontenc}
\usepackage{times}
\usepackage{booktabs}
\usepackage{rotating}
\usepackage{graphicx}
\usepackage{tikz}
\usepackage[section]{placeins} 
\usepackage[large, bf]{caption}
\usepackage{palatino}
\usepackage{textcomp}
\usepackage{longtable}
\usepackage{nicefrac}
\usepackage{adjustbox}	
\usepackage[hyphens]{url}
\usepackage{natbib}
\bibpunct{(}{)}{;}{a}{,}{,}
\usepackage{todonotes}

\usepackage{amsmath, amsfonts, amssymb, amsthm}

\usepackage{mathpazo} 
\usepackage{longtable}

\newtheorem{ass}{Assumption}

\def\urltilda{\kern -.15em\lower .7ex\hbox{\~{}}\kern .04em}

\usepackage{titlesec}

\titlespacing*{\section}{0pt}{1.5ex plus 1ex minus .2ex}{0.8ex plus .2ex}
\titlespacing*{\subsection}{0pt}{1.2ex plus 1ex minus .2ex}{0.8ex plus .2ex}

\usepackage{caption}
\usepackage{subcaption}

\usepackage{algorithm}
\usepackage{algpseudocode}

\usepackage{algorithm}
\usepackage{algpseudocode}

\usepackage[colorlinks=true, linkcolor=blue, citecolor=blue, urlcolor=blue]{hyperref}

\begin{document}
\title{Smiles in Profiles: Improving Efficiency While Reducing Disparities in Online Marketplaces}
\author{%
  Susan Athey\thanks{Stanford Graduate School of Business} \and 
  Dean Karlan\thanks{Kellogg School of Management, Northwestern University} \and 
  Emil Palikot\thanks{Northeastern University} \and 
  Yuan Yuan\thanks{Carnegie Mellon University; Stanford Graduate School of Business}
}
\date{\today}

\maketitle

\renewcommand{\thefootnote}{}
\footnotetext{We thank Kiva Microfunds for generously sharing data and discussing the research questions. We thank Herman Donner and Kristine Koutout at Stanford Graduate School of Business and Allison Koenecke at Cornell University for comments and suggestions. The Golub Capital Social Impact Lab at Stanford Graduate School of Business provided funding for this research. This research has been subject to review and approval by the Research Compliance Office at Stanford University, protocol number IRB-62442 and registered at the AEA RCT registry with the number 0010030.}
\renewcommand{\thefootnote}{\arabic{footnote}}

\begin{abstract}
\setstretch{1}

Online platforms often have conflicting goals: they face tradeoffs between increasing efficiency and reducing disparities, where the latter may relate to objectives such as the longer-term health of the marketplace or the organization's mission. We examine how participants' profile pictures shape this trade-off in the context of a peer-to-peer lending platform. We develop and apply an approach to estimate marketplace participants' preferences for different profile features, distinguishing between (i) "type" (e.g., gender, age) and (ii) "style" (e.g., smiling in the photo). Relative to type, style features are easier to change, and platforms may be more willing to encourage such changes. Our approach starts by using causal inference methods together with computer vision algorithms applied to observational data to identify type and style features of profiles that appear to affect demand for transactions. We further decompose type-based disparities into a component driven by demand for certain types and a component that arises because different types have different distributions of style features; we find that style differences often exacerbate type-based disparities. To improve internal validity, we then carry out two randomized survey experiments using generative models to create multiple versions of profile images that differ in one feature at a time. We then evaluate counterfactual platform policies based on the changeable profile features and identify approaches that can ameliorate the disparity-efficiency tension.

\vspace{2cm}
\textbf{JEL Classification:} J710, L1, C9

\textbf{Keywords:} Disparities, Decompositions, Digital Platforms, Computer Vision, Generative Models
\end{abstract}
\newpage

\section{Introduction}
Profile images are a central design feature of many online platforms and influence both user interactions and platform outcomes \citep{ert2016trust}. A well-established literature documents how profile images reveal socio-demographic characteristics, often enabling discrimination and leading to disparities across groups in outcomes \citep{pope2011s}. However, profile images are informative beyond the socio-demographic characteristics typically analyzed \citep{zhang2022makes}. We study how seemingly innocuous stylistic choices in profile images can influence both efficiency and across-group disparities on online platforms.

We distinguish between two categories of profile features:
fixed or difficult-to-change characteristics (\emph{type}, e.g., gender or age) and features that are easier to alter (\emph{style}, e.g., whether an individual smiles in the photo). The style features we consider are features that platform managers might encourage platform participants to change. If type and style choices are uncorrelated, two distinct sources of disparities might emerge on the platform: some types may transact more frequently, and some styles may be more successful. However, when type and style are positively correlated, disparities compound; when negatively correlated, they partially offset one another.

We analyze the type and style features of online profiles on Kiva, a non-profit micro-lending platform. On Kiva, individual lenders allocate capital to borrowers by selecting from a curated catalog of borrowing campaigns.\footnote{Technically, the loan is made to a microcredit institution and earmarked for the specific borrower.} In designing this marketplace, Kiva aims to balance efficiency, measured by the volume of transactions, with a notion of fairness, understood as the equitable distribution of capital among similar borrowers on the platform \citep{burke2022performance}. 

Fairness can be defined in various ways, depending on a platform’s objectives, the groups of interest, and the broader context \citep{kleinberg2016inherent,dwork2012fairness}. Implementing any fairness policy requires an empirical understanding of disparities  relevant groups. We focus on disparities by type, i.e., socio-demographic characteristics such as gender and age, that are commonly considered in fairness policies.

We develop a two-step approach to understanding how type and style profile features contribute to efficiency and type-based disparities, selecting features that are managerially relevant in the sense that the platform can design interventions based on them. In the first step, using observational data from the Kiva platform, we identify features that appear to matter to lenders in that they affect funding outcomes, and analyze the contribution of style features to type-based disparities. To extract profile image features, we apply an off-the-shelf machine learning algorithm that detects over one hundred features. By comparing the predictive performance of models trained with and without style features, we show that these features collectively predict funding outcomes. We further identify specific stylistic elements with a large and statistically significant impact on funding success, both unconditionally and after controlling for borrower characteristics and proxies for lender exposure. For example, a smile is associated with higher funding, while wearing sunglasses or featuring a body-shot-image correlates with lower funding outcomes.\footnote{A body-shot refers to an image where the person's body occupies a large portion of the frame.}

We next show that borrowers' style choices aggravate disparities between many types. First, style and type tend to be correlated - borrowers' from different socio-demographic groups systematically create different profiles. Male borrowers are less likely to smile and more likely to wear sunglasses than female borrowers, while young borrowers are more likely to both smile and wear sunglasses than older borrowers. Second, we carry out covariate decomposition of type-based disparities \citep{gelbach2016covariates} and show that style features jointly increase disparities between male and female or old and young borrowers. 

Estimates of the impact of profile features on outcomes from observational data rely on the assumption of unconfoundedness, that is, the assumption that conditional on other observable features of profiles, the assignment of style is as good as random, with no important omitted variables correlated with both the target feature and funding outcomes. This assumption is not directly testable, and there are reasons to question it in our context. Accordingly, we view our estimates from observational data as suggestive but not definitive, and we use our findings to prioritize certain type and style features to analyze further in recruited experiments.

The second step of our approach aims to provide internally valid estimates of the magnitude of the impact of profile features on funding decisions. We run two recruited experiments on Prolific.com, with 410 subjects in the first and 436 in the second, screened for English fluency and recent charitable giving to approximate the Kiva lender population. In each experiment, subjects make a series of pairwise choices between borrower profiles. We show them mostly artificial profiles, built from real images using Generative Adversarial Networks (GANs). GANs encode images into a latent space in which we can shift a single attribute while leaving the rest of the image unchanged, producing several variants of each original picture that differ in one profile feature and are otherwise visually identical. This is what allows the experimental estimates to be internally valid: within a pair, the only systematic difference between profiles is the feature under study.

The two experiments differ in the features they examine and in whether subjects' choices carry financial consequences. The first examines two style features (smile and body-shot) and one type feature (gender), using only hypothetical borrowers with AI-generated images. The second introduces financial stakes: subjects evaluate both hypothetical borrowers and real borrowers currently active on Kiva, and we allocate \$10 to a real borrower selected by each participant. This experiment analyzes three additional style features (sunglasses, glasses, and dark hair) and one additional type feature (age).

Across a range of empirical specifications, we find that a smile has a large and statistically significant positive effect on the probability of being selected by a subject. In contrast, wearing sunglasses or glasses has a negative and statistically significant impact. We also find suggestive evidence that body-shot has a negative effect and dark hair a positive effect, though these results are not statistically significant across all specifications. Finally, we find that experimental subjects prefer female borrowers over male borrowers, consistent with patterns observed in the Kiva observational data. The difference in selection probabilities between young and old profiles is small and not statistically significant, in contrast, to a statistically significant gap in favor of younger borrowers observed in Kiva data.

We then explore the mechanisms through which style features impact funding outcomes, distinguishing between monetary and non-monetary channels. To investigate the monetary channel, we examine whether style features predict loan repayment. Machine learning models trained on Kiva data show that adding style features does not improve repayment predictions, indicating they carry no financial information. To investigate the non-monetary channel, we estimate psychological traits from borrower images using the deep learning model of \cite{peterson2022deep}, trained on human judgments. To establish that style features causally shift these perceptions, we use GAN-generated images. We find that smiling raises predicted trustworthiness, happiness, and outgoingness by roughly one control-group standard deviation and lowers predicted dominance by about half. These findings are consistent with the hypothesis that lenders' decisions are shaped more by psychological perceptions of borrower images than by financial considerations.

Style features influence funding outcomes and contribute to disparities between borrowers. However, unlike fixed borrower characteristics, style features can be modified, making them a potential lever for platform policies aimed at reducing disparities or improving efficiency. We use the estimates of impact of image features on demand to examine counterfactual platform policies that modify the conditional distribution of style features in borrower profiles and adjust the probability of borrowers appearing in lenders' choice sets based on borrower characteristics. To assess these policies, we calibrate a model of lender demand using estimates from our recruited experiments. In our model, lenders are heterogenous with respect to style and type preference parameters. They, first, decide whether to participate and then choose one of the available borrowers. The choice set they observe depends on the platform policy.

A platform can act on style features in two ways: influence the photos borrowers post, or take submitted photos as given and adjust how prominently campaigns are shown. The two differ in their consequences. A style recommendation — encouraging borrowers to smile while avoiding body-shots — reduces disparities, as measured by a lower Gini coefficient and a smaller gender gap, and raises the total number of transactions. Increasing the visibility of campaigns with smiles and without body-shots — for example, ranking them higher in search — instead improves efficiency but widens disparities, because emphasizing these features disproportionately benefits borrowers of types that already receive higher funding. This second approach is also what a recommendation system trained on funding data would produce: it would prioritize profiles with the style attributes lenders favor. 

We examine a specific dimension of type-based inequity: the gender gap in favor of campaigns featuring female profiles.\footnote{Throughout, we use \emph{male} and \emph{female} to denote the gender classification assigned by the feature detection algorithm.} In observational data from the Kiva platform, campaigns featuring male profiles raise on average about \$30 less per day than those featuring female profiles, an unadjusted gap of approximately 25\% relative to the mean for female borrowers. The recruited experiment yields a consistent pattern: in pairwise comparisons of otherwise identical profiles, male profiles are selected approximately 32\% less often than female profiles.\footnote{Lenders may prefer campaigns with female profiles for several reasons. For example, extensive evidence suggests that women entrepreneurs who receive microfinance funding tend to use the funds effectively \citep{DESPALLIER2011758, AGGARWAL201555}. Additionally, lenders may seek to counteract gender discrimination in traditional entrepreneurial finance \citep{alesina2013women}.} The distribution of style features further amplifies this gap: 77\% of female borrowers have profiles with a smile, compared with 33\% of male borrowers, while 22\% of female borrowers have body-shot images, compared with 26\% of male borrowers. These differences in style feature distribution suggest that stylistic choices contribute to the observed gender gap in funding outcomes.

Our findings demonstrate that in marketplaces where users have preferences for certain profile image features, the correlation between type and style characteristics can matter for disparities and efficiency. Platforms seeking to balance these objectives need to account for this correlation before implementing policies based on profile images. 

The paper is organized as follows: Section \ref{lit_review} presents the related literature. Section \ref{context} describes how micro-lending platforms operate and provide institutional details about Kiva. Section \ref{offline_data} presents the observational data and its analysis. Section \ref{section_experiment} describes the design of the experiment and its results. Section \ref{sec:mechanisms} examines the mechanisms behind these effects, distinguishing monetary and non-monetary channels. Section \ref{simulations} focuses on counterfactual simulations, and Section \ref{conclusion} concludes.

\section{Literature Review}\label{lit_review}
Numerous papers document disparities in outcomes by race and gender (type features) in online platforms. Users with African-American-sounding names face higher cancellation rates on ride-sharing platforms \citep{ge2016racial} and Airbnb \citep{edelman2017racial}; drivers with Arabic or African-sounding names earn less on BlaBlaCar \citep{lambin2022impact}; Airbnb hosts with distinctively Asian names received fewer guests following the onset of the COVID-19 pandemic \citep{luca2024evolution}; non-Caucasian online profiles are shown fewer and different types of housing ads \citep{asplund2020auditing}; black and female NFT avatars are valued less \citep{yuan2024gender}; and women earn less on Lyft \citep{cook2021gender} and Airbnb \citep{davidson2023gender}. Specifically for platforms related to lending, several papers document disparities in funding outcomes by race, including on Prosper.com \citep{theseira2009competition, pope2011s} and Kickstarter.com \citep{younkin2018colorblind}. Beauty, which can be thought of as a function of both type (e.g., young) and style features (e.g., smiling), has also been shown to impact funding outcomes on Prosper.com \citep{ravina2008love}, as well as on Kiva \citep{park2019beauty}, the platform used in our study.

Another literature considers disparities by style features. For example, \cite{zhang2022makes} shows that Airbnb properties with verified photos have 9\% higher occupancy rates than those without verification, and \cite{zhang2025serving} shows that hosts who smile in their profile image have 3.5\% higher demand for their properties.\footnote{\cite{zhang2022makes} use Convolutional Neural Networks (CNN), the tool we use in this paper to identify image features, classify images as high or low quality, finding that including image quality in their linear regression model reduces the estimated treatment effect of verified photos by 41\%. They also define 12 image features based on the art and photography literature, which they sort into artistic categories like composition and color, that completely explain the difference between verified and unverified photos.} \cite{dupas2024keeping} shows that clothing and image background impact job interview chances. There is also evidence that different types choose different style features. For example, women write more, and more enthusiastically, in their profiles on OkCupid compared to men \citep{shishido2016tell}. Livestreamers' smiles and studio color design can affect their audience engagement and sales performance in the online live-streaming marketplace \citep{lin2021happiness, han2024unveiling}. In line with our results, \cite{haferkamp2012men} finds that women are less likely to choose body-shots than men in their online social media profile. 


Other papers evaluate how much of a disparity by type can be explained by other factors. Most closely resembling our own study, \cite{marchenko2019impact} evaluates whether features of the property listing \textit{text} on Airbnb, measured using natural language processing (NLP), can explain earnings disparities between minority and white male hosts. She finds that style features, specifically the subjectivity and polarity of the text, have at most a marginal impact on estimates after accounting for hard-to-change property features like location, property type, and number of bedrooms, which explain most of the disparity. Other studies find that immutable or difficult-to-change factors account for a portion of type-based disparities. For example, a Superhost designation and their rating, as well as guest ratings impact bi-directional racial disparities (i.e., preferences for own-race) \citep{zhang2022reducing}; number of guests accommodated, median home value, years of experience, and room type can account for more than half of the gender earning gap on Airbnb \citep{davidson2023gender}; reputation closes the gap between white and minority drivers on BlaBlaCar as minority drivers gain experience \citep{lambin2022impact}; facial femininity increases the disparity between women and men in whether potential customers online seek information about their tutoring services \citep{luo2024using}. We consider how different types making different choices in style features in profile images, features characterized by being easy to alter (particularly in comparison with other features listed in this paragraph), can explain disparities by type.

Furthermore, our paper uses a model to study how a policy encouraging all users to change mutable style features can both improve efficiency and decrease disparities compared to policies aimed directly at either objective. This result relates to the vast literature on improving the efficiency of algorithms, and recommendation systems in particular.\footnote{See, for example, \cite{lin2306can} for a recent survey on how large language models can improve recommendation system efficiency.} More relevantly to our paper, there is also evidence on how algorithms impact disparities by type. In one example, \cite{zhang2021can} shows that Airbnb's smart-pricing algorithm decreased the white-black host earnings gap by 71\%.
A more common theme in the literature is that algorithms may create or exacerbate disparities by type.\footnote{See \cite{williams2018algorithms} for a discussion of how a lack of data can cause algorithms to discriminate in a variety of ways; \cite{mehrabi2021survey} for a detailed description and categorization of different types of biases that can caused or exacerbated by algorithms.} For example, \cite{lambrecht2019algorithmic} shows that a gender-neutral ad for job opportunities in science, technology, engineering and math (STEM) fields was shown to fewer women than men by an algorithm optimizing for cost-effectiveness because young women are a more expensive demographic for ads. 

Current approaches for addressing algorithmic disparities by type generally intervene by changing either the training data, the algorithm, or the testing data post-processing \citep{mehrabi2021survey}.\footnote{\cite{chen2023bias} describe a similar categorization of interventions aimed at debiasing recommendation systems, namely a feedback loop involving data collection, model learning, and user serving.} Interventions that change training data use a technique such as reweighing or resampling the data to remove the disparity \citep{kamiran2012data}. Many approaches have been proposed that intervene at the algorithm level, such as algorithms that integrate fairness constraints \citep{naghiaei2022cpfair}, fairness regularizers \citep{berk2017convex, wang2024recommending}, and  fairness-efficiency trade-offs \citep{wang2021understanding}.
With respect to altering testing data post-processing, these interventions take the output of a model and alter it to reduce disparities through, for example, modifying embeddings that associate the words ``receptionist'' and ``female'' \citep{bolukbasi2016man}. Instead of following one of these three approaches, \cite{kleinberg2018algorithmic} proposes that interventions should be implemented post-estimation, where a policymaker can take the output of a model that optimizes efficiency and use it in a way that reduces disparities, for example, implementing minority-specific cut-offs for college admissions. We propose an intervention that leverages an existing managerial tool for Kiva and other platforms, style recommendations for profile images, to implement data-driven policy decisions that address disparities by type, disparities that exist given the training data, algorithm, and post-processing.

On Kiva, three papers have studied how to reduce disparities in terms of equitable funding outcomes, generally measured through funding to underfunded projects. Closest to our paper, \cite{burke2022performance} intervene outside of typical avenues by adding a 'slate' — a horizontal list of scrollable loans — of underfunded loans directly on the Kiva website, studying the impact on ``adds to basket'' (ATBs) to proxy for loans funded. They find that this additional slate of underfunded loans does not impact the ATBs on other slates, while attracting more than twice as many ATBs as the control slate from anonymous users. Two other papers study how algorithmic changes impact underfunded loans using collaborative filtering \citep{lee2014fairness}, and a combination of classification and the $\epsilon$-greedy algorithm \citep{hapek2021fairness}.

Another paper that bears similarities to ours is \cite{ludwig2024machine}, which uses a combination of machine learning to identify facial features in mug shots and humans on mTurk to label those features to generate testable hypotheses about disparities in judicial outcomes by facial features unrelated to type, like gender and age. They find that features previously studied in the literature, both type and other features that may be mutable (e.g., attractiveness, appearance of dominance), can explain about 22\% of the predicted variation in judge's detention rates, demonstrating that features identified through deep learning have predictive power beyond these previously studied features. \cite{sisodia2024generative} applies a similar methodology to identify impactful features of watches. Our approach is from the opposite direction, testing hypotheses suggested by our observational data analysis. 

One of the novel elements of this study is the use of generative AI, in the form of Generative Adversarial Networks (GANs), to create artificial profiles that vary a single feature of an image while holding other features fixed, in particular, type features like gender. Other approaches that have been used in the experimental economics literature to signal type features like gender and race include: names (e.g., \cite{bertrand2004emily}); images (e.g., \cite{andreoni2008beauty}), and subjects' physical identity in laboratory experiments (e.g., \cite{reuben2014stereotypes}). Generative AI tools such as GANs offer a different way to construct experimental stimuli. They allow the researcher to manipulate a specific visual attribute while preserving the rest of the image, making it possible to isolate the effect of features that are otherwise bundled together in naturally occurring profiles. Other work that uses GANs in an approach similar to the one described here include \cite{luo2024using} and \cite{sisodia2024generative}. 
\cite{dash2023review} surveys several applications of GANs to images, ranging from astronomy (e.g., to generate realistic simulations of deep space) to marketing (e.g., to generate many variations of a logo) to fashion design (e.g., to generate clothing designs).\footnote{See also \cite{jabbar2021survey} for additional applications of GANs in a wide variety of other domains.}


\section{The Setting, Institutional Background and Data
}\label{context}

Kiva is one of the most prominent online, non-profit, peer-to-peer microcredit platforms.\footnote{Zidisha, Lend with Care, and Lend a Hand are other major peer-to-peer microfinance platforms sharing many features with Kiva.}  Serving borrowers in more than 80 countries, Kiva has issued over 1.6 million loans, funded by more than 2 million lenders, totaling \$1.7 billion since its founding in 2005. On the Kiva marketplace, borrowers have individual profile pages featuring pictures that prospective lenders can browse when selecting borrowing campaigns to invest in. Kiva collaborates with local microcredit agencies to vet, curate, and promote borrowers. The platform aims to enhance efficiency, measured by the total number of transactions, while also reducing disparities, defined as achieving a more equitable distribution of funds across borrowers \citep{burke2022performance}.

\begin{figure}[!htb]
\centering
\caption{Kiva category page.}
\includegraphics[width=0.9\linewidth]{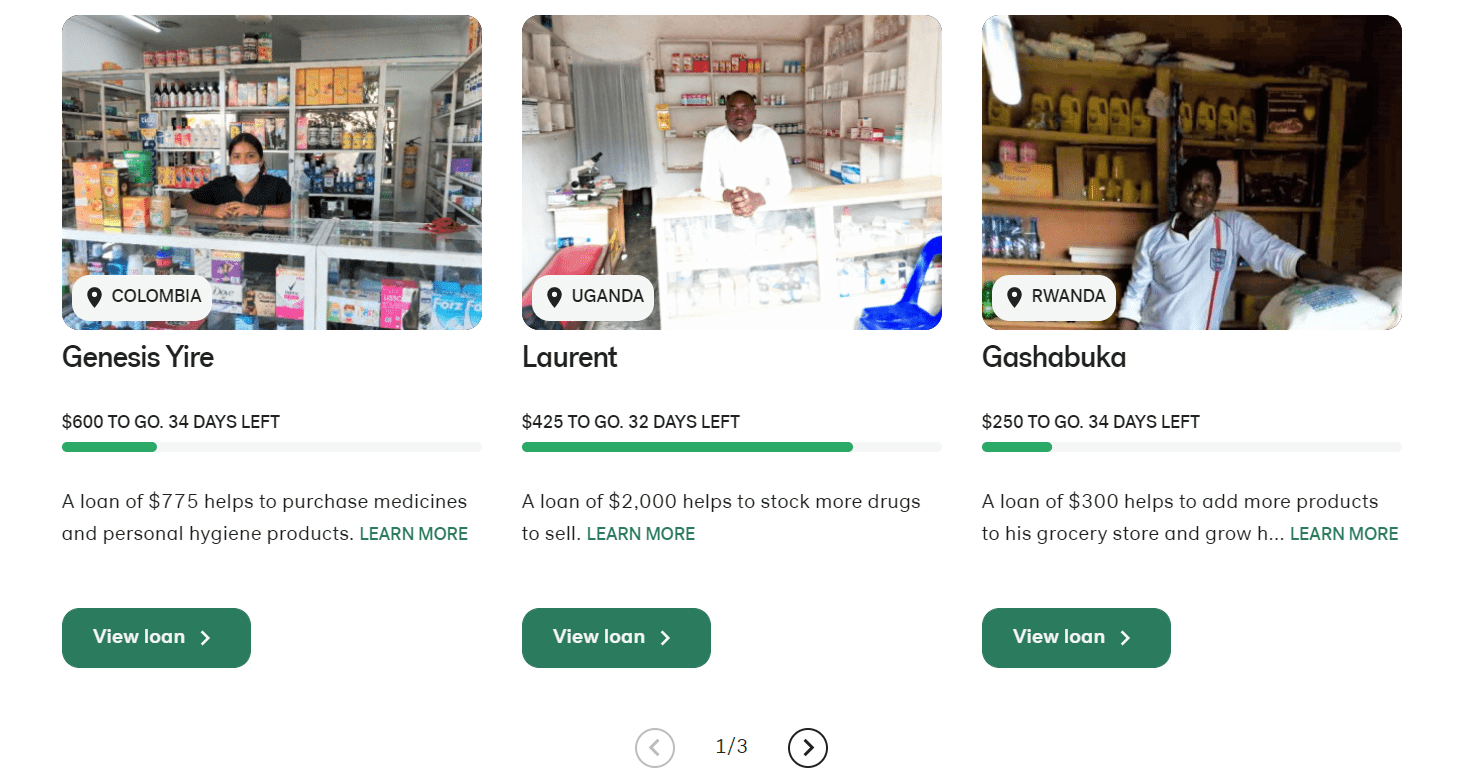}
\caption*{\footnotesize{\textit{Note: Screenshot from kiva.org collected on 3/3/2022.}}}\label{kiva_screenshot}
\end{figure}

Images play a significant role in the way lenders discover borrowers and help borrowers convey the reason they need a loan. When searching for prospective borrowers, a potential lender typically begins on the category page, as illustrated in Figure \ref{kiva_screenshot}. By clicking on "View loan," lenders can access more information about the loan's purpose and the geographical location of the borrower. These details inform lenders' investment decisions \citep{park2019beauty}.

Borrowers submit photographic images that can vary greatly in quality, content, and composition. Some images primarily feature the borrower, while others showcase the borrower's business. Additionally, the facial expressions of borrowers as well as technical elements like lighting and resolution can differ noticeably. To assist borrowers in presenting themselves effectively in this important application component, Kiva offers several recommendations. These include using high-resolution photos, ensuring a horizontal orientation, and incorporating both the business owner and their business in the background.\footnote{See \url{https://www.kivaushub.org/profile-photo}.}

\subsection{Kiva Data}\label{kiva_data}
We construct \emph{Kiva data} by combining three datasets: a publicly available dataset with loan characteristics and lending outcomes, a dataset that captures features in images associated with the borrowing campaigns, obtained using the methodology described in Appendix \ref{appendix_features}, and a dataset on repayments provided by Kiva.\footnote{See here: \url{https://www.kiva.org/build/data-snapshots} for the publicly available dataset.}

\paragraph{Public Available Data.} The publicly available dataset covers borrowing campaigns from April 2006 to May 2020, with over 500,000 observations. Each observation corresponds to a borrowing campaign, the primary unit of analysis. It includes key attributes such as sector, name of activity, country, loan amount, and repayment schedule. We also observe when the campaign was posted, from which we construct weekly and monthly time fixed effects and interaction terms between month and sector and month and country to control for sector- and country-specific fluctuations in loan availability and demand. We observe all borrowing campaigns that were active on Kiva at a point in time. From these, we construct measures of market conditions and competition faced by each campaign, including the total number of borrowing campaigns available at a given time, the number of concurrently listed borrowers from the same country and sector, and the number of borrowers of the same race and gender. At the time our data was collected, Kiva was displayed borrowers chronologically within categories. We thus include interactions of time-fixed effects and categories to proxy for exposure. We also construct a measure of lender supply, specifically the number of active lenders per week.
For funding outcomes, we focus primarily on money collected per day, as it captures how lenders allocate capital among competing campaigns. We also examine the number of days it took to raise the capital (campaigns generally stay active until they collect all funds), and the number of lenders that loaned money to the borrower.\footnote{In Appendix \ref{alt_outcome_appendix}, we expand the set of outcome variables and consider a constructed variable, which adjusts for differences across requested loans and funding success across categories.}

\paragraph{Image Data, Type-Style Classification, and Extraction.} Borrowers on Kiva can upload profile images, which are publicly visible to prospective lenders. We process these images using a Convolutional Neural Network (CNN) to extract structured visual attributes, which we classify into type and style features.\footnote{In the field of computer vision, it is customary to differentiate between mutable and immutable aspects of facial images. This distinction is often guided by frameworks such as those provided by the Facial Identification Scientific Working Group (FISWG), which offers comprehensive guidelines for facial comparison methodologies \url{https://www.fiswg.org/fiswg_facial_comparison_overview_and_methodology_guidelines_V1.0_20191025.pdf}}

We distinguish type and style features based on the effort required to change them. Type features, such as age, race, and facial structure, are intrinsic and biologically determined, making them difficult to alter without significant intervention.  Although borrowers might alter their appearance in a photograph, such a misrepresentation might incur psychological costs,\footnote{For example, there is robust evidence from the field \citep{abeler2014representative} and lab \cite{kajackaite2017incentives} that there are costs to lying.} and the intermediaries who screen borrowers might object. In contrast, style features are extrinsic and more easily modified, varying across images due to individual choices, environmental conditions, or temporary factors. Because style features can be adjusted with relatively low effort, they are particularly relevant for platform interventions, as platforms can influence them through guidance on image composition and presentation.\footnote{See (\url{https://www.kivaushub.org/profile-photo}) for Kiva guidelines for profile images highlighting image composition and style, and (\url{https://www.airbnb.com/resources/hosting-homes/a/how-to-take-a-great-airbnb-profile-photo-581}) for Airbnb's recommendations focused on facial expressions and composition.} In our analysis, we assume the platform, as the decision-maker, can identify which image attributes are managerially relevant and which are not. Some features may be deemed unsuitable for modification due to the substantial effort required from borrowers or the potential for such recommendations to be perceived as insensitive or politically incorrect.

Examples of features that we classify as style include: \emph{No Eyewear}, \emph{Sunglasses}, \emph{Smile}, \emph{Blurry}, \emph{Eyes Open}, \emph{Dark Hair}, \emph{Mouth Wide Open}, \emph{Harsh Lighting}, \emph{Flash}, \emph{Soft Lighting}, \emph{Outdoor}, \emph{Partially Visible Forehead}, \emph{Color Photo}, \emph{Posed Photo}, \emph{Flushed Face}, \emph{Top} (person’s face in the top part of the image), \emph{Right} (person’s face in the right part of the image), \emph{Bottle} (there is a bottle in the image), \emph{Chair} (there is a chair in the image), \emph{Person} (there is another person in the image), and \emph{Body-shot} (the body of the borrower occupies a substantial part of the image).

The CNN assigns probability scores to approximately 140 features, including object presence (e.g., \emph{cup}, \emph{chair}), technical aspects of the image (e.g., \emph{blurry}, \emph{flash}), facial expressions (e.g., \emph{smiling}, \emph{frowning}), and demographic characteristics such as \emph{race} and \emph{age}. We filter out features that appear in fewer than 0.01\% of images and remove highly correlated features (those with a Pearson correlation coefficient above 0.75, such as \emph{smiling} and \emph{frowning}). After these steps, we retain 55 key features for analysis (full list in Appendix~\ref{appendix_features}). Throughout the paper, we use \emph{italics} when referring to demographic features predicted using CNN.

CNN-based feature extraction introduces the potential for misclassification, particularly for complex or ambiguous image attributes. To assess the accuracy of our image-based features, we conduct an audit study comparing algorithmic predictions with human-annotated labels (Appendix~\ref{app:measurement_error}). The audit reveals systematic non-classical errors, with a higher incidence of false negatives than false positives, implying that the algorithm is more likely to miss features that are present than to falsely detect them. To quantify and correct for this bias, we implement a Simulated-Extrapolation (SIMEX) procedure that uses the feature-specific false-positive and false-negative rates estimated in the audit. The resulting SIMEX-adjusted estimates provide a measurement-error-corrected benchmark, allowing us to assess how classification errors affect the magnitude of the estimated treatment effects.

\paragraph{Loan Repayment Data and Defaults.} The repayment dataset spans 2006–2016, covering approximately 420,000 borrowing campaigns. In this dataset, each observation corresponds to an individual lender’s contribution to a borrower. We aggregate these data to construct an indicator for whether a campaign was fully repaid, meaning all lenders who contributed received their funds back; 95\% of campaigns are fully repaid. Defaulted loans are categorized as follows: (i) borrower default (75\%), where the borrower fails to repay; (ii) microfinance partner default (23\%), where the intermediary organization managing the loan defaults; and (iii) joint default (2\%), where both the borrower and the microfinance partner default. Our main analysis pools all defaults regardless of cause; Appendix \ref{default_types} repeats it separately by default category.

\paragraph{Data Merging, Platform Adjustments, and Considerations.} We merge the three datasets to create a panel covering 2006–2016, enabling us to track borrowing campaigns, funding dynamics, and repayment outcomes over time. Since Kiva has undergone multiple platform design changes during this period, we include time-fixed effects in all our analyses to account for structural shifts in borrower composition, lender behavior, and macroeconomic conditions.

Several limitations should be noted. For instance, while we observe lender funding decisions, we do not have direct data on how lenders search for or browse campaigns, which limits our ability to account for visibility effects. Additionally, our repayment data extends only through 2016, restricting our analysis of long-term repayment patterns.

Table \ref{sum_stats_main} presents summary statistics for the key variables, and a full list of covariates is provided in Appendix \ref{sum_stats_all}.

\begin{table}[!htbp] \centering 
  \caption{Summary statistics of the main variables} 
  \label{sum_stats_main} 
\begin{tabular}{@{\extracolsep{5pt}}lccccc} 
\\[-1.8ex]\hline 
\hline \\[-1.8ex] 
Statistic & \multicolumn{1}{c}{N} & \multicolumn{1}{c}{Mean} & \multicolumn{1}{c}{St. Dev.} & \multicolumn{1}{c}{Min} & \multicolumn{1}{c}{Max} \\ 
\hline \\[-1.8ex]
cash per day & 420,908 & 108.040 & 150.314 & 0.833 & 756.250 \\ 
days to raise & 420,908 & 13.204 & 11.015 & 1 & 39 \\ 
loan amount & 420,908 & 800.411 & 1,000.102 & 25 & 50,000 \\ 
default & 420,908 & 0.050 & 0.218 & 0 & 1 \\ 
number of lenders & 420,908 & 0.012 & 0.015 & 0.001 & 1.000 \\ 
number of competitors & 420,908 & 0.088 & 0.164 & 0.004 & 1.000 \\ 
\emph{male} & 420,908 & 0.478 & 0.330 & 0.003 & 0.999 \\ 
\emph{youth} & 420,908 & 0.264 & 0.211 & 0.000 & 0.982 \\ 
\emph{smiling} & 420,908 & 0.549 & 0.177 & 0.013 & 0.966 \\ 
\emph{body-shot} & 420,769 & 0.735 & 0.441 & 0 & 1 \\ 
\emph{glasses} & 420,908 & 0.197 & 0.398 & 0 & 1 \\ 
\emph{sunglasses} & 420,908 & 0.354 & 0.478 & 0 & 1 \\ 
\emph{dark hair} & 420,908 & 0.499 & 0.500 & 0 & 1 \\   \hline \\[-1.8ex] 
\end{tabular} 
\caption*{\footnotesize{\textit{Note: Summary statistics of selected variables. Cash per day and days to raise are Winsorized at the top 97th percentile. Cash per day and loan amount are in USD; male and smile take the value of 1 when CNN predicted probability is above 0.5 and zero otherwise. No. competitors is the number of borrowing campaigns from the same sector and country posted concurrently. Both the number of lenders and the number of competitors are standardized by the maximum.}}}
\end{table}

\section{Funding Outcomes and Profile Features in Kiva Data}\label{offline_data}

This section presents the first step in our two-step approach to understanding how type and style features contribute to platform efficiency and type-based disparities. Using observational data from Kiva, we document disparities in funding outcomes, identify style features that appear to affect funding, and quantify how style differences contribute to type-based disparities. The end product is a prioritized list of style features that we examine further in the recruited experiments of Section \ref{section_experiment}.

\subsection{Disparities in Funding Outcomes}\label{inequity_offline}

Funding outcomes on Kiva are unequal in the pooled sample, within contemporaneous lender choice sets, and across borrower types. We document each in turn.

\begin{figure}[htp]
  \caption{Days to raise: histogram and Lorenz curve}\label{fig:days_to_raise}
  \centering
  \includegraphics[width=.39\textwidth]{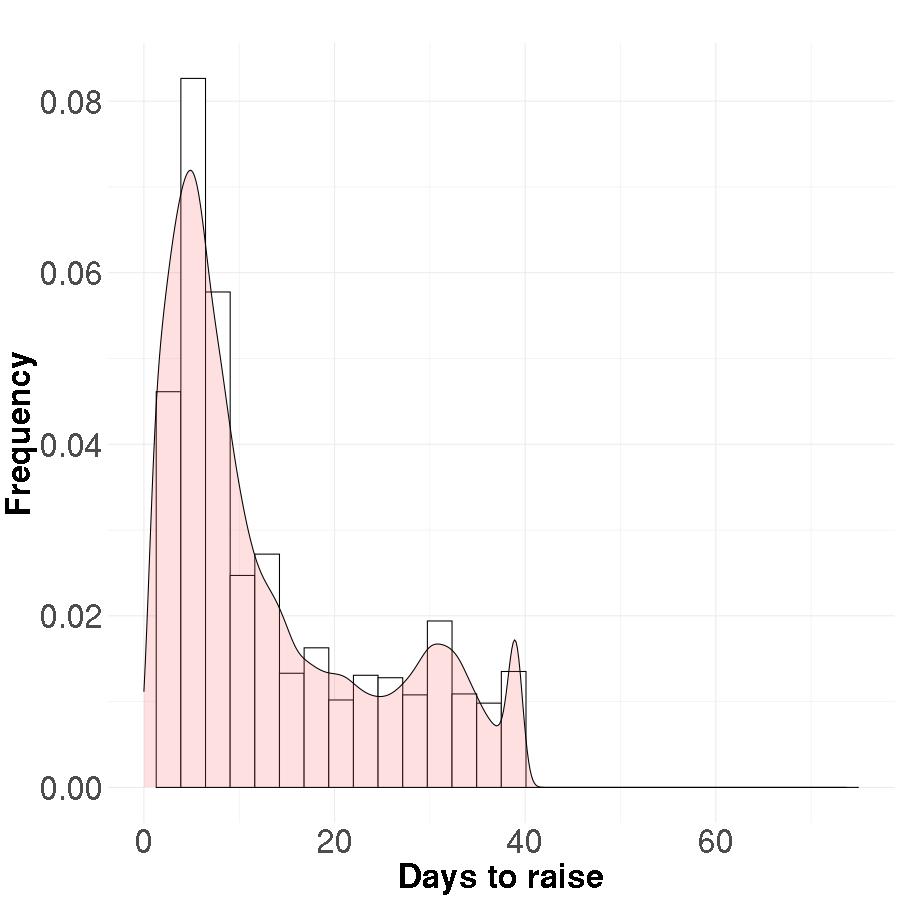}\hfill
  \includegraphics[width=.39\textwidth]{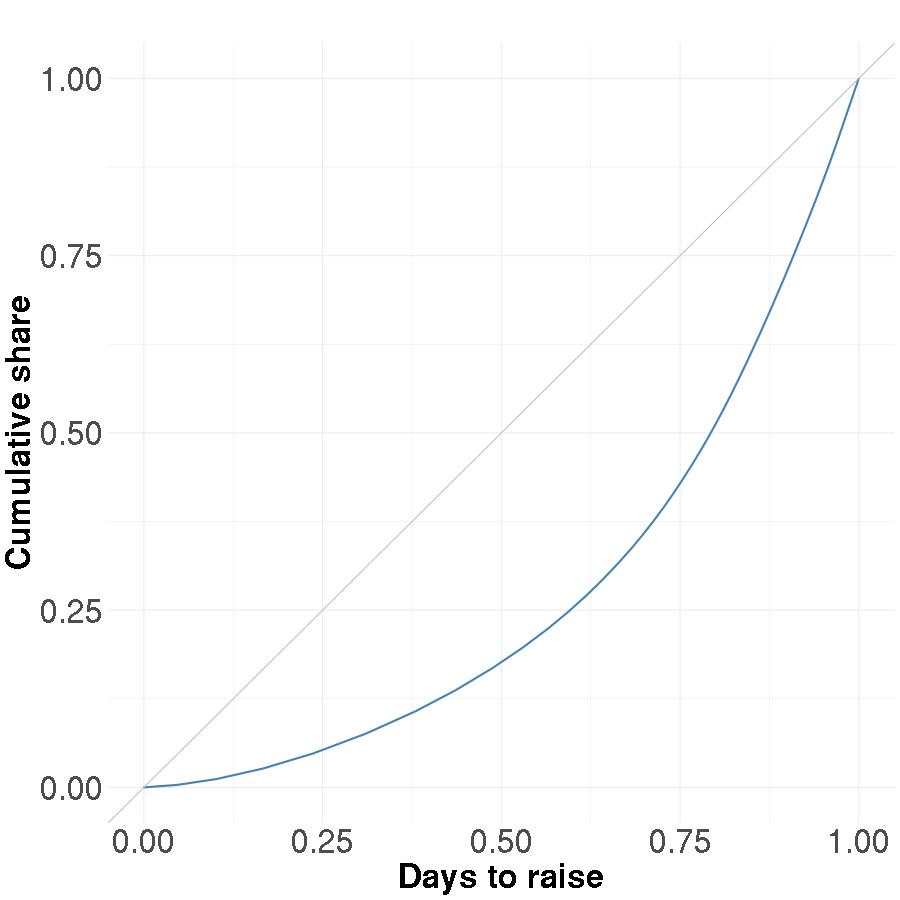}
  \caption*{\footnotesize{\textit{Note: Left panel: histogram of days to raise. The fitted density curve is shown in pink. Right panel: Lorenz curve of days to raise.}}}
  \end{figure}

  \paragraph{Days to raise.}
  Figure \ref{fig:days_to_raise} reports the distribution of days to raise
  -- the number of days a campaign needs to collect its full requested amount.
  Some campaigns fund within hours; others remain open for over a month. The
  Lorenz curve in the right panel lies well below the 45-degree line. Funding
  speed is far from equal.

\begin{figure}[htp]
  \caption{Cash per day: histogram and Lorenz curve}\label{fig:cash_per_day}
  \centering
  \includegraphics[width=.39\textwidth]{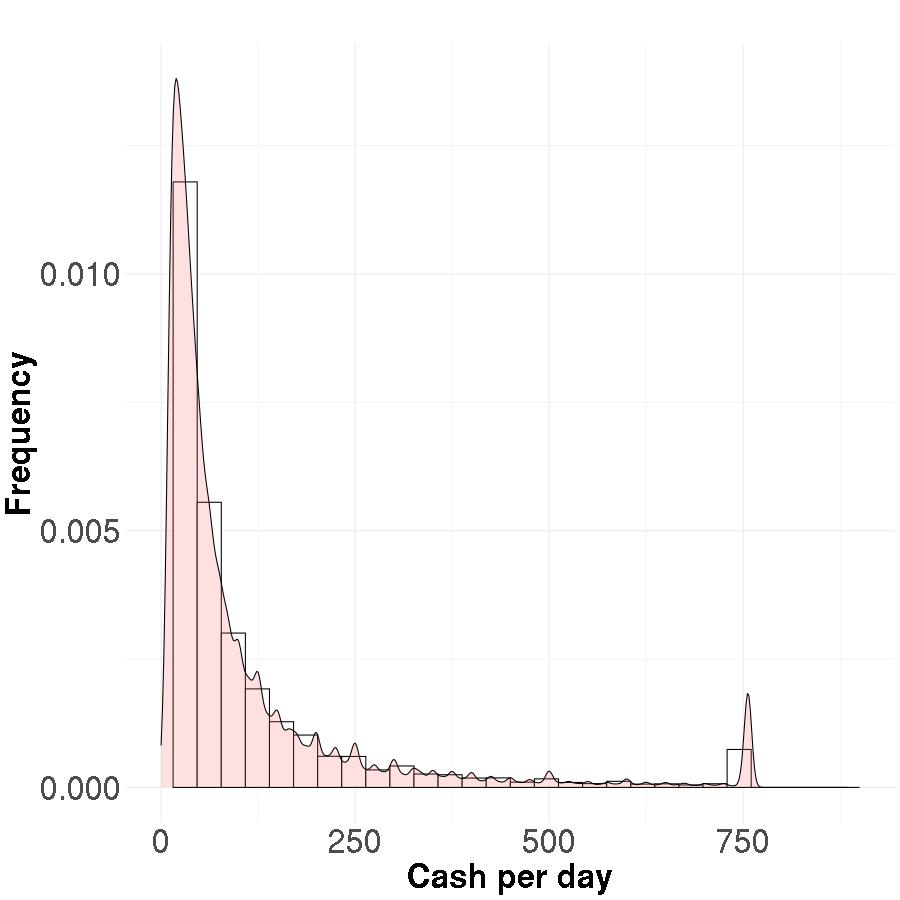}\hfill
  \includegraphics[width=.39\textwidth]{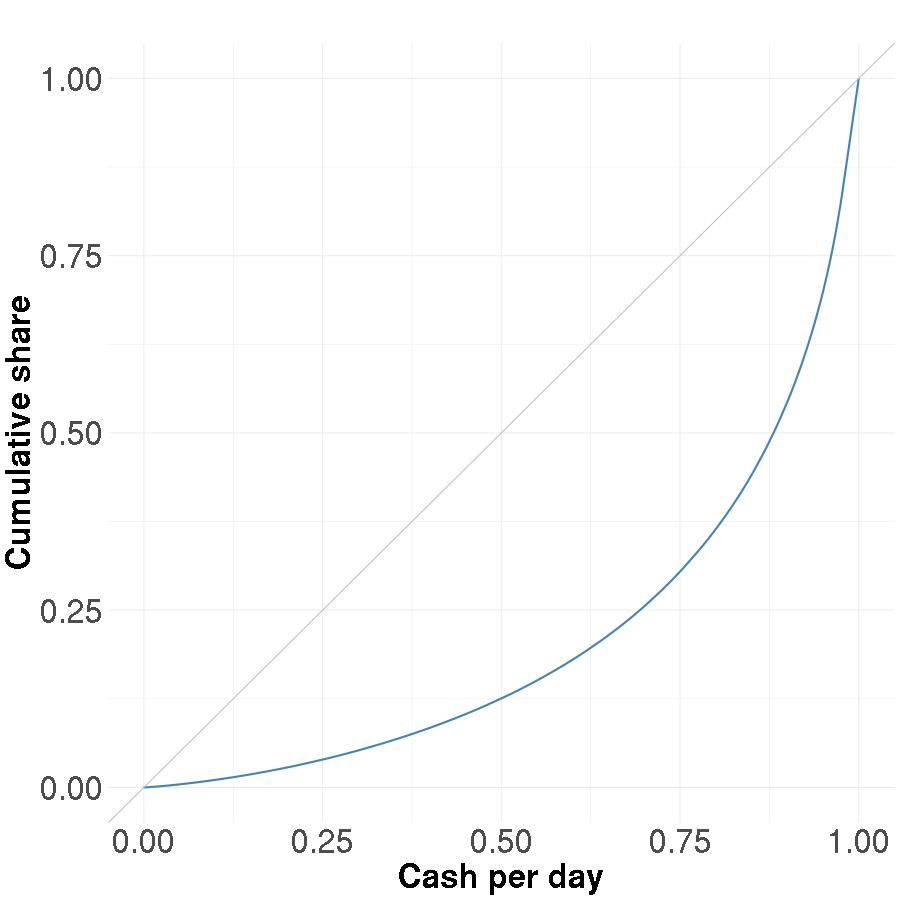}
  \caption*{\footnotesize{\textit{Note: Left panel: histogram of cash per day capped at 1250 USD. The fitted density curve is shown in pink. Right panel: Lorenz curve of cash per day.}}}
  \end{figure}

\paragraph{Cash per day.}
Larger loans mechanically take longer to fund, so days to raise mixes campaign size with funding speed. We scale by loan size and define cash per day: dollars raised per active day on the platform. Figure \ref{fig:cash_per_day} reports its distribution. The mean is \$108, with substantial mass below \$10 per day and substantial mass above \$400. The Lorenz curve in the right panel shows even greater inequality than for days to raise.

\begin{figure}[h]
  \caption{Cash per day distribution within weeks: Gini coefficient and share of the bottom tertile.}\label{fig:market_lvl_inequities}
  \centering
  \includegraphics[width=.39\textwidth]{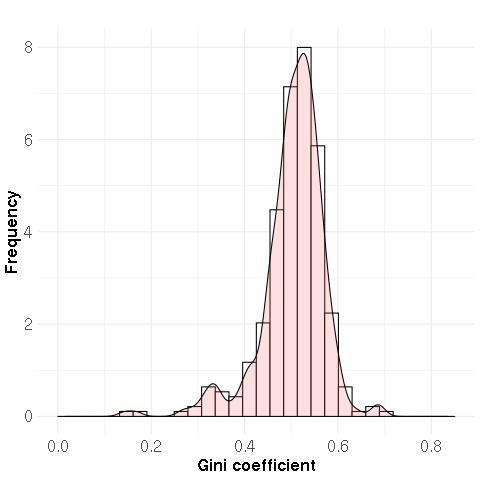}\hfill
  \includegraphics[width=.39\textwidth]{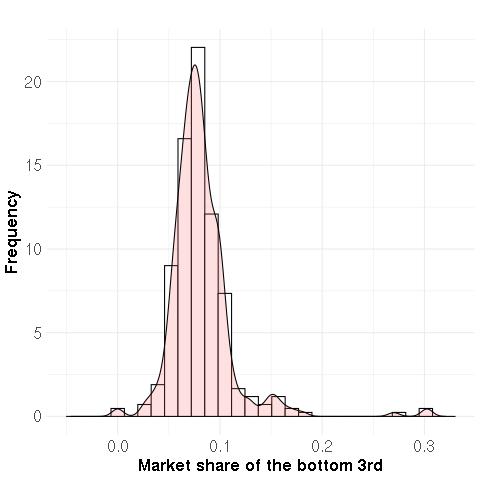}
  \caption*{\footnotesize{\textit{Note: Statistics in both panels are computed on a weekly basis. Left panel -- Gini coefficients of weekly distributions of cash collected per day. Right panel -- weekly sums of cash
   collected per day by the 33\% lowest performing borrowers.}}}
  \end{figure}

  \paragraph{Within-week disparities.}
  The pooled data span ten years, so secular shifts in lender supply
  or borrower entry could account for some of the dispersion. To net out these
  aggregate fluctuations, we compute disparity statistics within weekly cohorts
  of borrowers. Campaigns listed in the same week approximate the choice set
  facing a contemporaneous lender.\footnote{The approximation is imperfect: a
  campaign posted Monday and fully funded by Wednesday is invisible to a lender
  arriving Friday. The average week contains 450 active campaigns and over
  \$400{,}000 in pledged loans.} We summarize within-week dispersion with two
  statistics: the Gini coefficient of cash per day, and the dollar share
  collected by the bottom tertile of borrowers.\footnote{The Gini coefficient is
  $\text{Gini} = \frac{\sum_{j=1}^{n}\sum_{j'=1}^{n}|x_j - x_{j'}|}{2n^{2}\bar{x}}$,
  where $x_j$ is borrower $j$'s outcome, $n$ is the number of borrowers active
  that week, and $\bar{x}$ is the weekly mean. A Gini of zero indicates perfect
  equality; a bottom-tertile share of $33\%$ indicates equal distribution across
  tertiles.} Figure \ref{fig:market_lvl_inequities} plots both series across the
  sample period. In every week, the bottom tertile collects far less than
  one-third of the dollars, and the weekly Gini remains persistently elevated.
  Disparities do not wash out within the contemporaneous market.

  \paragraph{Disparities across borrower types.}
  The disparities also align with demographic lines. Male borrowers raise \$30.2 less per
  day than female borrowers (SE 0.9); seniors raise \$57.2 less than younger borrowers
  (SE 4.7); and Black borrowers raise \$12.1 less than non-Black borrowers
  (SE 1.4). Relative to the sample mean of \$108 per day, these gaps are
  large -- roughly $28\%$, $53\%$, and $11\%$ of mean daily fundraising,
  respectively. The remainder of the paper decomposes each type-level gap into
  the share attributable to differences in style.
\subsection{Style Features and Funding Outcomes}\label{style_funding}

Disparities in borrower outcomes can arise due to differences in borrowers' types and their style. Style features are central to this analysis because a platform can design interventions to modify them. We first ask whether style features in profile images jointly predict funding outcomes.

We train three gradient-boosted machine (GBM) models of cash per day \citep{friedman2001greedy}: a mean-only benchmark, a model with style features only, and a full model using all variables in the Kiva data.\footnote{Appendix \ref{pred_model} compares GBM accuracy to other predictive models.} We split the data 70:30 into train and test sets. Table \ref{tab:image_pred} reports test-set mean squared error. Adding style features improves predictive performance over the mean model, and the full model improves further on the style-only model. Appendix \ref{GBM_diagnostics} provides GBM diagnostics and extends the analysis to an outcome that adjusts for category-level differences in cash per day and loan size; style features remain predictive within business categories.

\begin{table}[h] \centering 
  \caption{Image Features as Predictors of Cash per day.} 
  \label{tab:image_pred} 
\resizebox{0.30\textwidth}{!}{%
\begin{tabular}{lrr}
\\[-1.8ex]\hline 
\hline \\[-1.8ex] 
Specification & MSE & SE\\
\midrule
Mean & 22841 & 202\\
Style features & 17559 & 159\\
Full model & 14846 & 132\\
\bottomrule
\end{tabular}
}
\caption*{\footnotesize{\textit{Note: Test set performance of a GBM trained using all available covariates (full model), models with only image style features, and a mean model. Mean squared errors are in the second column. Standard errors of MSE are in the third column.}}}
\end{table}

\paragraph{Specific style features.} Results presented in Table \ref{tab:image_pred} show that style features are predictive of funding outcomes. However, if Kiva is to design platform policies around style features, we need to identify individual features with substantial effects. We want to know: "What would happen if a profile were presented with a change in one characteristic and remained unchanged otherwise." In other words, we want to know the average treatment effect (ATE) of a specific feature in an image.

To estimate ATEs we use the Augmented Inverse Propensity Weighing (\emph{AIPW}) estimator \citep{robins1994estimation,glynn2010introduction}. AIPW is a doubly robust method: it adjusts for covariates in the outcome model and the propensity score. We use the \emph{grf} implementation of the AIPW estimator \citep{athey2019generalized}. We consider a rich set of covariates: we control for \emph{(i.)} the requested amount of loan, the week in which the fundraising campaign was posted on Kiva, sector and country of the business, repayment schedule, time fixed effects and interaction terms between the month and sector and the month and country; \emph{(ii.)} all type and other style features, \emph{(iii.)} measures of the other loans available at the time on Kiva (we observe the entire choice set available to the lender), specifically, the total number of loans available and the number of loans from borrowers of the same race and of the same gender, \emph{(iv.)} the number of lenders active on Kiva in the specific week to account for inter-temporal differences in the supply of money. In the estimation of the effect of each feature, we drop other features that are very highly correlated, where it may be difficult to hold the highly correlated features fixed while changing the target feature. Then, we interpret our results as the treatment effect of the relevant feature and other features that covary strongly with it.

Some style features are observed only in specific subsets of the data, such as particular sectors or countries. As a result, it is not possible to credibly estimate the impact of these features on borrowers outside these subsets. To address this limitation, we estimate a propensity score function and exclude cases where either the treatment group (borrowers with the feature in the Kiva data) or the control group (borrowers without the feature) falls outside the range of common support. Specifically, we drop features for which a large probability mass of either group exceeds 0.9 or falls below 0.1. Appendix \ref{density_out} provides density plots illustrating examples of the features removed through this process. Furthermore, we also report estimates adjusted for overlap following the methodology of \cite{li2018balancing}.

\begin{figure}
    \centering
    \caption{Estimates of the Average Treatment Effect of Selected Style Features}

    \includegraphics[scale = 0.6]{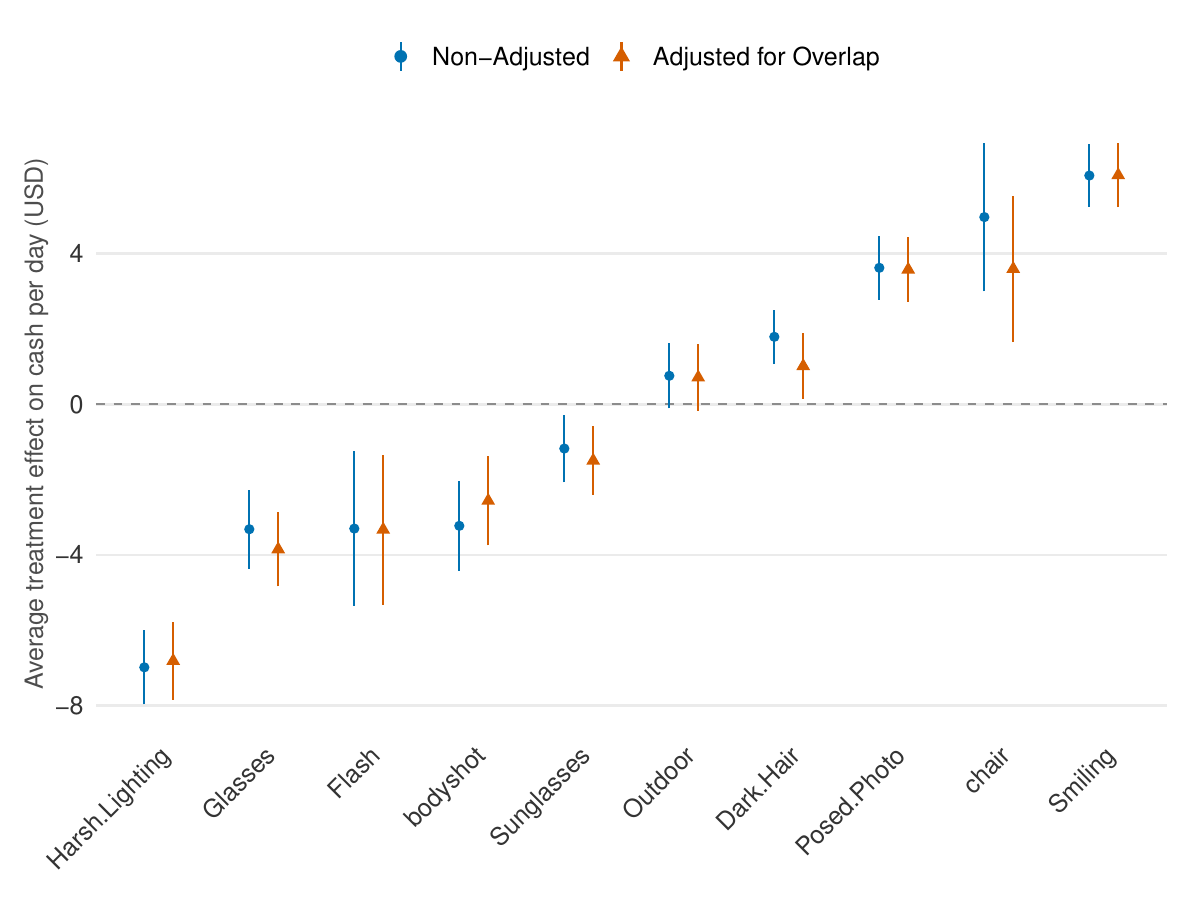}
    \caption*{\footnotesize{\textit{Note: Estimates of the average treatment effect of selected features on cash collected per day with 95\% confidence interval. The propensity and outcome model was estimated using Regression Forest. We transform the treatment variable to a binary variable that takes the value of one when the predicted probability of the feature is above 0.5 and zero otherwise.}}}\label{fig:ATE_cash}
\end{figure}

Figure \ref{fig:ATE_cash} shows estimates of average treatment effects on cash per day for selected features. We find that several features have negative ATE, like \emph{Body-shot} and \emph{Sunglasses}, while others like \emph{Posed Photo} or \emph{Smiling} have positive effects.

From the perspective of a causal diagram or directed acyclic graph (DAG), certain image-extracted features may, in principle, be \emph{caused} by a given \emph{treatment feature}. In such cases, these features would moderate the effect of the treatment feature and should not be included as controls in an AIPW estimator. For instance, if \emph{Sunglasses} is the treatment feature, then \emph{Bags under eyes} may only be observed when \emph{Sunglasses} = 0, making it inappropriate as a control. Similarly, if \emph{Lighting} is the treatment feature, other features, such as \emph{Age}, may only be detected by the algorithm under favorable lighting conditions. To assess the robustness of our estimates to this concern, we conduct a sensitivity analysis in Appendix~\ref{ate_robust_mediators}, where we exclude features potentially subject to this type of mechanism. The results indicate that our treatment effect estimates remain stable.

In the estimation of the ATE, we are assuming unconfoundedness. While we control for a rich set of borrower and market characteristics, we may still be missing some variables that might be correlated with the treatment variable and also influence lenders' decisions. Thus, our estimates should be interpreted as suggestive rather than definitive. The analysis of the observational data constitutes the first step in our approach; we use the results to select features to be tested in the recruited experiment, which has greater internal validity (omitted variable bias eliminated, in principle) at some cost to external validity (given the artificial context of the experiment).

\subsection{Style Features Impact Disparities Between Types.}\label{corr_type_style}

Borrowers' style choices can aggravate or mitigate inequities across types. When borrowers with type features associated with high outcomes also choose attractive style features, disparities widen; when borrowers with less desirable type features choose attractive styles, disparities narrow. This section documents the correlation between types and styles and quantifies how much of the observed disparity between types is accounted for by differences in style.

\begin{figure}
    \centering
    \caption{Correlation Between Selected Type and Style Features.}
    \includegraphics[scale = 0.5]{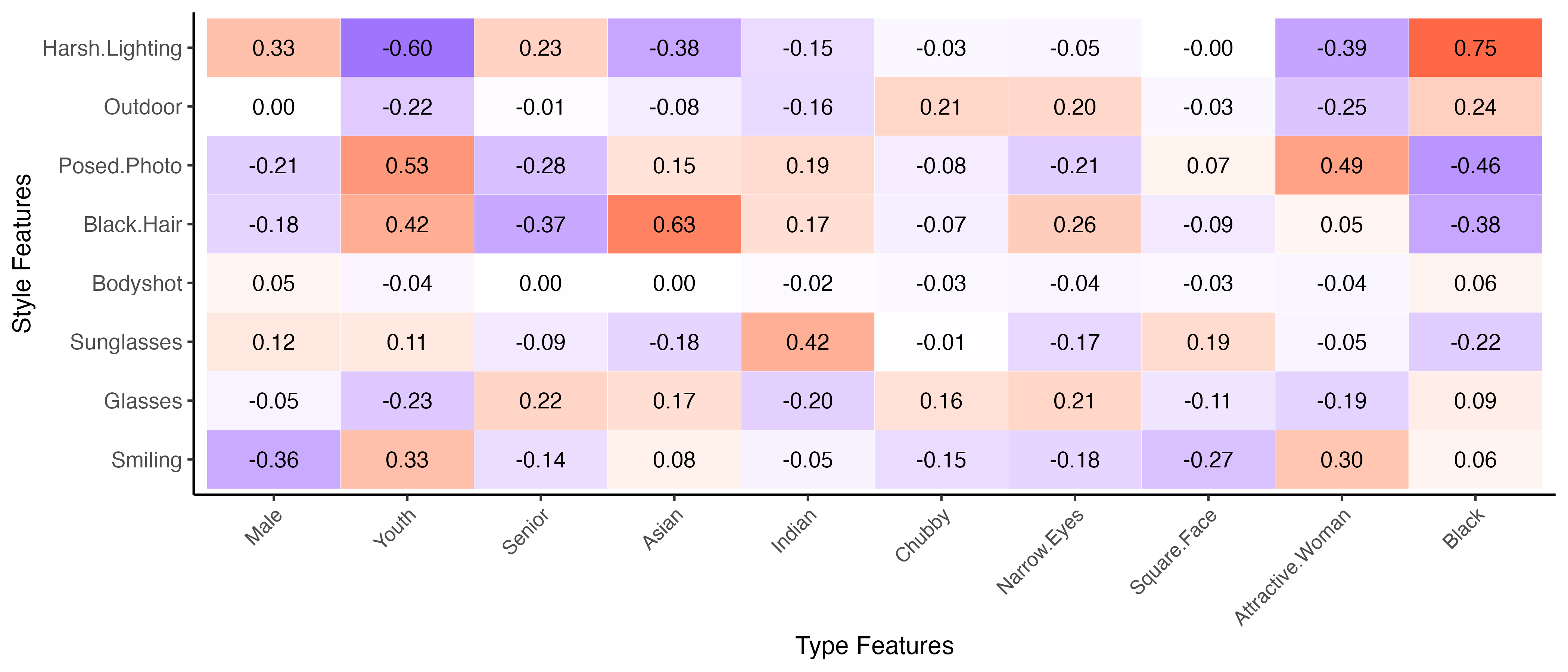}
    \caption*{\footnotesize{\textit{Note: Pearson correlation coefficient between selected type features in columns and style features in rows.}}}\label{fig:correlogram}
\end{figure}

Figure~\ref{fig:correlogram} shows that several type and style features are correlated: \emph{Smiling}, for example, is less common among \emph{Male} and \emph{Senior} borrowers and more common among \emph{Youth}. These unadjusted correlations are suggestive but not sufficient: to assess whether style choices exacerbate type-based disparities, we need to show that the more attractive style features are systematically more prevalent among borrowers whose types already lead to better funding outcomes. We do this with a \emph{Gelbach decomposition} \citep{gelbach2016covariates}, which measures how the coefficient on a type feature changes as additional covariates are introduced.

Table~\ref{tab:corr_type_style} reports the decomposition for selected type features. \emph{Coefficient base} is the coefficient on the type feature in a univariate regression of cash per day; \emph{Coefficient full} is the coefficient from a model adjusting for all variables in the Kiva data; \emph{Delta style} isolates the contribution of style features to the change between the two. Style aggravates a disparity when \emph{Delta style} shares the sign of \emph{Coefficient base} and mitigates it otherwise.
 
\begin{table}[!htbp] \centering 
  \caption{Gelbach decomposition of type-based disparities in cash per day} 
  \label{tab:corr_type_style} 
\resizebox{0.9\textwidth}{!}{
\begin{tabular}{lccccc}
\\[-1.8ex]\hline 
\hline \\[-1.8ex] 
Feature & Coefficient base & Std. error base & Coefficient full & Std. error full & Delta style\\
\midrule
\textit{Male} & -26.722 & 0.796 & -32.642 & 0.701 & -3.697\\
\textit{Bald} & -91.250 & 3.000 & -88.048 & 3.155 & -12.402\\
\textit{Chubby} & -2.372 & 1.110 & 24.536 & 1.219 & 2.120\\
\textit{Narrow Eyes} & -11.400 & 1.135 & -10.069 & 1.136 & 0.134\\
\textit{Square Face} & -146.834 & 5.061 & -25.030 & 5.667 & -28.291\\
\textit{Senior} & -39.873 & 2.404 & -61.925 & 2.918 & -18.884\\
\textit{Attractive Woman} & 28.299 & 1.356 & 90.617 & 1.530 & 19.571\\
\textit{Asian} & 17.322 & 0.817 & -7.157 & 0.888 & -2.164\\
\textit{Indian} & -15.520 & 2.115 & 16.708 & 2.375 & -12.808\\
\bottomrule
\end{tabular}
}
\caption*{\footnotesize{\textit{Note: Gelbach decomposition of selected type features \citep{gelbach2016covariates}. \emph{Coefficient base} is the coefficient from a univariate regression of cash per day on the selected type feature; Coefficient full is the coefficient from a model adjusting for all covariates in Kiva data; Delta style is the contribution of style features to the change. We use the R implementation by \cite{gelbachimplementation}.}}}
\end{table}
Style choices aggravate disparities for \emph{male}, \emph{bald}, \emph{square face}, \emph{senior}, \emph{attractive woman}, and \emph{Indian} types and mitigate them for \emph{chubby}, \emph{narrow eyes}, and \emph{Asian} types. The aggravating contributions are substantial in magnitude: \emph{Delta style} reaches USD 28.3 for \emph{square face}, USD 19.6 for \emph{attractive woman}, and USD 18.9 for \emph{senior}, with further substantial contributions for \emph{Indian} (USD 12.8) and \emph{bald} (USD 12.4); the largest mitigating contribution is USD 2.2 for \emph{Asian}. The gender gap, which also reflects an aggravating contribution from style features, receives a full decomposition across all groups of covariates in Appendix~\ref{app:gender_decomp}.

\section{Recruited Experiments}\label{section_experiment}

Our analysis of the Kiva data identifies a set of candidate style features, those that appear to move funding outcomes and to contribute to type-based disparities. Those estimates rest on an unconfoundedness assumption that we cannot test. In the second step of our approach, we run two conjoint experiments in which recruited participants choose between pairs of fundraising campaigns, and we estimate how individual profile features shift the probability that a campaign is chosen. In the experiments, we use Generative Adversarial Networks (GANs) to build counterfactual profile images: pairs of images that are identical except for a single feature. Holding the rest of the image fixed by construction enables us to get causal estimates.

\subsection{Counterfactual Profile Images from GANs}

In Kiva data, a style feature never varies on its own. Borrowers who smile also tend to be younger, to be female, and to differ in lighting, pose, and dozens of other attributes. Covariate adjustment addresses this only as far as the observed controls reach. A clean estimate of the effect of one feature requires the counterfactual image: the same profile, the same person, with the feature switched off and nothing else changed. Such pairs of images do not exist in the collection of Kiva photographs.

We construct them. Generative adversarial networks \citep{goodfellow2014generative} learn the distribution of a class of images and draw new samples from it. Social scientists have used them to generate realistic images for hypothesis generation \citep{ludwig2024machine} and to produce synthetic datasets \citep{athey2024using}. We put them to a different use, as a source of experimental stimuli. For each base image we generate a set of variants that differ in one targeted feature and are otherwise identical, and we use matched variants as the two arms of a randomized choice. The treatment is the feature; the counterfactual is the same image without it.

\begin{figure}[htp]
  \caption{Counterfactual variation in \emph{Smile} and \emph{Male}}\label{smile_var}
  \centering
  \includegraphics[scale=0.165]{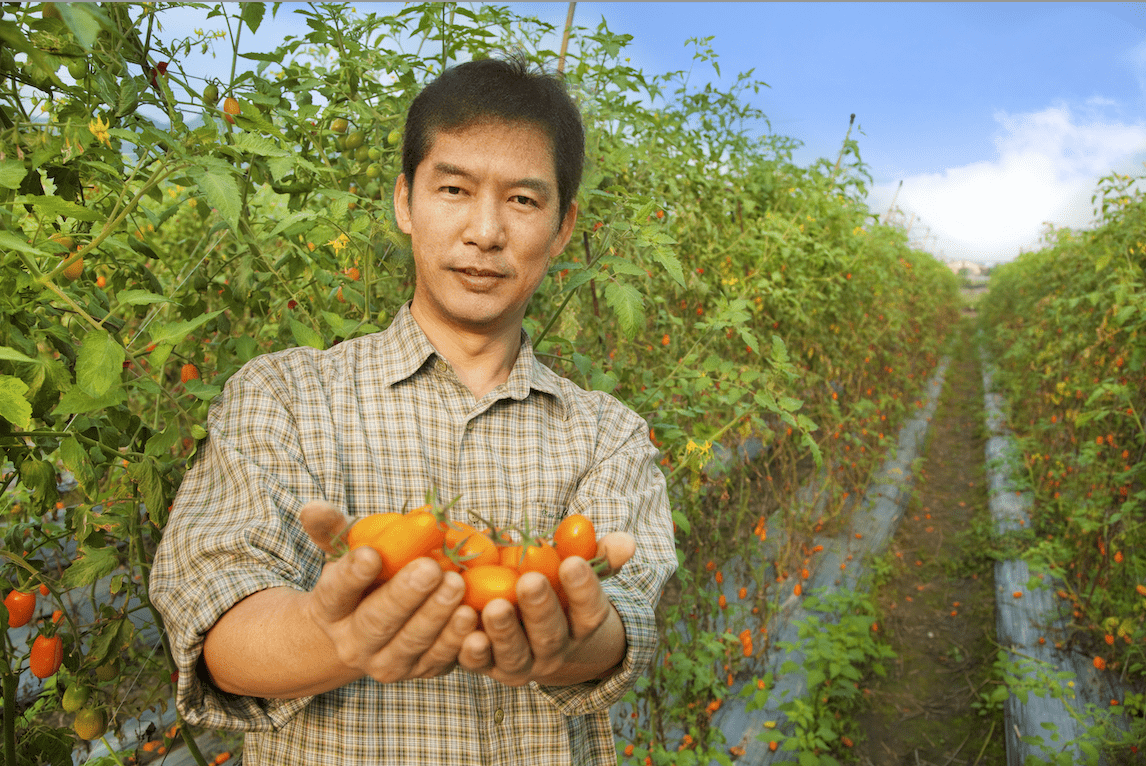}
  \hfill
  \includegraphics[scale=0.158]{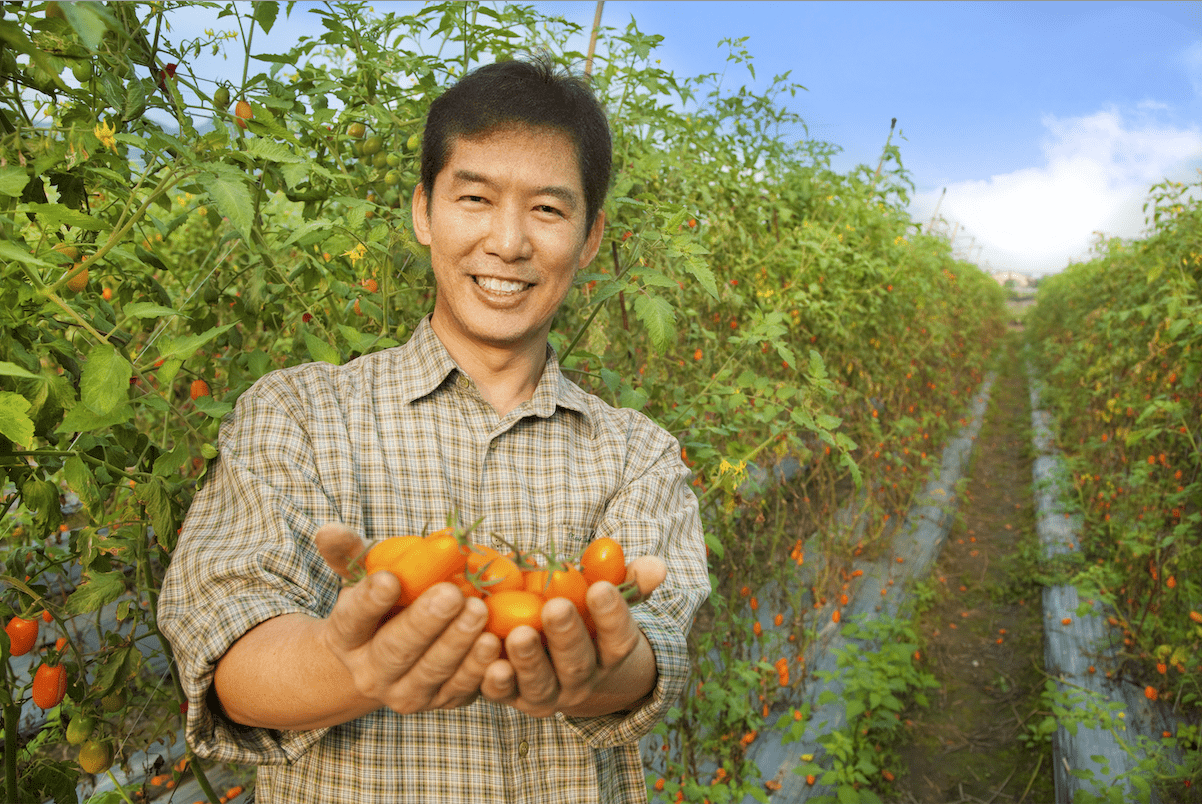}
  \includegraphics[scale=0.315]{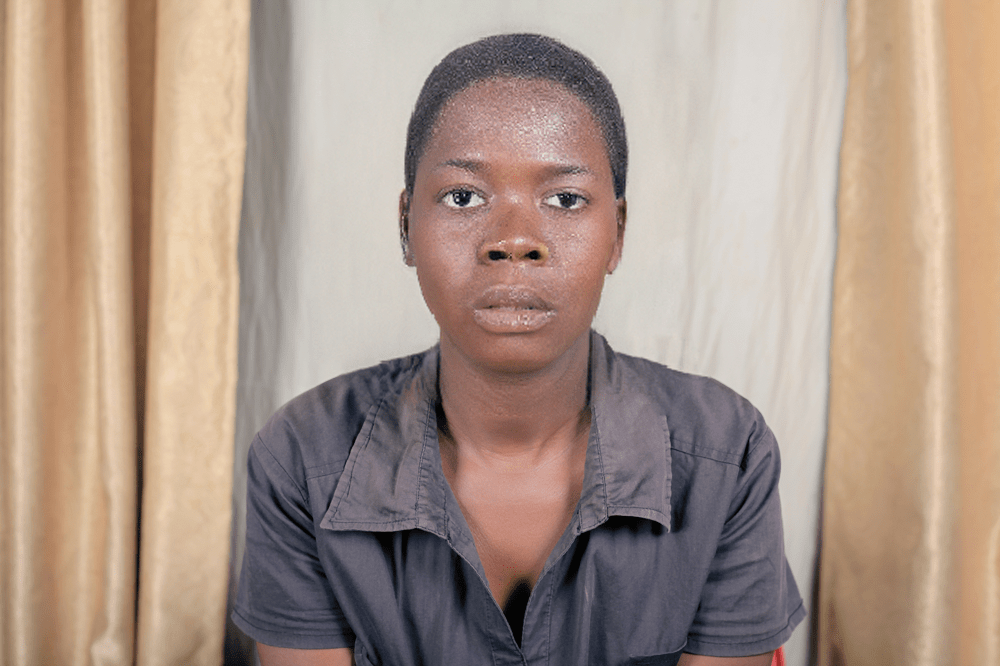}
  \hfill
  \includegraphics[scale=0.315]{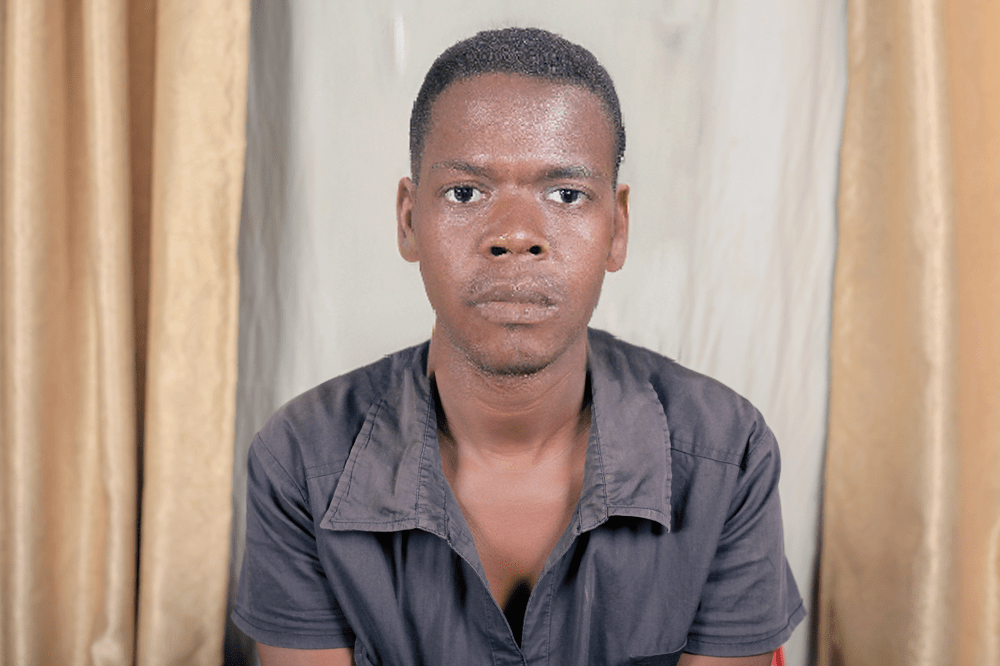}
  \caption*{\footnotesize{\textit{Note: Two pairs of GAN-generated profile images. Within each pair the images are identical except for one feature: \emph{Smile} in the top pair and \emph{Male} in the bottom pair.}}}
\end{figure}

The procedure works in the latent space of the generative model. We encode a photograph into the model's latent representation, identify the direction along which the targeted feature varies, for example, the smile direction, move the encoded image along it, and regenerate. The output is the original profile with one attribute added or removed and the rest of the face, pose, and background preserved. A few features that the latent-space method handles poorly, body-shot and sunglasses, are instead added to GAN-generated images with standard image editing. Appendix~\ref{appendix_gans} gives the construction in full, including how we recover direction vectors for features without an off-the-shelf one. Figure~\ref{smile_var} shows two matched pairs.

Two facts make the generated images suitable as stimuli. First, they are realistic: \citet{shen2021study} find that viewers cannot distinguish GAN-generated faces from photographs. Second, they evoke the same kind of reactions as real borrower images. In Appendix~\ref{GAN_emo_check} we score both sets of images on ten perceived psychological traits and find overlapping distributions with similar means and standard deviations. Using generated rather than real profiles also allows us to study the impact of style features without using images of actual Kiva users; who might object to it and whose consent is generally hard to obtain. We license photographs from Shutterstock that match the visual characteristics of Kiva profiles and train the model on them, so every hypothetical borrower shown to a participant is fabricated.

\subsection{Experimental Design}

\begin{figure}[htp]
\centering
\caption{Example of a Choice Instance}\label{choice_example}
\includegraphics[scale=0.5]{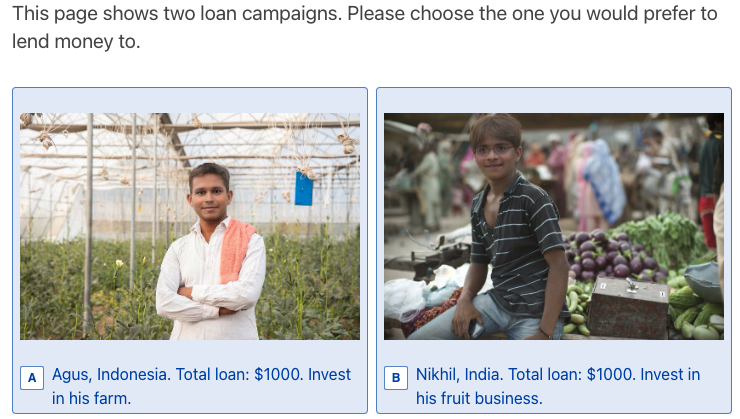}
\caption*{\footnotesize{\textit{Note: An example of a choice instance shown to participants. Each participant selected the fundraising campaign she preferred from the displayed pair.}}}
\end{figure}

Both experiments use a conjoint design: a participant sees a pair of fundraising campaigns and selects the one she prefers. Figure~\ref{choice_example} shows one such choice.

Experiment~1 studies smile and body-shot, the two style features with substantial treatment-effect estimates in Kiva data, together with gender. We include gender because it is correlated with both: male borrowers smile less and use body-shot images more, so a gender gap and a style gap are entangled in the field. From $20$ Shutterstock base images we generate eight variants each---every combination of smile, body-shot, and male---and pair them so that the two profiles a participant compares come from different base images and differ in at least one feature. Each participant sees six pairs; we keep the count low so that fatigue does not erode the later choices \citep{hainmueller_hopkins_yamamoto_2014}. Participants are assigned at random to $15$ such protocols.

Experiment~2 adds financial stakes and widens the feature set. It studies three further style features: sunglasses, glasses, and dark hair, and one further type feature, age, again selected for their estimated effects in Kiva data. Each participant sees nine pairs. Eight pairs are generated profiles, as in Experiment~1; the ninth shows two real Kiva borrowers, drawn at random through the Kiva API. Participants are told that some profiles are real borrowers active on Kiva.org and that we will lend \$10 to one real borrower they select.\footnote{The funds from repaid loans where recycled to other Kiva borrowers, which is the default setting on the platform.} The incentive ties a participant's choice to a real outcome and encourages deliberation. It is not meant to reproduce a financial investment: repayments do not accrue to the participant, consistent with the charitable-giving experiments this design follows \citep{eckel1996altruism}.

We ran both experiments on Prolific.com, recruiting $410$ participants in the first and $436$ in the second; the difference reflects non-consent and dropout before completion. Participants were fluent in English and had given to a charitable cause in the past year. Attention checks confirm data quality (Appendix~\ref{attention_checks}); covariate balance across treatments, subject demographics, and mean outcomes by base image are reported in Appendices~\ref{exp_diagnostics} and~\ref{sum_stats_recruited}.

\subsection{Estimation}

A participant chooses between two profiles, so the outcome is binary and its mean is one-half. We model the choice with a conditional logit. Let lender $i$'s utility from profile $j$ depend on the profile's features, a profile fixed effect $\mu_{j}$, and an idiosyncratic shock $\epsilon_{ij}$. For Experiment~1,
\begin{equation}
 u_{ij}=\alpha \cdot male_{j}+ \beta \cdot smile_{j}+ \gamma \cdot bodyshot_{j}+\mu_{j}+\epsilon_{ij},
\end{equation}
where $male_{j}$, $smile_{j}$, and $bodyshot_{j}$ indicate the corresponding features of profile $j$. When $\epsilon_{ij}$ follows a type-I extreme value distribution, the probability that $i$ chooses $j$ over the alternative $j'$ is
\begin{equation}\label{eq:logit}
P_{ij}=\frac{\exp(\alpha \cdot male_{j}+\beta \cdot smile_{j}+\gamma \cdot bodyshot_{j}+\mu_{j})}{\sum_{k\in\{j,j'\}}\exp(\alpha \cdot male_{k}+ \beta \cdot smile_{k}+\gamma \cdot bodyshot_{k}+\mu_{k})}.
\end{equation}
We estimate $\alpha$, $\beta$, and $\gamma$ by conditional maximum likelihood. The same specification, with the Experiment~2 features in place of $male$, $smile$, and $bodyshot$, applies to the second experiment.

Each coefficient is a within-pair object: it measures how much a feature shifts the chance that a profile is chosen when that profile is paired with an otherwise identical profile without the feature. Averaged across base images, this is the average treatment effect of the feature on selection. The profile fixed effects $\mu_{j}$ absorb everything the base image carries, so the estimate is identified by the GAN-induced variation alone.

\subsection{Results}

Table~\ref{results_ATE} reports the estimates. Columns~(1)--(3) cover Experiment~1 and columns~(4)--(6) Experiment~2. Within each experiment, the first column is the conditional logit of equation~\eqref{eq:logit}, the second adds subject characteristics, and the third is a logit with subject fixed effects estimated on the restricted sample. Profile fixed effects enter every column.

\begin{table}[!htbp] \centering
  \caption{Average Treatment Effects Estimates from Experiments 1 and 2}
  \label{results_ATE}
  \resizebox{\textwidth}{!}{%
\begin{tabular}{@{\extracolsep{5pt}}lcccccc}
\\[-1.8ex]\hline
\hline \\[-1.8ex]
 & \multicolumn{3}{c}{\textit{Experiment 1}} & \multicolumn{3}{c}{\textit{Experiment 2}} \\
\cline{2-4} \cline{5-7}
\\[-1.8ex] & Conditional Logit & + Covariates & + Fixed Effects & Conditional Logit & + Covariates & + Fixed Effects \\
\\[-1.8ex] & (1) & (2) & (3) & (4) & (5) & (6) \\
\hline \\[-1.8ex]
 Male & $-$0.385$^{***}$ (0.079) & $-$0.299$^{***}$ (0.083) & $-$0.243$^{***}$ (0.062) &  &  &  \\
 Smile & 0.298$^{***}$ (0.074) & 0.326$^{***}$ (0.078) & 0.160$^{**}$ (0.069) &  &  &  \\
  Body-shot & $-$0.191$^{**}$ (0.079) & $-$0.118 (0.086) & $-$0.152$^{**}$ (0.068) &  &  &  \\
  Age &  &  &  & $-$0.009 (0.053) & $-$0.012 (0.055) & 0.080 (0.052) \\
Dark Hair &  &  &  & 0.061 (0.053) & 0.067 (0.055) & 0.148$^{***}$ (0.051) \\
Glasses &  &  &  & $-$0.244$^{***}$ (0.070) & $-$0.251$^{***}$ (0.072) & $-$0.263$^{***}$ (0.066) \\
  Sunglasses &  &  &  & $-$0.430$^{***}$ (0.068) & $-$0.429$^{***}$ (0.070) & $-$0.400$^{***}$ (0.066) \\
\hline \\[-1.8ex]
Image FE & x & x & x & x & x & x \\
Subject's Characteristics &  & x & x &  & x & x \\
Restricted Sample &  &  & x &  &  & x \\
Observations &  4,920 & 4,428 & 4,428 & 6,864 & 6,608 & 6,608 \\
\hline
\hline \\[-1.8ex]
\end{tabular}
}
\caption*{\footnotesize{\textit{Note: Estimates of logistic regression models: conditional logit (Columns 1, 4), conditional logit with subject covariates (Columns 2, 5), and logit with subject fixed effects (Columns 3, 6). Borrower profile fixed effects are included in all regressions. Experiment 1 variables: Male, Smile, Body-shot. Experiment 2 variables: Age, Dark Hair, Glasses, Sunglasses. $^{*}$p$<$0.1; $^{**}$p$<$0.05; $^{***}$p$<$0.01.}}}
\end{table}

In Experiment~1, smile and male are large and significant in every specification. The conditional-logit estimate for smile, $0.298$, implies that a smiling profile is chosen about $57\%$ of the time against an otherwise identical non-smiling profile. The estimate for male, $-0.385$, runs the other way: a male profile is chosen about $40\%$ of the time against an otherwise identical female profile, roughly $32\%$ less often. Body-shot enters negatively, as in Kiva data, but loses significance once subject characteristics or profile fixed effects are added; we read it as suggestive.

In Experiment~2, sunglasses and glasses both lower selection and are significant throughout. Their conditional-logit estimates, $-0.430$ and $-0.244$, imply selection probabilities of about $39\%$ and $44\%$ against a bare-faced profile. Dark hair and age are small and, with one exception, statistically indistinguishable from zero; their standard errors are tight enough to rule out effects approaching the magnitude of smile or gender. The experiments thus confirm four candidate style features: smile, sunglasses, glasses, and, more weakly, body-shot, and show that one type feature, gender, moves choices sharply while another, age, does not.

There are two important limitations of the experimental design. First, the generated profiles are built from Shutterstock photographs, not from Kiva borrower images, and the two sets may differ on dimensions we do not measure; the trait comparison in Appendix~\ref{GAN_emo_check} shows overlapping distributions but cannot rule out every difference. Second, the participants are recruited to resemble Kiva lenders on observable margins, English-speaking, in a developed country, active in charitable giving, but they are likely not Kiva lenders, and their preferences may differ. Neither concern affects the internal validity of the experimental estimates; both bear on how far those estimates carry to the platform, which is why we pair them with the observational analysis rather than rely on either alone.

\subsection{Observational versus Experimental Estimates}

The experiments and the observational analysis answer the same question under different identifying assumptions. The observational estimates are based in lender behavior on the Kiva platform, while the experimental estimates are based on choices made by recruited participants in a controlled setting. The observational estimates rely on the unconfoundedness assumption; the experimental estimates rely on random assignment (satisfied by design). In addition, the observational estimates use machine-detected visual features, which are subject to non-classical measurement error, whereas the variation in features across experimental treatments is generated algorithmically, changing one feature at the time.

Across the main style features, the two approaches produce estimates with the same sign, although the experimental estimates are larger in magnitude. The gender gap is the clearest example. In the Kiva data, campaigns with male profiles raise about \$30 less per day than those with female profiles, an unadjusted gap of approximately a quarter of the average amount raised by female borrowers. In the experiment, male profiles are chosen about $32\%$ less often.   Appendix~\ref{app:comparison_estimates} reports the observational, experimental, and SIMEX-adjusted estimates side by side. The SIMEX-adjusted estimates lie between the observational and experimental estimates for sunglasses and glasses, fall below both for dark hair, and exceed both for smile. The field data provide external validity, the experiments internal validity, and the gap between them can be partly explained by the measurement error in the machine-detected features.

\section{Mechanisms: Monetary and Non-Monetary Channels}
\label{sec:mechanisms}

Style features may influence lenders' decisions through monetary and non-monetary channels. The monetary channel operates if these features signal repayment probability; the non-monetary channel operates if they evoke psychological reactions. We examine each in turn.

\subsection{Style Features and Repayment Probability}
\label{sec:monetary}

If lenders rationally use style features to assess creditworthiness, these features should predict repayment. We test this by comparing the out-of-sample performance of three gradient-boosted machine (GBM) specifications: a mean-only baseline, a specification that adds style features, and a specification that includes all available covariates.

\begin{table}
\centering
\caption{Image Features as Predictors of Repayment Probability.}
\label{tab:image_pred_def}
\resizebox{0.3\textwidth}{!}{%
\begin{tabular}{lcc}
\toprule\toprule
Specification & MSE & SE\\
\midrule
Mean model & 0.047 & 0.001 \\
Style features & 0.046 & 0.001 \\
Full model & 0.046 & 0.001 \\
\bottomrule\bottomrule
\end{tabular}
}
\caption*{\footnotesize{\textit{Note: Out-of-sample performance of three gradient-boosted machine (GBM) specifications: a model with only an intercept (Mean model), a model using image style features (Style features), and a model using all available covariates (Full model). Models are trained on a random 70\% of the sample and evaluated on the remaining 30\%. The second column reports the mean squared error; the third reports its standard error.}}}
\end{table}

Table \ref{tab:image_pred_def} reports the results. Adding style features does not improve predictive accuracy relative to the baseline, and including the full set of covariates yields no further gain. More broadly, none of the features available in Kiva data meaningfully improve predictions of default. Style features thus appear uninformative about repayment risk. Appendix \ref{default_types} corroborates this conclusion using disaggregated default categories. These results imply that rational lenders should not rely on style features to forecast loan outcomes.\footnote{This does not preclude the possibility that lenders systematically misinterpret the information contained in style features, as in \cite{esponda2016berk, bohren2023inaccurate}.}

\subsection{Style Features Shift Predicted Psychological Traits.}\label{psych_traits}

Style features may also evoke psychological reactions that impact lenders' choices. A style feature could shift how a viewer perceives the borrower: smiling makes a face look trustworthy, glasses make it look smart, and those perceptions could then affect funding preferences. The prerequisite for any such channel is that style features in fact move perceived traits. We measure perception using the One Million Impressions model of \citet{peterson2022deep}, which predicts scores for ten attributes, \textit{trustworthy, attractive, dominant, smart, happy, familiar, outgoing, well-groomed, healthy,} and \textit{privileged}, from a face image.\footnote{The model  is a deep network trained on $1{,}020{,}000$ ratings collected from $4{,}157$ Amazon Mechanical Turk workers (predominantly White, North American, mean age 39) who scored $1{,}004$ synthetic faces. These are not estimates of any borrower's actual character or Kiva lender's perception of it; they are predictions of the average impression a US, English-speaking online viewer would form on first sight.}

We estimate the effect of three style features: \emph{smiling, glasses}, and \emph{sunglasses}, on each of the ten traits using our GAN generated images.\footnote{We combine images from both recruited experiments with additional 350 images generated in an analogous way and varying the considered style features.}  For each (feature, trait) pair we run a regression to estimate the within-base-image effect of style on the (feature, trait) score. Formally, we estimate:
\begin{equation}
m_{jb} \;=\; \beta\,t_{jb} \;+\; \mu_b \;+\; \epsilon_{jb},
\label{eq:apath}
\end{equation}
where $b$ indexes a base image, $j$ indexes the image variants, $m_{jb}$ is the predicted trait score for variant $j$ of image $b$, $t_{jb} \in \{0,1\}$ indicates whether variant $j$ of image $b$ displays the style feature in question (e.g., smiling versus not smiling), and $\mu_b$ is a base-image fixed effect. The coefficient $\beta$ is identified by within-image contrasts: the difference in predicted trait between, say, a smiling and a non-smiling image of the same borrower. We report the estimate of $\beta$ in control-group standard deviations (Cohen's $d$), so a coefficient of $1.0$ means the style change shifts the trait by as much as moving across the control distribution by one standard deviation. The GAN procedure varies $t_{jb}$ exogenously within a base image, so $\beta$ identifies the effect on $m_{jb}$ of the GAN-induced manipulation of the targeted feature. The specification imposes the same functional-form assumptions used in our analysis of the experimental selection outcomes—no interaction between the base image and the style feature, with base-specific intercepts.

\begin{figure}
\centering
\caption{Impact of Style Features on Psychological Traits}
\includegraphics[width=\textwidth]{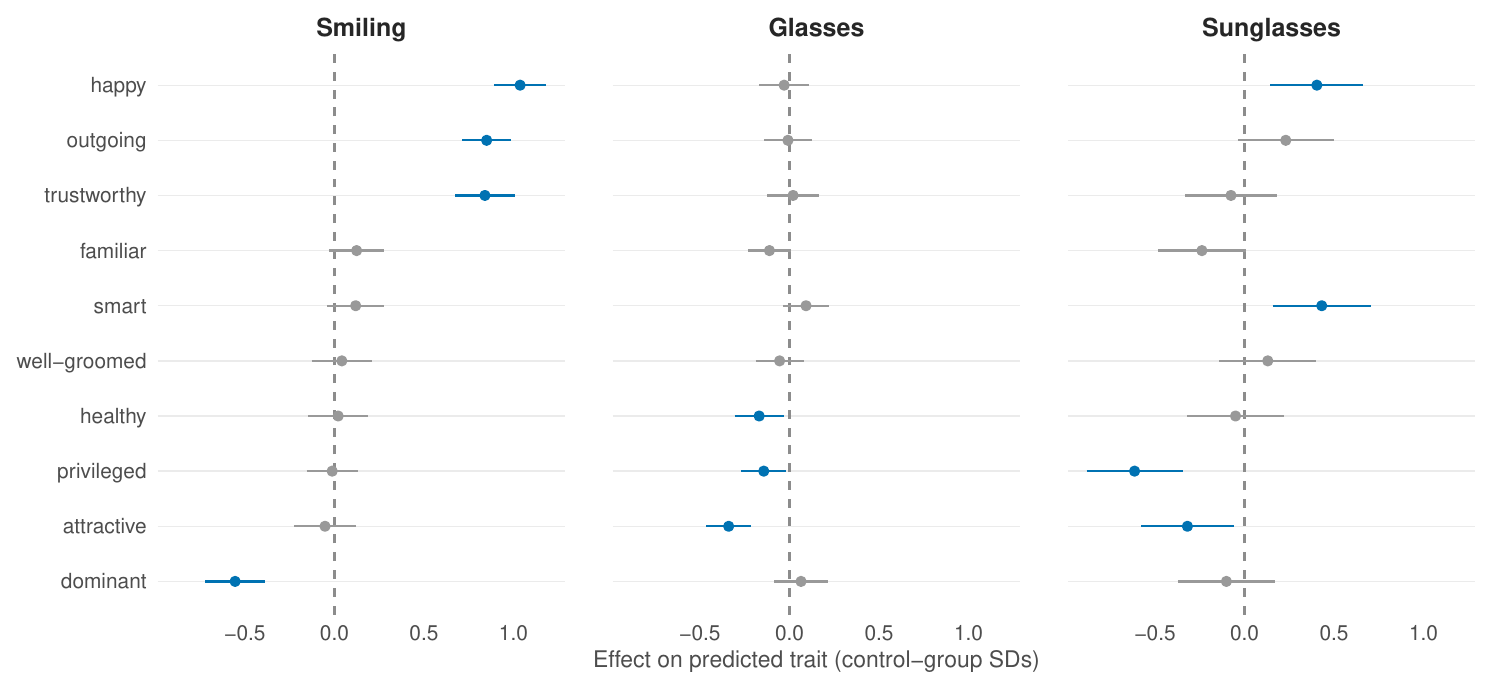}
\caption*{\footnotesize{\textit{Note: Each point is the $\beta$ coefficient estimate from Equation \ref{eq:apath}; horizontal bars give $95\%$ confidence intervals.
Effects are standardized by the standard deviation of the trait among
control-group (untreated) images, so the units are control-group
standard deviations. Filled (blue) points are statistically significant
at the $5\%$ level. }}}
\label{fig:apath_forest}
\end{figure}

Figure~\ref{fig:apath_forest} reports the estimates and
$95\%$ confidence intervals for all 30 (feature, trait) pairs. \emph{Smiling}
produces the largest shifts, raising predicted happiness, trustworthiness, and the outgoing score by roughly one control-group SD while lowering predicted dominance by about half an SD. \emph{Glasses} reduce attractiveness, healthiness, and privilege by less than 0.4 SD. \emph{Sunglasses} raise predicted smartness and happiness while lowering predicted privilege and attractiveness.

The GAN-induced exogenous variation pins down the causal impact of style features on estimated psychological traits. The GAN procedure does not, however, separately vary the style feature and the trait it shifts. Estimating how much of the impact of style features on funding outcomes is mediated by the trait channel, and how much operates through other channels, therefore requires additional assumptions \citep{imai2010general}. In Appendix \ref{sec:mediation_appendix}, we carry out the mediation analysis using Kiva observational data rather than GAN images to increase sample size. Under the (strong) assumptions required for mediation analysis, we find that the three style features differ in how much of their funding effect operates through measured trait channels. A significant share of the smile effect runs through perceived trustworthiness and dominance. The glasses effect reflects sizable but offsetting contributions: positive through trustworthiness and dominance, negative through attractiveness and familiarity. The sunglasses penalty operates almost entirely outside the trait channels we measure.

\section{Efficiency-Disparity tradeoff: counterfactual simulations}\label{simulations}
There are many different platform policies that could exploit our finding that certain style features impact funding outcomes, and type-based disparities in particular. In this section, we propose several such policies, simulate counterfactual outcomes, and evaluate their impact on disparities and efficiency. To do that, we consider a simplified model of interactions on Kiva characterized by the parameters from the recruited experiment. 

Although our approach is stylized, our findings provide insight into which types of policies are likely to be effective. In practice, our method can help prioritize policies for randomized experiments. The policies we study, policies based on style features, are particularly relevant from a managerial perspective, as platforms like Kiva often provide guidelines for style choices, given that these features are easily modifiable.

\subsection{A model of a micro-lending platform}

\paragraph{Pool of borrowers.} The pool of available borrowers is a set of borrowing campaigns from which Kiva selects a subset to display to lenders. The pool of borrowing campaigns can be summarized by a vector of profiles $\mathbf{x}$, where each element is a profile $x_{i} = (type_{i}, style_{i},\eta_{i})$, which describes features of the borrower $i$; the first element corresponds to the borrower's type and we consider two dimensions male or female and young and old. The style features encompass smile, body-shot, sunglasses, and Dark hair, and $\eta_{i}$ is a fixed effect which summarizes all other characteristics of the borrower. 

The pool of borrowers is exogenously determined and the joint distribution of borrowers' characteristics is denoted as $G$. The expected pool of borrowers is denoted as $\tilde{x}$.

\paragraph{Policy and markets.} A market is a set of borrowers shown to a lender. Platform policy $\mathbb{H}$ transforms the joint distribution of borrowers' characteristics from $G$ to $H$. Specifically, the policy defines $\mathbf{E}_{H}\left[style|type, \eta \right]$  the conditional probability of style features in the pool of borrowers. Additionally, a policy applies the probability of being shown to lenders $h: (\eta, type, style) \rightarrow [0,1]$ to the pool of borrowers. Thus, a policy can be summarized as $\mathbb{H} = \left\{\mathbf{E}_{H}\left[style|type, \eta \right], \mathbf{h} \right\}$. The expected pool of borrowers shown to lenders under the policy $\mathbb{H}$ is denoted as $\tilde{x}^{H}$.

The policies that we consider have two elements. First, they can impact the distribution of style features in the pool of borrowers. Examples of this include advice on profile creation, a protocol that requires borrowers to upload several images and selects the most compliant one, or behavioral interventions that nudge borrowers to create compliant profiles. Second, a policy can modify the probabilities with which borrowers in the pool appear in the market as a function of image features and other borrowers' characteristics.

\paragraph{Lenders.}
Lenders, indexed by $j$, are heterogeneous with respect to their preference parameters $\boldsymbol{\alpha}_{j}$ (for type) and $\boldsymbol{\beta}_{j}$ (for style).\footnote{ $\boldsymbol{\alpha}_{j}$ and $\boldsymbol{\beta}_{j}$ are vectors with $\boldsymbol{\alpha}_{j}$ having two elements for male and old and $\boldsymbol{\beta}_{j}$ having four elements for smile, body-shot, sunglasses, and Dark hair. } Preference parameters are random variables that are realized before a lender sees any profiles. Each parameter is drawn from a normal distribution centered at the corresponding average treatment effect estimate from the recruited experiment, and the standard deviation is equal to the standard error of the estimate. After the parameters are determined, the lender decides whether to participate based on the expected utility of participating, computed before the realization of the idiosyncratic shocks $\epsilon$:
\begin{equation}\label{util_max}
  V_{j}(\tilde{x}^{H}) \;=\; \mathbb{E}_{\epsilon}\!\left[\,\max_{i \in \tilde{x}^{H} \cup \{o\}}\!\big(\boldsymbol{\alpha}_{j} \cdot type_{i} + \boldsymbol{\beta}_{j} \cdot style_{i} + \eta_{i} + \epsilon_{ij}\big)\right],
\end{equation}
where $\tilde{x}^{H}$ is the vector of borrowers active on the platform under policy $\mathbb{H}$, $o$ denotes the outside option with utility $u_{oj} = \omega + \epsilon_{oj}$, and $\epsilon_{ij}$ is a random utility parameter, independent across lenders and borrowers, GEV distributed. If $V_{j}(\tilde{x}^{H}) > \delta$, where $\delta$ is the cost of participating, the lender participates; otherwise, the lender stays out. Because $V_{j}(\tilde{x}^{H})$ depends on the policy $\mathbb{H}$ through $\tilde{x}^{H}$, the distribution of preference parameters among participating lenders will differ across policies.

Lenders who decide to participate observe the realized choice set of borrowers and choose the option that maximizes their utility. The utility associated with choosing one of the borrowers is
\begin{equation}\label{util_choice}
  u_{ij} = \boldsymbol{\alpha}_{j} \cdot type_{i} + \boldsymbol{\beta}_{j} \cdot style_{i} + \eta_{i} + \epsilon_{ij},
\end{equation}
and the utility from the outside option is $u_{oj} = \omega + \epsilon_{oj}$. Lenders choose the option that maximizes their utility.

To summarize, we assume the following timing: (1) the pool of available borrowers is exogenously determined; (2) the platform chooses policy $\mathbb{H}$; (3) lenders arrive, their preference parameters are realized, and they decide whether to participate (they know which policy the platform chose); (4) lenders who decided to participate choose between borrowers or the outside option.

\subsection{Implementation} 
\paragraph{Markets.}

We consider a pool of 22 borrowing campaigns. The distribution of style features conditioned on considered type features and overall profile attractiveness is based on Kiva data.\footnote{At the time our data was collected, Kiva's policy was based on when the borrower posted the campaign. Thus, assuming that arrival time is independent of characteristics, a lender sees each borrower in the pool with equal probability. In reality, this is an approximation because campaigns that reach their funding outcomes are removed from the platform. Thus, the less attractive campaigns stay longer on the platform, so lenders have a higher chance of observing them. As a consequence, the distribution of type and style features that we observe in Kiva data might differ from the distribution of the pool of borrowers that arrive to Kiva.}    First,  campaigns' fixed effects take values of fixed effects estimated in the recruited experiments. We assign these values randomly. Second, we train a GBM model of cash per day, and predict the outcome net of the type and style features of interest. We compute the distribution of types in Kiva data across deciles of predicted cash per day, and match the deciles of the fixed effect to the decile of the predicted cash per day:

\[
\mathbf{E}_{G}\left[ type |D(\hat{\eta})\right] = \mathbf{E}_{K}\left[type | D(\eta_{k})\right],
\]
where $K$ stands for distribution in Kiva data, $D(\cdot)$ is the decile of the fixed effect and  $\eta_{k}$ is the fixed effect from Kiva data. $\mathbf{E}_{K}\left[type | D(\eta_{k})\right]$ is the conditional distribution of type profiles in Kiva data; thus, for example, the share of male borrowers with the fixed effect in the first decile of fixed effects estimated from the recruited experiment equals the share of male borrowers in the lowest decile of Kiva data fixed effects. This way, we get the distribution of types across residual profile attractiveness. Next, we compute the distribution of style in Kiva data across the types and the deciles of predicted cash per day, and match on both the deciles and the types.

\[
\mathbf{E}_{G}\left[style|type, D(\hat{\eta}) \right] = \mathbf{E}_{K}\left[style|type , D(\eta_{k}) \right].
\]

Thus, we allow the distribution of the style to differ across fixed effects and type.

The number of borrowers in a market will depend on the policy, but in all cases, it will be a subset of the pool of borrowers.

\paragraph{Lenders' preferences.}  We assume that lenders' preferences $(\alpha_{j}, \beta_{j})$ are parameters drawn from distributions estimated using experimental data, such that $\alpha_{j} \sim N(\alpha,sd_{\alpha})$, where $\alpha$ is the estimate of the average treatment effect and $sd_{\alpha}$ is its standard error, and $\epsilon_{ij}$ is a random utility parameter, which is iid across lenders and borrowers, GEV distributed.\footnote{We use $sd_\alpha$, the standard error of the estimated average treatment effect, to calibrate cross-lender dispersion. This is not a direct measure of population heterogeneity. This conflates estimation uncertainty with population heterogeneity and likely understates true dispersion in lender preferences; the simulations should therefore be read as informative about the ranking of policies.} We set the utility from choosing the outside option to one (the highest fixed effect estimated in the experiment is 0.64).

We assume that the cost of participating $\delta$ is fixed and the same for all borrowers. We set the cost at $2.5$, resulting in approximately half of the lenders choosing to participate. 

\paragraph{Outcome metrics.} We propose two metrics of disparities: first, to capture the overall distribution of outcomes, we use the Gini coefficient defined as \[\text{Gini} = \frac{\sum_{j=1}^{n}\sum_{j'=1}^{n}|x_j - x_{j'}|}{2n^{2}\bar{x}} \]where $x_{j}$ is the outcome of borrower $j$ and $x_{j'}$ of borrower $j'$, $n$ is the number of borrowers and $\bar{x}$ the average outcome. We consider all borrowers in the pool. Second, we consider the gender disparity, defined as the share of lenders that choose a male borrower amongst the lenders who decided to participate and did not choose an outside option. We standardize this metric by the share of male borrowers in the pool. We measure efficiency as the share of lenders that chose a borrower instead of an outside option. 

The type of policy, and thus the expected set of borrowers shown to lenders, will impact the number and the type of active lenders. To capture this, we report the number of active lenders that have high style preference parameters. We define high as above the mean of the distribution for each parameter. Thus, a high smiling type is the lender who cares more than a typical lender that a borrower smiles in the profile image.
 
\paragraph{Market outcomes.} To determine market outcomes, we simulate markets and choices by lenders. Based on the distribution of outcomes, we compute disparity and efficiency metrics. Each simulation proceeds in three steps: first, we simulate the pool of borrowers. Then, we construct markets from the pool of borrowers. A policy determines $h(\eta_{i}, type{i}, style_{i})$, the probability that a borrower in a pool appears in the market. A market is constructed per lender. This means that in one simulation there is one pool of borrowers, from which borrowers are sampled for each lender. 

Finally, we simulate lenders' preferences, their entry decisions and their choices.. We perform 100 simulations of 2000 lenders' choices for each policy. We use the outcomes to compute our metrics of disparity and efficiency. We consider all borrowers in the pool, irrespective of whether they were shown to lenders or not. Appendix \ref{algo_cf} presents the algorithm that we used.\footnote{In this analysis, we assume that lenders' preferences are stable across different platform policies. In Appendix \ref{coef_stability}, we exploit a natural experiment in the form of Kiva landing page redesign to provide support of this assumption. The website redesign introduced borrowers categories in place of a list where all borrowers would be displayed together. We find that the impact of smile on cash per day was similar before and after the website redesign; the difference is not statistically significant.}

\subsection{Counterfactual policies}
\paragraph{Baseline.} The \emph{Baseline} policy represents the existing policy on Kiva. In the \emph{Baseline} policy, the platform shows 10 borrowers to each lender, and each borrower in the pool is assigned an equal probability of being included in the market.

\paragraph{Naive Recommendation.} In the \emph{Naive Recommendation} policy, we show what happens when a platform oversamples profiles with attractive style. The platform selects all borrowers with attractive style features and randomly selects 10 to include in the market. The selection is done such that the platform starts by selecting profiles that match all 4 style criteria (smile, no sunglasses, Dark Hair, and no Body-shots). If there are fewer than 10 such profiles, the platform relaxes the requirements until there are 10 profiles. If there are more than 10 profiles, the platform selects 10 of them at random.

\paragraph{Style Recommendation.} In the \emph{Style Recommendation} policy, the platform recommends that all borrowers follow style guidelines. In practice, we assume that previously non-compliant borrowers become compliant with a probability of 75\%.\footnote{Such a profile feature recommendation can be implemented in various ways, for example, through behavioral nudges or a script requiring that several images need to be uploaded from which platform selects the ones to be shown to lenders.} After a pool of borrowers is determined, the platform assigns all borrowers an equal probability of being included in the market.\footnote{Note that this policy requires that borrowers comply with the policy recommendation. In the analysis, we assume that 25\% of the borrowers do not adhere to the recommendation. A particular type of non-compliance in the case of smiling might be that the smile created by the borrower does not appear genuine. In Appendix \ref{fake_smiles_section}, we develop an additional algorithm that distinguishes between fake and genuine smiles and apply it to the Kiva observational data. We show that only genuine smiles lead to higher outcomes. Consequently, the policy will be less effective if some of the newly added smile's are perceived as non-genuine. This analysis highlights the importance of clear instructions and a well-designed system that supports borrowers when they create profiles.A computer vision algorithm showcased in Appendix \ref{fake_smiles_section} could be a component of such a system, where borrowers could be prompted if their smile is at risk of being perceived as not genuine.}

\paragraph{Low-type Support.} The \emph{Low-type Support} policy promotes borrowers predicted to have low funding outcomes based on their types, by ensuring they are always included in the market. We focus on gender in this application. Practically, the approach is analogous to \emph{Naive Recommendation}: when the number of male campaigns is above ten, the platform samples randomly from them. Otherwise, the platform includes all male profiles and fills in other slots by randomly selecting from available profiles. In expectation, there are some female profiles included in the market.

\paragraph{Restrict Competition.} In the \emph{Restrict Competition} policy, the platform promotes fairness by reducing the competition between borrowers. To implement this, the platform randomly selects five borrowers from the pool to form the market (instead of ten).

All policies that we propose in expectation give non-zero probabilities of being included in the market to any borrower. 

\subsection{Results}

\begin{figure}[ht]
\centering
  \caption{Disparity-Efficiency Tradeoff}
\begin{minipage}[b]{0.45\textwidth}
    \centering
    \textbf{Gini coefficient}
    \includegraphics[width=\textwidth]{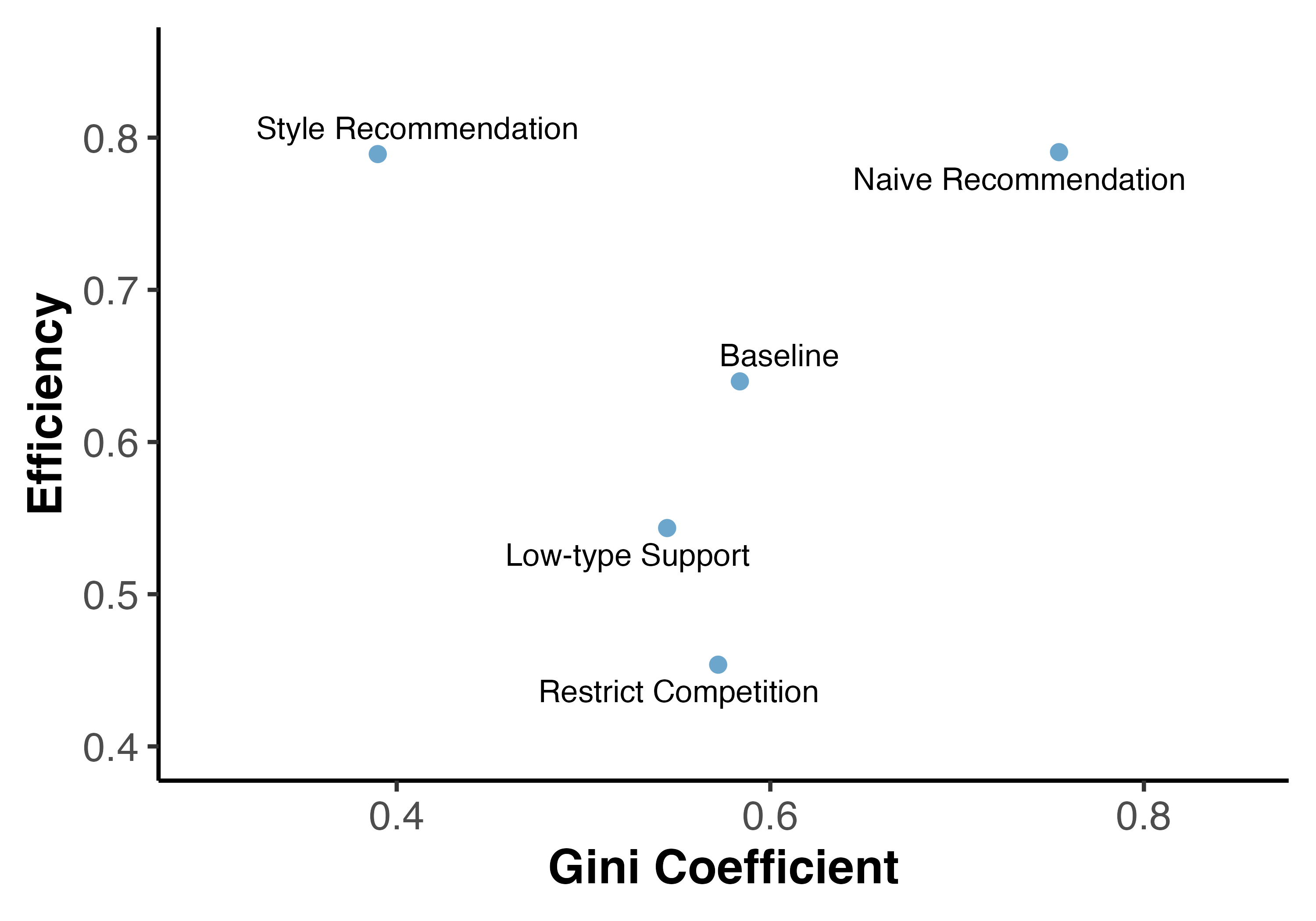}
\end{minipage}
\hfill
\begin{minipage}[b]{0.45\textwidth}
    \centering
    \textbf{Gender disparity}
    \includegraphics[width=\textwidth]{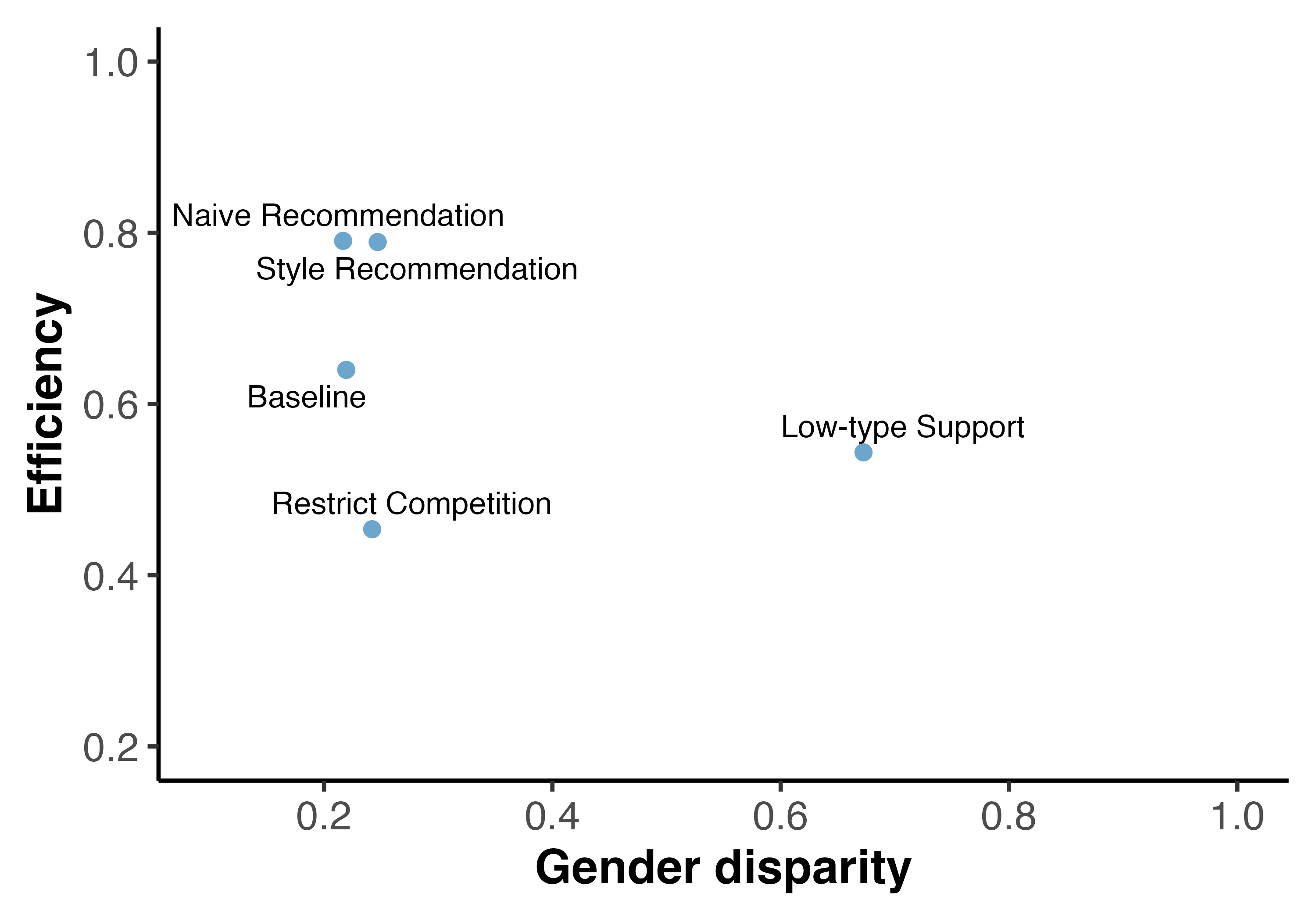}
\end{minipage}
  \caption*{\footnotesize{\textit{Note: Each point represents the mean of 100 simulations with 2000 lenders each. The vertical axis reports the share of lenders choosing one of the borrowers rather than the outside option (efficiency). Left panel: horizontal axis is the Gini coefficient. Right panel: horizontal axis is the share of lenders choosing a borrower with a male profile, normalized by the share of male profiles in the borrower pool.}}}
    \label{fig:tradeoff_1}
\end{figure}

Figure \ref{fig:tradeoff_1} presents the results from simulations of the proposed policies. In the left panel, the horizontal axis shows the mean of Gini coefficients across all simulations of each policy. The vertical axis shows the mean share of lenders choosing one of the borrowers rather than the outside option.

We find that the proposed policies considerably impact both metrics, disparity and efficiency. First, in the Baseline policy, the Gini coefficient is around 0.59 and efficiency is 0.64. Both the Naive Recommendation and Style Recommendation substantially increase the share of lenders choosing one of the borrowers rather than the outside option. The Naive Recommendation policy has a slightly higher impact on efficiency. Low-type Support and Restrict Competition have a strong negative impact on efficiency. In particular, restricting the number of alternatives substantially increases the share of lenders choosing an outside option.

Naive Recommendation is the only policy that increases disparities; the Gini coefficient rises to 0.74 under this policy. Low-type Support and Restrict Competition lead to a small reduction in disparities. Style Recommendation substantially reduces disparities.

The right panel of Figure \ref{fig:tradeoff_1} shows the impact of our counterfactual policies on the gender gap and efficiency. Low-type Support stands out: by actively including more male profiles in a market, it has a strong effect on reducing the gender gap. Both Style Recommendation and Restrict Competition have minor effects on reducing the gender gap. Naive Recommendation increases the gender gap.

Naive Recommendation and Low-Type Support operate by altering the conditional probability of borrower inclusion in the market. Naive Recommendation aims to increase efficiency by prioritizing attractive profiles, while Low-Type Support seeks to reduce disparities by increasing impressions for underperforming borrowers. However, both policies have unintended consequences. Favoring attractive profiles exacerbates disparities, as desirable style features often coincide with advantageous type characteristics, compounding inequalities. Conversely, prioritizing borrowers from an underperforming type results in more frequent exposure to profiles with less attractive style features.

Style Recommendation, in contrast, adjusts the conditional distribution of style across type and so improves efficiency and reduces disparities at the same time.

When desirable style features are positively correlated with type characteristics that improve funding outcomes, platform policies that amplify these features reinforce inequities. In contrast, policies that redistribute attractive style features among borrowers with low type characteristics promote a more equitable allocation of funding. Moreover, increasing the overall prevalence of desirable style features enhances both equity and efficiency.

\subsection{Impact of counterfactual policies on type of lenders active in the market}

\begin{table} 
    \centering
    \caption{Impact of Platform Policies on the Type of Active Lenders} 
    \label{tab:policy_comparison} 
    \resizebox{0.85\textwidth}{!}{%
\begin{tabular}{@{\extracolsep{5pt}}lccccc}
\\[-1.8ex]\hline
\hline \\[-1.8ex]
& \textbf{Baseline} & \textbf{Style Rec.} & \textbf{Naive Rec.} & \textbf{Restrict Comp.} & \textbf{Low-Type Supp.} \\
\hline \\[-1.8ex]
\multicolumn{6}{l}{\textit{Share of Active Lenders with High Preference for Feature}} \\
\hline \\[-1.8ex]
Share Smile & 0.24 & 0.44 & 0.27 & 0.57 & 0.28 \\
Share Body-shot & 0.12 & 0.31 & 0.15 & 0.34 & 0.16 \\
Share Sunglasses & 0.14 & 0.31 & 0.16 & 0.35 & 0.17 \\
Share Hair & 0.17 & 0.39 & 0.20 & 0.46 & 0.23 \\
\hline \\[-1.8ex]
\multicolumn{6}{l}{\textit{The Number of Active Lenders with High Preference for Feature}} \\
\hline \\[-1.8ex]
Smile & 3.00 & 7.03 & 4.31 & 5.37 & 3.22 \\
Body-shot & 1.53 & 4.88 & 2.41 & 3.15 & 1.80 \\
Sunglasses & 1.74 & 4.93 & 2.60 & 3.30 & 1.98 \\
Hair & 2.12 & 6.19 & 3.23 & 4.29 & 2.59 \\
\hline
\hline \\[-1.8ex]
\end{tabular}
    }
    \caption*{\footnotesize{\textit{Note: The top section shows the share of lenders that have high preference for the specific feature in the population of active lenders. High preference is defined as the preference that is above the mean in the distribution. Active lenders are those that choose one of the borrowers. The bottom section shows the total number of active lenders who have a high preference. These numbers are out of 100 lenders. Thus, 19 in the case of Baseline and Smile preference means that out of 100 lenders who arrived on the platform, 19 decided to participate, selected one of the borrowers, and have a high preference for a smile.}}}
\end{table}

Table \ref{tab:policy_comparison} reports how counterfactual policies 
shift the composition of active lenders---those who select a borrower 
rather than the outside option. The top panel reports the share of 
active lenders with above-mean preferences for each visual feature; 
the bottom panel scales these shares by participation, expressing 
counts per 100 arriving lenders.

Two patterns emerge. First, every alternative policy raises the share 
of active lenders with above-mean preferences for each feature, but 
the magnitudes differ sharply. Restrict Competition produces 
the largest shifts: the share with above-mean smile preferences rises 
from 0.24 to 0.57, and the corresponding shares for body-shot, 
sunglasses, and hair more than double. Style Recommendation 
produces smaller but substantial increases (smile: 0.24 to 0.44; 
hair: 0.17 to 0.39). Naive Recommendation and Low-Type 
Support move the shares only marginally.

Second, similar share shifts can reflect different participation 
dynamics. Style Recommendation yields the largest absolute 
counts of high-preference active lenders across all four features 
(e.g., 7.03 with above-mean smile preferences, up from 3.00). 
Restrict Competition, despite generating the highest shares, 
produces lower absolute counts (5.37 for smile), consistent with 
concentrating participation among lenders aligned with the smaller 
choice set rather than expanding the active pool. Low-Type 
Support and Naive Recommendation produce only modest 
increases in absolute counts.

Platform policies therefore affect both borrower-level allocations and lender selection into participation. Style Recommendation broadens and visually selects the active pool; Restrict Competition concentrates participation among visually selective lenders without comparably expanding it; Low-Type Support and Naive Recommendation produce only modest
compositional shifts.

\section{Conclusion} \label{conclusion}

This paper studies how the type and style features of profile images shape funding outcomes and disparities on Kiva, a large peer-to-peer microfinance platform. The type-based gaps are large: campaigns with male profiles raise about 25\% less per day than campaigns with female profiles in the observational data, and are selected about 32\% less often in otherwise-identical pairwise comparisons in the recruited experiment. These gaps reflect both fixed socio-demographic characteristics and malleable style choices, such as whether a borrower smiles, wears sunglasses, or appears in a body-shot, that are themselves correlated with type. Decomposing type-based disparities, we show that differences in style contribute to funding gaps between many demographic groups. Because style, unlike type, can be changed at low cost, it gives platforms a practical lever for improving efficiency while reducing disparities.

Our empirical results based on observational data may suffer from omitted variable bias. To address this, we use the observational results to prioritize features for a sequence of two randomized experiments with recruited subjects, who choose between pairs of borrower profiles. These experiments rest on a methodological contribution of the paper: a procedure for generating experimental stimuli with Generative Adversarial Networks. For each borrower photograph, we construct counterfactual variants---images that are identical except for a single, exogenously varied feature, such as a smile or sunglasses. The experiments produce internally valid estimates. We confirm that style features move selection probabilities: smiling increases the likelihood of selection, while wearing sunglasses or glasses reduces it.

Exploring the mechanisms behind these effects, we find that style features do not improve predictions of loan repayment probability, suggesting that rational lenders should not use these cues to assess creditworthiness. Instead, they impact perceived attributes such as trustworthiness and dominance, implying that psychological considerations drive lender decisions more than financial considerations.  

Our findings have implications for platform design. Because style features can be modified, they offer a potential lever for reducing disparities while maintaining or improving efficiency. Counterfactual simulations suggest that style recommendations—such as encouraging borrowers to smile while avoiding body-shots—can reduce disparities and increase transaction volume. On the other hand, increasing the visibility of borrowers with preferred style traits may improve efficiency but exacerbate disparities by disproportionately benefiting borrowers from advantaged types.  

Under the Style Recommendation policy, which assumes that 75\% of previously non-compliant borrowers adopt the recommended profile features, the share of arriving lenders who fund a loan, rather than choosing the outside option of not lending, increases by 14 percentage points, or roughly 22\% relative to the baseline. At platform scale, Kiva funded approximately \$188~million in loans in 2024, drawing on a community of more than two million cumulative lenders and a typical lender contribution of \$25.\footnote{Source: Kiva annual reports,\url{https://www.kiva.org/about/finances}.} The implied effect is therefore on the order of tens of millions of dollars in additional annual disbursements and tens of thousands of additional funded loans. Substantial gains remain even under more conservative compliance assumptions. Moreover, because profile-photo guidance is inexpensive to provide, these effects are large relative to implementation costs. The magnitudes are also comparable to those from related interventions on similar platforms: \citet{zhang2025serving} find that smiling in a host's profile photo increases Airbnb bookings by about 3.5\% on average and by more than 8\% for male hosts. The effect in our setting is larger, consistent with the fact that the policy bundles several profile-image features rather than altering a single visual cue. Low-cost profile-image guidance can therefore be an empirically meaningful intervention on online platforms.

More broadly, our results show that disparities in online marketplaces arise from both fixed socio-demographic characteristics and malleable style choices correlated with type. Encouraging the adoption of beneficial style features among underrepresented groups can mitigate disparities while maintaining efficiency. However, the appropriateness of style recommendations must be carefully considered in context.  

While our study provides a framework for identifying platform policies that balance fairness and efficiency, the effectiveness of specific interventions depends on lender responses and borrower compliance. Our findings can inform the design of future randomized experiments testing style-based interventions in real-world settings. More broadly, they highlight a challenge in algorithmic decision-making: when predictive models incorporate style features correlated with type, they may reinforce disparities even if the features themselves appear neutral. Future research should examine whether similar mechanisms operate in other online marketplaces, such as hiring platforms and social networks, where style choices may systematically influence economic and social outcomes.
\newpage



\newpage{}
\setcounter{page}{1}
\gdef\thepage{A\arabic{page}}
\appendix
\section*{Appendix}

\section{Feature Detection Algorithms}\label{appendix_features}

This Appendix A provides a detailed explanation of how we use computer vision algorithms to extract features from our image data. Specifically, we employ different algorithms, outlined in Appendix A, for object and feature detection using the Mask-RCNN model. The extracted image features were merged with the Kiva dataset for use in the analysis..

\begin{figure}[!htb]
\centering
\caption{The Mask R-CNN framework}
\label{maskrcnn_flow}
\includegraphics[width=0.7\linewidth]{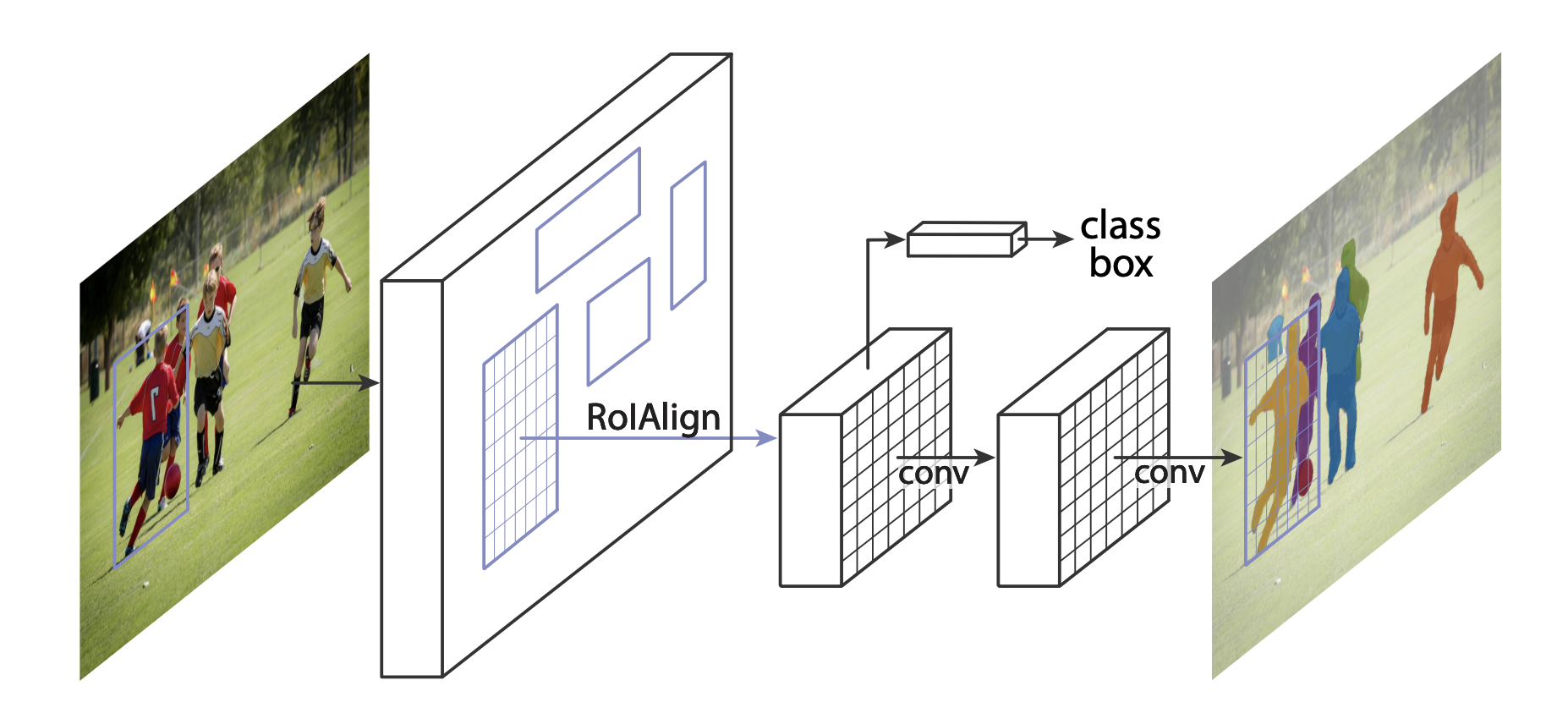}
\\[0.4em]
{\footnotesize\textit{Note:} Reproduced from \citet{he2017maskrcnn}. Available at \url{https://arxiv.org/pdf/1703.06870.pdf}.}
\end{figure}

\paragraph{Mask-RCNN.}
To systematically extract image features, we use Mask R-CNN, an object detection algorithm developed by Facebook. As illustrated in Figure \ref{maskrcnn_flow}, Mask R-CNN processes an input image and produces a “package” for each detected object, which includes the object’s class label, bounding box, and mask. These predictions are jointly optimized through a single loss function.

\begin{figure}[!htb]
\centering
\caption{An example outcome of image detection using Mask R-CNN}
\label{maskrcnn}
\includegraphics[width=0.7\linewidth]{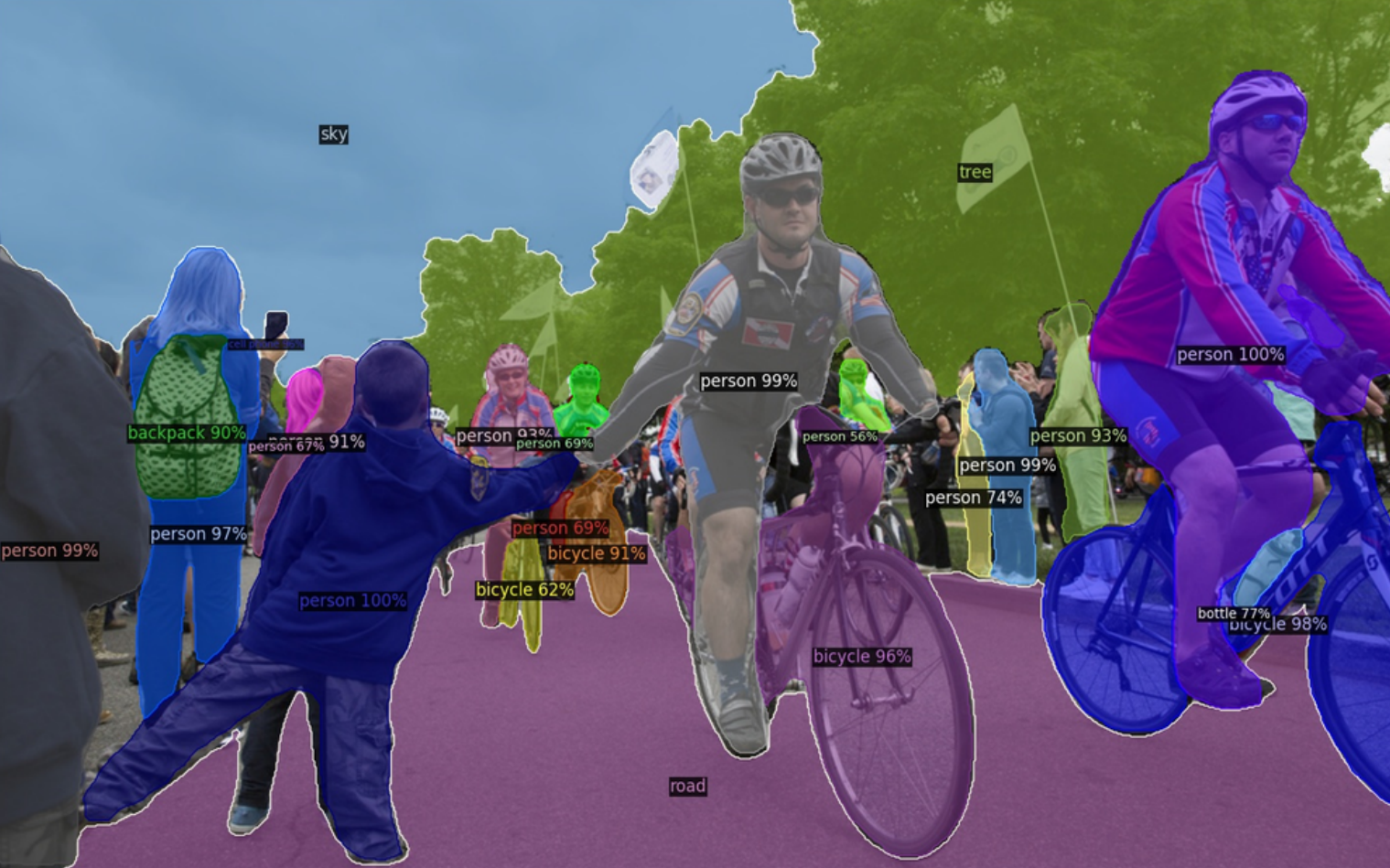}
\\[0.4em]
\begin{minipage}{0.85\linewidth}
{\footnotesize\textit{Note:} Each detected object is given a label, placed on a mask, and assigned a corresponding probability score. Image generated using the Detectron2 framework \citep{wu2019detectron2}. Available at \url{https://github.com/facebookresearch/detectron2}.}
\end{minipage}
\end{figure}

\paragraph{Object detection.}
We apply this pre-trained model to generate a confidence score for each detected object, ranging from 0 to 1. The score represents the algorithm's confidence in the presence of specific features, such as a tree, person, animal, or digital item. Figure \ref{maskrcnn} illustrates the resulting output. We also use this algorithm to detect full-body human figures.\footnote{https://github.com/facebookresearch/detectron2}
 
\paragraph{Facial feature classification.}
We detect facial features using the \emph{face-classification} algorithm that takes in one face image and outputs a face embedding vector, evaluated by a pre-trained neural network.\footnote{https://github.com/wondonghyeon/face-classification} Then, the embedding vector, as well as the feature labels, enter another neural network model (Multi-layer Perceptron). This model takes in one facial embedding vector and assigns a score for each unique facial feature such as \emph{race}, \emph{gender}, \emph{smile}, etc. It is a supervised learning process, and the training label is pre-annotated. 

The features that we obtain from images can be informally classified into three categories: (i) technical aspects of the image (e.g., \emph{blurry}, \emph{flash}, \emph{harsh light}), (ii) personal characteristics (e.g., \emph{straight hair}, \emph{eyes open}, \emph{pale skin}), (iii) objects in the image (e.g., \emph{chair}, \emph{clock}). 

Image and personal characteristics ( e.g., race, age, hair color, facial shape, eyes/nose characteristics) are detected by FaceNet model which was pre-trained and tested on the large dataset CelebA with over 200,000 facial images. The algorithm detects the person's face and then identifies its features.

\section{Summary statistics of Kiva data}\label{sum_stats_all}

We present summary statistics for the complete set of variables in Kiva data, organized by thematic groups.
{\scriptsize
\renewcommand{\arraystretch}{0.95}
\setlength{\tabcolsep}{4pt}

\newcommand{\panelhead}[1]{%
  \addlinespace[2pt]
  \multicolumn{7}{@{}l}{\textbf{\textit{#1}}}\\*
  \addlinespace[1pt]
  \cmidrule(l{0pt}r{0pt}){1-7}\\*[-2pt]%
}

\begin{longtable}{@{\extracolsep{\fill}}l*{6}{r}@{}}
\caption[Summary statistics of Kiva data]{Summary statistics of Kiva data.}\label{tab:sumstats_full}\\
\toprule
Variable & Mean & St.\,Dev. & Min & Pctl(25) & Pctl(75) & Max \\
\midrule
\endfirsthead

\multicolumn{7}{c}{\tablename\ \thetable\ -- \textit{Continued from previous page}}\\
\toprule
Variable & Mean & St.\,Dev. & Min & Pctl(25) & Pctl(75) & Max \\
\midrule
\endhead

\midrule \multicolumn{7}{r}{\textit{Continued on next page}}\\
\endfoot

\bottomrule
\multicolumn{7}{p{0.95\linewidth}}{\footnotesize\textit{Note:}
Sample size is $N=420{,}908$. Type and style indicators in Panels~C--I are CNN-predicted probabilities
in $[0,1]$, except \emph{Glasses}, \emph{Sunglasses}, and \emph{Bodyshot}, which are binarized at the $0.5$ threshold. Detected-object
variables in Panel~J report the count of detected instances per image. \emph{Face position
--- top} and \emph{Face position --- right} are bounding-box pixel coordinates from the
face detector. Panel~J omits detected objects with mean below 0.005 (max count $\leq 31$):
airplane, backpack, baseball bat, baseball glove, bear, broccoli, bus, carrot, cat, donut,
elephant, fire hydrant, fork, frisbee, giraffe, hair drier, horse, hot dog, keyboard, kite,
knife, laptop, microwave, mouse, orange, oven, parking meter, pizza, sandwich, scissors,
sink, skateboard, skis, snowboard, spoon, stop sign, surfboard, teddy bear, tennis racket,
toaster, toilet, toothbrush, traffic light, train, wine glass, zebra.}\\
\endlastfoot

\panelhead{Panel A. Funding outcomes}
Cash per day (USD)              & 108.04  & 150.31     & 0.83  & 25.00  & 116.67 & 756.25 \\
Days to raise                   & 13.20   & 11.02      & 1     & 5      & 20     & 39     \\
Lenders per campaign            & 17.40   & 22.30      & 0     & 4      & 22     & 546    \\
Number of lenders (normalized)  & 0.012   & 0.015      & 0.001 & 0.005  & 0.015  & 1.000  \\
Default                         & 0.050   & 0.218      & 0     & 0      & 0      & 1      \\
\addlinespace

\panelhead{Panel B. Loan and market characteristics}
Loan amount (USD)               & 800.41  & 1{,}000.10 & 25    & 275    & 950    & 50{,}000 \\
Borrower count                  & 1.96    & 3.17       & 1     & 1      & 1      & 50     \\
Number of competitors (norm.)   & 0.088   & 0.164      & 0.004 & 0.008  & 0.073  & 1.000  \\
Same race-gender count          & 11.14   & 22.25      & 1     & 1      & 8      & 163    \\
Same race-gender share          & 0.685   & 0.294      & 0.004 & 0.500  & 1.000  & 1.000  \\
Total lenders (per week)        & 6.88    & 66.12      & 0     & 0      & 0      & 1{,}871 \\
\addlinespace

\panelhead{Panel C. Type features --- gender and race}
Male                            & 0.478   & 0.330      & 0.003 & 0.176  & 0.823  & 0.999  \\
Asian                           & 0.191   & 0.261      & 0.000 & 0.016  & 0.266  & 0.995  \\
White                           & 0.219   & 0.265      & 0.001 & 0.031  & 0.323  & 0.999  \\
Black                           & 0.167   & 0.281      & 0.000 & 0.006  & 0.148  & 0.990  \\
Indian                          & 0.061   & 0.098      & 0.000 & 0.009  & 0.066  & 0.962  \\
\addlinespace

\panelhead{Panel D. Type features --- age}
Baby                            & 0.004   & 0.003      & 0.000 & 0.002  & 0.006  & 0.067  \\
Child                           & 0.073   & 0.056      & 0.001 & 0.034  & 0.095  & 0.609  \\
Youth                           & 0.264   & 0.211      & 0.000 & 0.092  & 0.391  & 0.982  \\
Middle aged                     & 0.084   & 0.093      & 0.000 & 0.026  & 0.104  & 0.898  \\
Senior                          & 0.041   & 0.079      & 0.000 & 0.004  & 0.039  & 0.950  \\
\addlinespace

\panelhead{Panel E. Type features --- facial structure}
Chubby                          & 0.339   & 0.190      & 0.012 & 0.185  & 0.466  & 0.972  \\
Attractive woman                & 0.125   & 0.151      & 0.001 & 0.028  & 0.158  & 0.989  \\
Square face                     & 0.019   & 0.041      & 0.000 & 0.002  & 0.015  & 0.759  \\
Round face                      & 0.201   & 0.155      & 0.002 & 0.078  & 0.287  & 0.908  \\
Big nose                        & 0.730   & 0.190      & 0.042 & 0.606  & 0.886  & 0.998  \\
Big lips                        & 0.586   & 0.215      & 0.014 & 0.425  & 0.766  & 0.986  \\
Narrow eyes                     & 0.588   & 0.204      & 0.031 & 0.431  & 0.755  & 0.992  \\
Arched eyebrows                 & 0.451   & 0.213      & 0.004 & 0.282  & 0.618  & 0.978  \\
Bald                            & 0.037   & 0.073      & 0.000 & 0.004  & 0.030  & 0.835  \\
\addlinespace

\panelhead{Panel F. Type features --- hair}
Dark hair & 420,908 & 0.499 & 0.500 & 0 & 0 & 1 \\
Black hair                      & 0.388   & 0.242      & 0.000 & 0.171  & 0.589  & 0.970  \\
Blond hair                      & 0.007   & 0.029      & 0.000 & 0.001  & 0.004  & 0.943  \\
Brown hair                      & 0.405   & 0.156      & 0.012 & 0.288  & 0.517  & 0.919  \\
Curly hair                      & 0.394   & 0.155      & 0.031 & 0.275  & 0.499  & 0.932  \\
Wavy hair                       & 0.226   & 0.170      & 0.004 & 0.095  & 0.312  & 0.991  \\
Straight hair                   & 0.606   & 0.178      & 0.034 & 0.489  & 0.741  & 0.982  \\
Receding hairline               & 0.205   & 0.235      & 0.000 & 0.039  & 0.282  & 0.995  \\
Bangs                           & 0.171   & 0.171      & 0.001 & 0.052  & 0.229  & 0.993  \\
Sideburns                       & 0.145   & 0.195      & 0.001 & 0.025  & 0.169  & 0.977  \\
Mustache                        & 0.072   & 0.160      & 0.000 & 0.004  & 0.046  & 0.998  \\
\addlinespace

\panelhead{Panel G. Style features --- facial expression and eyewear}
Smiling                         & 0.549   & 0.177      & 0.013 & 0.424  & 0.685  & 0.966  \\
Mouth closed                    & 0.303   & 0.146      & 0.018 & 0.193  & 0.390  & 0.944  \\
Mouth wide open                 & 0.057   & 0.040      & 0.002 & 0.030  & 0.072  & 0.516  \\
Eyes open                       & 0.871   & 0.073      & 0.338 & 0.834  & 0.925  & 0.991  \\
No Eyewear                      & 0.865   & 0.148      & 0.007 & 0.830  & 0.959  & 1.000  \\
Sunglasses                      & 0.020   & 0.012      & 0.005 & 0.011  & 0.026  & 0.043  \\
\addlinespace

\panelhead{Panel H. Style features --- image composition and lighting}
Blurry                          & 0.162   & 0.095      & 0.006 & 0.090  & 0.214  & 0.758  \\
Harsh lighting                  & 0.339   & 0.165      & 0.031 & 0.217  & 0.430  & 0.930  \\
Flash                           & 0.245   & 0.126      & 0.010 & 0.148  & 0.322  & 0.855  \\
Soft lighting                   & 0.677   & 0.090      & 0.222 & 0.623  & 0.742  & 0.943  \\
Outdoor                         & 0.447   & 0.140      & 0.045 & 0.343  & 0.545  & 0.914  \\
Color photo                     & 0.948   & 0.026      & 0.632 & 0.935  & 0.966  & 0.997  \\
Posed photo                     & 0.486   & 0.132      & 0.069 & 0.391  & 0.581  & 0.925  \\
Bodyshot                        & 0.735   & 0.441      & 0     & 0      & 1      & 1      \\
Partially visible forehead      & 0.094   & 0.090      & 0.001 & 0.032  & 0.125  & 0.834  \\
Face position --- top (px)      & 157.56  & 106.75     & 0     & 80     & 204    & 1{,}598 \\
Face position --- right (px)    & 410.08  & 174.18     & 29    & 271    & 534    & 960    \\
\addlinespace

\panelhead{Panel I. Style features --- skin appearance}
Bags under eyes                 & 0.586   & 0.170      & 0.016 & 0.468  & 0.717  & 0.967  \\
Rosy cheeks                     & 0.122   & 0.069      & 0.011 & 0.072  & 0.155  & 0.729  \\
Shiny skin                      & 0.215   & 0.121      & 0.004 & 0.121  & 0.288  & 0.808  \\
Pale skin                       & 0.334   & 0.171      & 0.014 & 0.192  & 0.460  & 0.908  \\
Strong nose-mouth lines         & 0.611   & 0.172      & 0.026 & 0.496  & 0.746  & 0.966  \\
Flushed face                    & 0.102   & 0.050      & 0.009 & 0.067  & 0.126  & 0.573  \\
\addlinespace

\panelhead{Panel J. Detected objects (count per image)}
Person                          & 2.118   & 3.002      & 0     & 1      & 2      & 39     \\
Bottle                          & 0.503   & 2.259      & 0     & 0      & 0      & 99     \\
Chair                           & 0.125   & 0.498      & 0     & 0      & 0      & 24     \\
Bowl                            & 0.100   & 0.480      & 0     & 0      & 0      & 18     \\
Cup                             & 0.059   & 0.338      & 0     & 0      & 0      & 26     \\
Cow                             & 0.050   & 0.328      & 0     & 0      & 0      & 13     \\
Handbag                         & 0.048   & 0.264      & 0     & 0      & 0      & 8      \\
Book                            & 0.045   & 0.425      & 0     & 0      & 0      & 53     \\
Dining table                    & 0.038   & 0.206      & 0     & 0      & 0      & 6      \\
Car                             & 0.037   & 0.330      & 0     & 0      & 0      & 15     \\
Potted plant                    & 0.032   & 0.261      & 0     & 0      & 0      & 34     \\
Banana                          & 0.022   & 0.301      & 0     & 0      & 0      & 20     \\
Motorcycle                      & 0.022   & 0.181      & 0     & 0      & 0      & 8      \\
Bird                            & 0.019   & 0.267      & 0     & 0      & 0      & 24     \\
Refrigerator                    & 0.017   & 0.143      & 0     & 0      & 0      & 5      \\
Sheep                           & 0.015   & 0.213      & 0     & 0      & 0      & 16     \\
TV                              & 0.015   & 0.139      & 0     & 0      & 0      & 5      \\
Vase                            & 0.015   & 0.152      & 0     & 0      & 0      & 12     \\
Boat                            & 0.014   & 0.214      & 0     & 0      & 0      & 14     \\
Bicycle                         & 0.012   & 0.131      & 0     & 0      & 0      & 15     \\
Clock                           & 0.012   & 0.116      & 0     & 0      & 0      & 22     \\
Tie                             & 0.012   & 0.121      & 0     & 0      & 0      & 7      \\
Remote                          & 0.011   & 0.110      & 0     & 0      & 0      & 4      \\
Apple                           & 0.010   & 0.159      & 0     & 0      & 0      & 22     \\
Umbrella                        & 0.010   & 0.119      & 0     & 0      & 0      & 6      \\
Cell phone                      & 0.009   & 0.115      & 0     & 0      & 0      & 14     \\
Dog                             & 0.009   & 0.101      & 0     & 0      & 0      & 5      \\
Bed                             & 0.008   & 0.090      & 0     & 0      & 0      & 3      \\
Bench                           & 0.008   & 0.099      & 0     & 0      & 0      & 7      \\
Suitcase                        & 0.008   & 0.108      & 0     & 0      & 0      & 10     \\
Cake                            & 0.006   & 0.086      & 0     & 0      & 0      & 6      \\
Couch                           & 0.006   & 0.091      & 0     & 0      & 0      & 4      \\
Truck                           & 0.006   & 0.083      & 0     & 0      & 0      & 6      \\
Sports ball                     & 0.005   & 0.074      & 0     & 0      & 0      & 7      \\

\end{longtable}
\normalsize}

\section{Measurement Error in Visual Features and Its Implications}
\label{app:measurement_error}

This appendix examines the role of measurement error in the machine learning (ML)–generated visual features used throughout the analysis. Because our empirical strategy relies on these features as regressors, any misclassification may introduce bias in estimated treatment effects. We proceed in two steps. First, we conduct an audit of the ML-generated features by comparing algorithmic predictions to human-labeled ground truth. This allows us to quantify feature-specific false positive and false negative rates. Second, we use these estimated misclassification rates to implement a Monte Carlo simulation extrapolation (\emph{MC-SIMEX}) procedure, which evaluates how estimated treatment effects evolve as additional noise is introduced and extrapolates back to a counterfactual setting with no measurement error.

The goal of this analysis is threefold. First, we assess the sensitivity of our estimates to increasing levels of misclassification. Second, we construct corrected estimates that account for the estimated error structure. Third, we repeat the simulation of counterfactual policies to assess whether the Style Recommendation policies relax the disparity-efficiency tradeoff after the SIMEX correction.

\subsection{Audit of CNN Features}\label{appendix:audit}

To quantify misclassification in the ML-generated visual features, we conduct an audit study comparing algorithmic predictions to human-coded ground truth. Using a sample of borrower images annotated by a third-party provider, we estimate feature-specific false-positive and false-negative rates, which we subsequently use in the measurement-error correction analysis.

We drew a stratified sample of 2,300 borrower images. Independent annotations were obtained from MolarData, a third-party data annotation provider.\footnote{MolarData(\url{https://www.molardata.com/})}. Annotators label the visual features used in the experimental analysis: \textit{male}, \textit{smiling}, \textit{youth}, \textit{glasses}, \textit{sunglasses}, and \textit{dark hair}. In the annotation guidelines, age is coded to mirror the observed Kiva age distribution: annotators label age groups in ten-year intervals. We define \textit{youth} as individuals below the annotated age of 20, and \textit{dark hair} as the darkest category on a five-point hair-color scale. For each feature, annotators indicate whether the the feature is present or absent, or that the image cannot be audited for technical reasons, such as low resolution, ambiguity, or other image defects. Table~\ref{tab:annot_summary} reports summary statistics for the annotated sample.

\begin{table}[!htbp] \centering
  \caption{Summary Statistics of Annotation Data}
  \label{tab:annot_summary}
  \scriptsize
  \setlength{\tabcolsep}{3pt}
  \resizebox{0.8\textwidth}{!}{%
\begin{tabular}{@{\extracolsep{4pt}}p{2.5cm}ccccccc}
\\[-1.8ex]\hline
\hline \\[-1.8ex]
Statistic & N & Mean & Min & Pctl(25) & Median & Pctl(75) & Max \\
\hline \\[-1.8ex]
Smiling & 1,505 & 0.3714 & 0.0000 & 0.0000 & 0.0000 & 1.0000 & 1.0000 \\
Glasses & 2,297 & 0.0313 & 0.0000 & 0.0000 & 0.0000 & 0.0000 & 1.0000 \\
Sunglasses & 2,297 & 0.0126 & 0.0000 & 0.0000 & 0.0000 & 0.0000 & 1.0000 \\
Male & 2,297 & 0.2904 & 0.0000 & 0.0000 & 0.0000 & 1.0000 & 1.0000 \\
Youth & 2,297 & 0.0556 & 0.0000 & 0.0000 & 0.0000 & 0.0000 & 1.0000 \\
Dark Hair & 2,221 & 0.9878 & 0.0000 & 1.0000 & 1.0000 & 1.0000 & 1.0000 \\
\hline
\hline \\[-1.8ex]
\end{tabular}%
}
\caption*{\footnotesize{\textit{Note:} For the smiling annotation, annotators were allowed to indicate that they were "unsure" whether an image showed a smile, likewise for Dark Hair Color; these unsure cases are recorded as missing. The uncertain or unavailable annotations and are excluded from the misclassification-matrix calculations.}}
\end{table}

We treat human labels as ground truth and compare the ML features against it. Table~\ref{tab:audit_misclass} reports the resulting misclassification matrices, expressed as row percentages. Several patterns emerge. First, misclassification is highly asymmetric for some features. For example, the \textit{male} indicator exhibits a very low false-negative rate but a substantial false-positive rate, indicating that the algorithm tends to over-predict male classifications. Second, some features are measured with relatively high precision: \textit{glasses} has a very low false-positive rate, although a nontrivial share of true positives is missed. Third, other features display more balanced but non-negligible two-sided misclassification, including \textit{smiling} and \textit{Dark hair}. Finally, features such as \textit{youth} and especially \textit{sunglasses} are measured more noisily, consistent with the difficulty of visually identifying these attributes and their low prevalence in the sample.

\begin{table}[!htbp] \centering 
  \caption{Human--CNN Misclassification Rates (Row Percentages)} 
  \label{tab:audit_misclass} 
\footnotesize
\setlength{\tabcolsep}{6pt}
\renewcommand{\arraystretch}{1.1}

\begin{minipage}[t]{0.32\textwidth}
\centering
\textit{Panel A. Male}\\[0.3em]
\begin{tabular}{lcc}
\hline\hline
 & \multicolumn{2}{c}{CNN} \\
Human & 0 & 1 \\
\hline
0 & 68.3\% & 31.7\% \\
1 & 0.8\% & 99.2\% \\
\hline\hline
\end{tabular}
\end{minipage}
\hfill
\begin{minipage}[t]{0.32\textwidth}
\centering
\textit{Panel B. Smiling}\\[0.3em]
\begin{tabular}{lcc}
\hline\hline
 & \multicolumn{2}{c}{CNN} \\
Human & 0 & 1 \\
\hline
0 & 74.1\% & 25.9\% \\
1 & 29.4\% & 70.6\% \\
\hline\hline
\end{tabular}
\end{minipage}
\hfill
\begin{minipage}[t]{0.32\textwidth}
\centering
\textit{Panel C. Youth}\\[0.3em]
\begin{tabular}{lcc}
\hline\hline
 & \multicolumn{2}{c}{CNN} \\
Human & 0 & 1 \\
\hline
0 & 81.3\% & 18.7\% \\
1 & 40.1\% & 59.9\% \\
\hline\hline
\end{tabular}
\end{minipage}

\vspace{1em}

\begin{minipage}[t]{0.32\textwidth}
\centering
\textit{Panel D. Glasses}\\[0.3em]
\begin{tabular}{lcc}
\hline\hline
 & \multicolumn{2}{c}{CNN} \\
Human & 0 & 1 \\
\hline
0 & 98.2\% & 1.8\% \\
1 & 30.6\% & 69.4\% \\
\hline\hline
\end{tabular}
\end{minipage}
\hfill
\begin{minipage}[t]{0.32\textwidth}
\centering
\textit{Panel E. Dark Hair}\\[0.3em]
\begin{tabular}{lcc}
\hline\hline
 & \multicolumn{2}{c}{CNN} \\
Human & 0 & 1 \\
\hline
0 & 73.3\% & 26.7\% \\
1 & 28.0\% & 72.0\% \\
\hline\hline
\end{tabular}
\end{minipage}
\hfill
\begin{minipage}[t]{0.32\textwidth}
\centering
\textit{Panel F. Sunglasses}\\[0.3em]
\begin{tabular}{lcc}
\hline\hline
 & \multicolumn{2}{c}{CNN} \\
Human & 0 & 1 \\
\hline
0 & 67.8\% & 32.2\% \\
1 & 33.3\% & 66.7\% \\
\hline\hline
\end{tabular}
\end{minipage}

\caption*{\footnotesize{\textit{Note:} Entries report row percentages, with human annotations treated as the benchmark. Each panel shows the feature-specific misclassification matrix used to recover false-positive and false-negative rates for the SIMEX correction.}}
\end{table}

Overall, the audit reveals that measurement error is substantial and varies systematically across features. These feature-specific false-positive and false-negative rates form the basis for the SIMEX correction implemented in the next section.

\subsection{Measurement-Error Correction via MC-SIMEX}
\label{app:simex}

To assess how measurement error in the ML-generated features affects our estimates, we implement a Monte Carlo simulation extrapolation (MC-SIMEX) procedure \citep{cook1994simulation, carroll2006measurement}. The key idea of SIMEX is to use the empirically estimated misclassification rates from the audit to simulate additional measurement error, evaluate how estimated treatment effects change as noise increases, and then extrapolate back to a counterfactual setting with no measurement error.

Concretely, for each feature, we begin with the observed (noisy) binary indicator and its associated false-positive and false-negative rates estimated from the audit. We then generate a sequence of perturbed treatment variables by introducing additional misclassification, scaled by a parameter $\lambda \geq 0$, which governs the intensity of the added noise. For each value of $\lambda$, we draw multiple Monte Carlo realizations of the perturbed treatment and re-estimate the corresponding average treatment effect using the same specification as in the main analysis. This yields a sequence of estimates that traces how the estimated effect evolves as measurement error is progressively amplified. In the final step, we fit a smooth function to the relationship between the estimated treatment effect and the noise level $\lambda$, and extrapolate this function to $\lambda = -1$, which corresponds to the hypothetical case of no measurement error. This extrapolation produces a SIMEX-corrected estimate of the treatment effect.

Table~\ref{tab:simex_correction} reports the results of the MC-SIMEX procedure for the main visual features. For each feature, the table presents the estimated treatment effect at increasing levels of simulated misclassification, indexed by $\lambda$, along with the corresponding SIMEX-corrected estimate obtained via extrapolation. The column $\lambda=0$ corresponds to the baseline observational estimate, while higher values of $\lambda$ reflect progressively noisier versions of the treatment variable constructed using the audit-based misclassification rates.

\begin{table}[!htbp] \centering
\caption{MC-SIMEX correction of observational ATE estimates for selected profile features}
\label{tab:simex_correction}
\resizebox{0.8\textwidth}{!}{
\begin{tabular}{lccccc}
\toprule
Feature & $\lambda = 0$ & $\lambda = 0.25$ & $\lambda = 0.50$ & $\lambda = 1.00$ & Corrected \\
\midrule
\textit{Smiling} & 12.256 (0.358) & 8.754 (0.370) & 6.310 (0.374) & 3.828 (0.377) & 34.772\\
\textit{Glasses} & -8.583 (0.360) & -9.057 (0.362) & -9.293 (0.365) & -9.995 (0.369) & -6.857\\
\textit{Sunglasses} & -9.587 (0.366) & -6.406 (0.372) & -4.760 (0.375) & -2.904 (0.377) & -29.104\\
\textit{Dark Hair} & 0.296 (0.363) & 0.621 (0.376) & 0.295 (0.378) & 0.374 (0.378) & -0.266\\
\textit{Male} & -9.941 (0.261) & -4.916 (0.337) & -0.998 (0.353) & 0.383 (0.347) & -49.987\\
\textit{Youth} & -8.350 (0.291) & -7.835 (0.372) & -6.709 (0.385) & -5.211 (0.387) & -11.279\\
\bottomrule
\end{tabular}
}
\caption*{\footnotesize{\textit{Note:} For each feature and each noise level $\lambda$, entries in columns $\lambda=0$, $\lambda=0.25$, $\lambda=0.50$, and $\lambda=1.00$ report the mean ATE across all available Monte Carlo replications from both runs; mean standard errors across replications are reported in parentheses. The final column reports the average of the SIMEX-corrected estimates.}}
\end{table}

Two main patterns emerge. First, for several features, including \textit{smiling} and \textit{youth}, the estimated effects attenuate smoothly toward zero as additional noise is introduced. For example, the estimated effect of \textit{smiling} declines by nearly 70\% between $\lambda=0$ and $\lambda=1$, while remaining statistically significant throughout. This pattern is consistent with measurement error biasing estimates toward zero, suggesting that the baseline estimates understate the magnitude of the true effect.

Second, for other features, the response to measurement error is more complex, reflecting the non-classical nature of misclassification. For instance, the \textit{male} indicator shows sensitivity: the estimated effect shrinks and changes sign as $\lambda$ increases, indicating that inference for this feature is sensitive to misclassification. 

The final column reports SIMEX-corrected estimates. For features exhibiting smooth attenuation (e.g., \textit{smiling} and \textit{youth}), the corrected estimates are larger in magnitude, consistent with classical attenuation bias. In contrast, for features with unstable or non-monotonic paths (notably \textit{male}), the extrapolated corrections are highly sensitive and should be interpreted with caution. Measurement error therefore matters in different ways across features: it attenuates some effects and shifts the sign of others.

\subsection{Counterfactual Policy Simulations with SIMEX-Adjusted Coefficients}
\label{app:simex_policies}

As a final exercise, we re-estimate the policy counterfactuals using the SIMEX-adjusted coefficients in place of the original observational parameter estimates. This addresses the concern that measurement error in the ML-generated visual features may attenuate the estimated importance of those attributes and therefore understate the effects of interventions that operate through them.

\begin{figure}[!htbp]
\centering
\caption{Efficiency--disparity tradeoff under SIMEX-adjusted coefficients}
\includegraphics[width=0.5\textwidth]{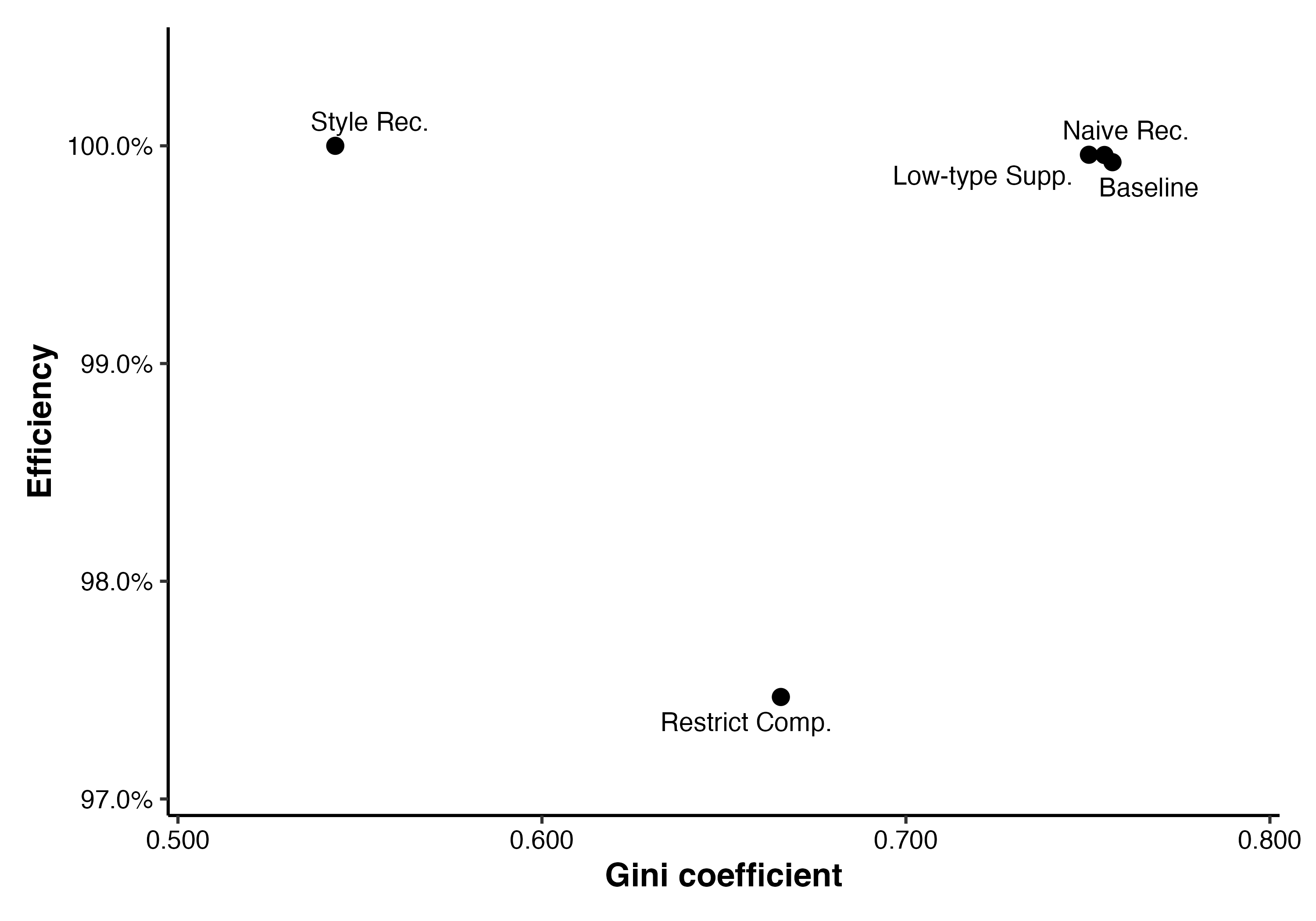}
\label{fig:simex_policies_gini}
\caption*{\footnotesize{\textit{Note:} Simulated policy outcomes using SIMEX-corrected coefficients in place of the baseline observational estimates. The horizontal axis shows disparity, measured by the Gini coefficient of allocation outcomes, while the vertical axis reports efficiency, defined as the share of lending activity occurring within the platform. Each point corresponds to a policy intervention.}}
\end{figure}

Figure~\ref{fig:simex_policies_gini} reports the resulting efficiency--disparity frontier, where disparity is measured by the Gini coefficient of allocation outcomes. Relative to the baseline simulations, the qualitative ranking of policies is unchanged. 

The most notable result is that Style Recommendation continues to reduce disparity substantially while preserving essentially the same level of efficiency. By contrast, Restrict Competition also lowers disparity, but at a somewhat greater efficiency cost. The remaining policies are closer to the baseline and therefore deliver more limited gains in this specification. The differences between policies on the efficiency score are minor.

\section{Choice of the predictive model}\label{pred_model}
In this section, we present a comparison of the accuracy of the GBM to other predictive models. We consider several predictive models over three specifications and determine the model to be used in the baseline analysis. 

We analyze the performance of models predicting cash per day. We consider the following models: Linear Regression, LASSO, Random Forrest (grf), and Boosted Random Forrest (grf and gbm). All models (except for LM) are tuned for the task at hand, we report the performance of the selected best (lowest MSE) model. All models are trained using a 70\% sample of Kiva data and tested on the 30\%.

We consider three specifications differing by the number of covariates: (A) covariates include: details of the loan including amount, repayment scheme, \emph{sector}, \emph{country}, etc. and weekly dummies, (B) details of the photo including both type and style characteristics, (C) total number of active lenders in this \emph{week}*\emph{sector}, total number of competitors in this \emph{week}*\emph{sector}, number of competitors of the same \emph{race} and \emph{gender}, and interaction of \emph{week} and \emph{sector}, and interaction of \emph{week} and \emph{country}. For boosted Forrest we also add a 4th specification where we have a sufficient representation of \emph{week}* \emph{sector} (D) \citep{johannemann2019sufficient}. Table \ref{tab:test_set_perf} presents results.

\begin{table}[!ht]
    \centering
      \caption{Comparison of the test-set predictive performance of selected models} 
    \begin{tabular}{llll}
\\[-1.8ex]\hline 
\hline \\[-1.8ex] 
        Model & Specification & MSE & SE \\ \hline
        Linear regression & A & 13840 & 159 \\ 
        Linear regression & B & 13466 & 155 \\ 
        Linear regression & C & 13565 & 166 \\ \hline
        LASSO & A & 13797 & 161 \\ 
        LASSO & B & 13379 & 157 \\ 
        LASSO & C & 13183 & 156 \\ \hline
        Random forest & A & 13930 & 163 \\
        Random forest & B & 13530 & 145 \\ 
        Random forest & C & 13099 & 157 \\ \hline
        Boosted forest (gbm) & A & 12235 & 156 \\ 
        Boosted forest (gbm) & B & 11477 & 141 \\ 
        Boosted forest (gbm) & C & 10929 & 157 \\ 
        Boosted forest (gbm) & D & 11406 & 173 \\ \hline
        Boosted forest (grf) & A & 12665 & 147 \\ 
        Boosted forest (grf) & B & 12003 & 149 \\ 
        Boosted forest (grf) & C & 11777 & 139 \\ 
        Boosted forest (grf) & D & 11962 & 177 \\ 
        \\[-1.8ex]\hline 
\hline \\[-1.8ex] 
    \end{tabular}
        \caption*{\footnotesize{\textit{Note: Test set performance of selected predictive models with different sets of covariates.}}}\label{tab:test_set_perf}
\end{table}

We conclude that Boosted Forrest has the best test-set predictive performance across all specifications and we decide to use it as a baseline model for the predictive tasks throughout the paper. \emph{GBM} implementation of the Boosted Forrest has better performance than \emph{GRF}, the difference is moderately small. Sufficient representation does not improve models' performance and will not be used in the predictive tasks.

\section{Model Diagnostics}\label{GBM_diagnostics}

In Appendix \ref{GBM_diagnostics}, we expand the set of outcome variables and consider a constructed variable, which adjusts for systematic differences in cash per day and total loan amounts requested across business categories. It also provides diagnostics for the GBM models.

\subsection{Alternative outcome variable}\label{alt_outcome_appendix}

To account for heterogeneity in the supply-side motivations of lenders on Kiva, we introduce a new outcome variable that captures the cash collected by a borrower relative to the total funds requested, adjusted for differences within and across categories. This adjustment reflects that lenders' preferences and funding behaviors vary both by borrower characteristics and by the categories Kiva uses to segment loans. Specifically, we compute the outcome variable as:

\[
\text{Adjusted Outcome} = \left( \frac{\text{Cash per Day}}{\text{Category Average Cash per Day}} \right) \cdot \left( \frac{\text{Category Average Requested}}{\text{Platform Average Requested}} \right).
\]

This formula incorporates two key components. First, the ratio of the cash collected per day to the category average cash collected per day accounts for differences within categories, capturing how well a specific loan is performing relative to others in the same category. Second, the ratio of the category average requested to the overall platform average requested adjusts for differences across categories, recognizing that funding needs and typical loan amounts can vary significantly across different types of projects (e.g., drilling a town’s well versus supporting a seed entrepreneur). The resulting variable measures lender behavior net of intra-category and inter-category differences.

\subsection{Diagnostics for the GBM models}
Table \ref{tab:image_pred_def_new_outcome} shows test-set MSE for the three GBM models: a constant model, a model with style features, and a full model. Note that differences in outcomes across categories are already partly accounted for in the outcome variable. 

\begin{table} \centering 
  \caption{Image features as predictors of \emph{adjusted outcome}.} 
  \label{tab:image_pred_def_new_outcome} 
\resizebox{0.3\textwidth}{!}{%
\begin{tabular}{lcc}
\toprule
Specification & MSE & SE\\
\midrule
Mean model & 1.977 & 0.018 \\
Style features & 1.512 & 0.014\\
Full model & 1.281 & 0.012\\ 
\bottomrule
\end{tabular}
}
\caption*{\footnotesize{\textit{Note: Test set performance of a gradient boosted machine (GBM) trained using all available covariates (full model) and simplified model using image style features (Style features) and a model with only an intercept (Mean model). Models trained on 70\% of data and tested on 30\%. Mean squared errors are in the second column. Standard errors of MSE are in the third column.}}}
\end{table}

We find that including style features in the predictive model of the adjusted outcome improves the predictive accuracy as measured by the MSE. The difference between the MSE of the mean model and the model with style features is statistically significant.

\begin{figure}[ht]
\centering
\caption{Histograms of error terms across model specifications and outcomes}
\label{fig:diagnostics_histograms}
\includegraphics[width=.32\textwidth]{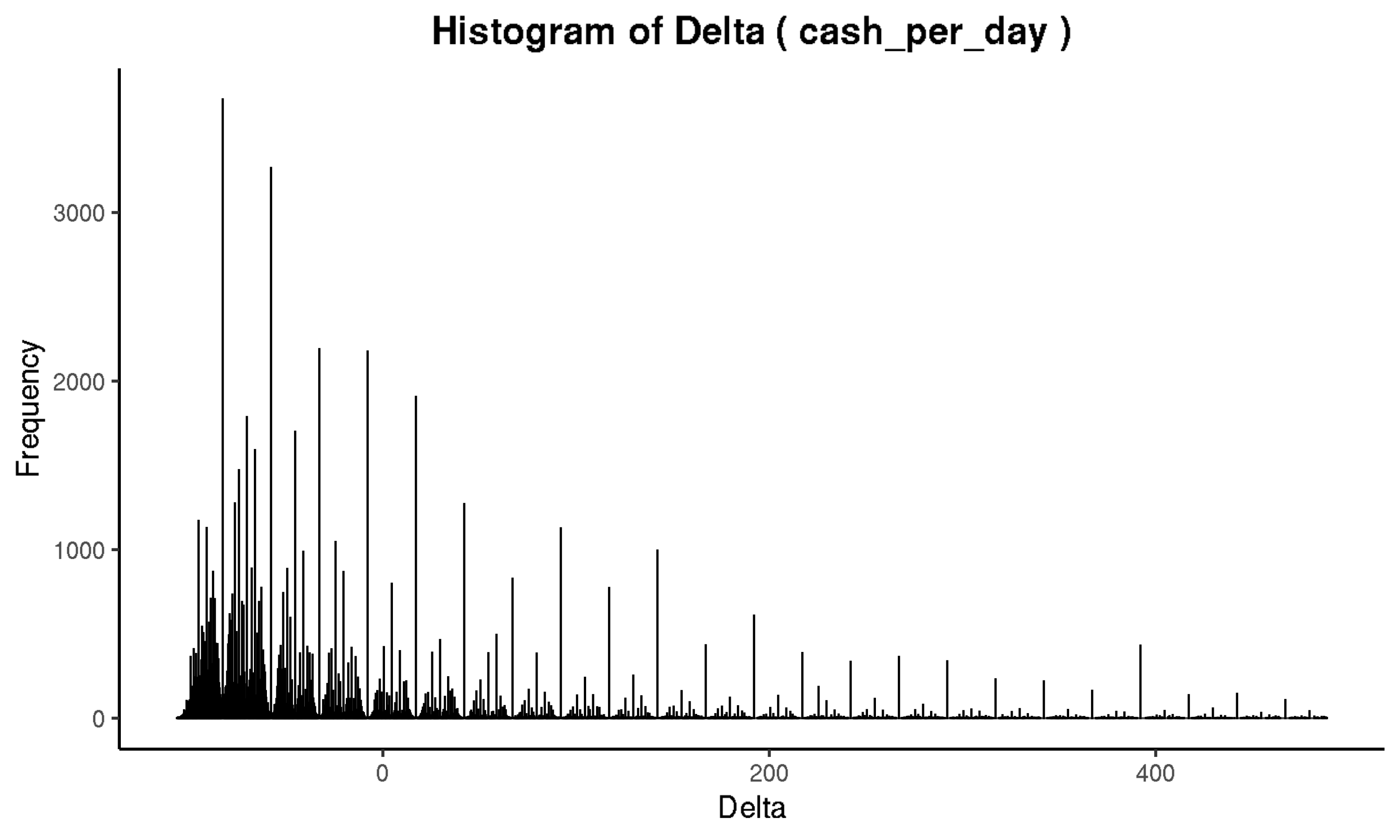}\hfill
\includegraphics[width=.32\textwidth]{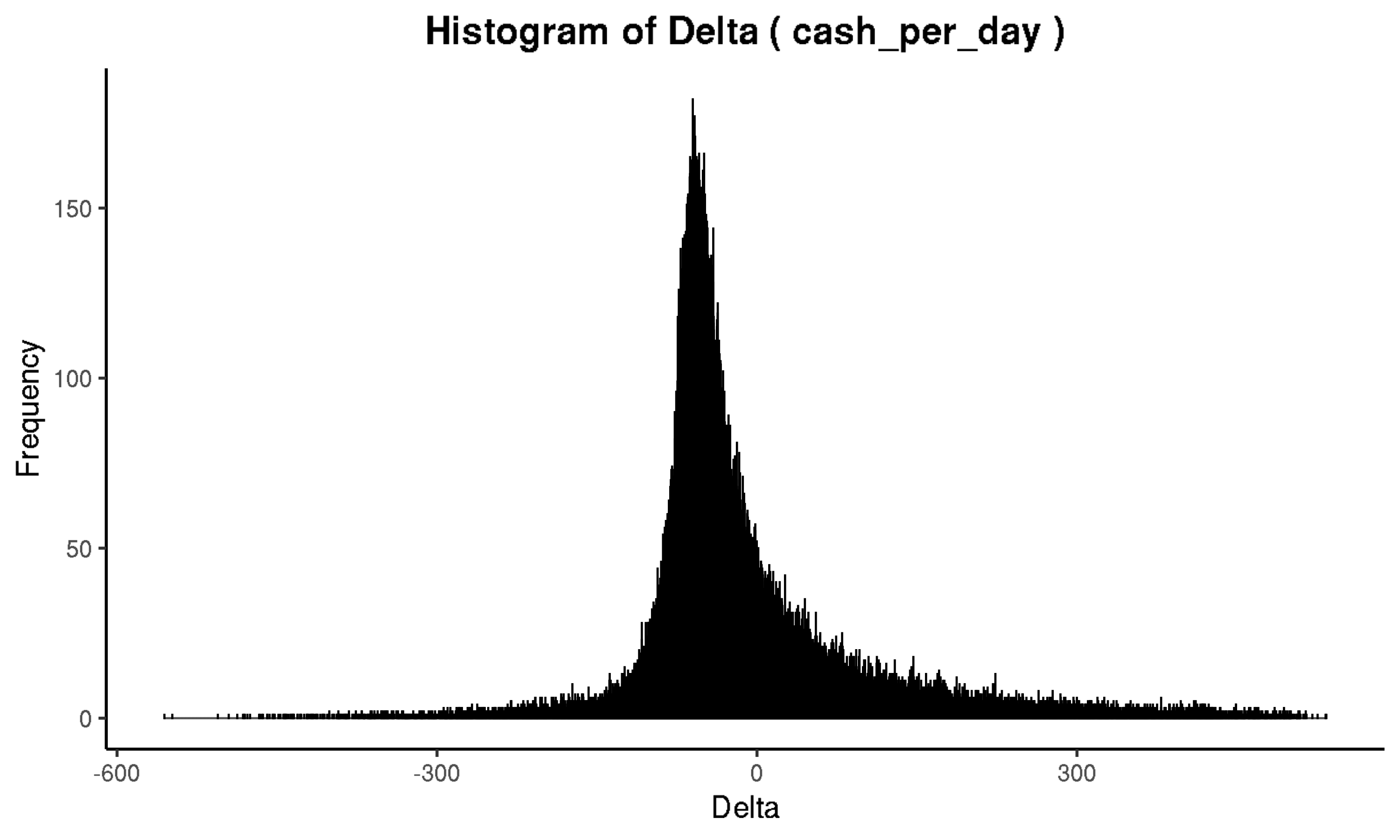}\hfill
\includegraphics[width=.32\textwidth]{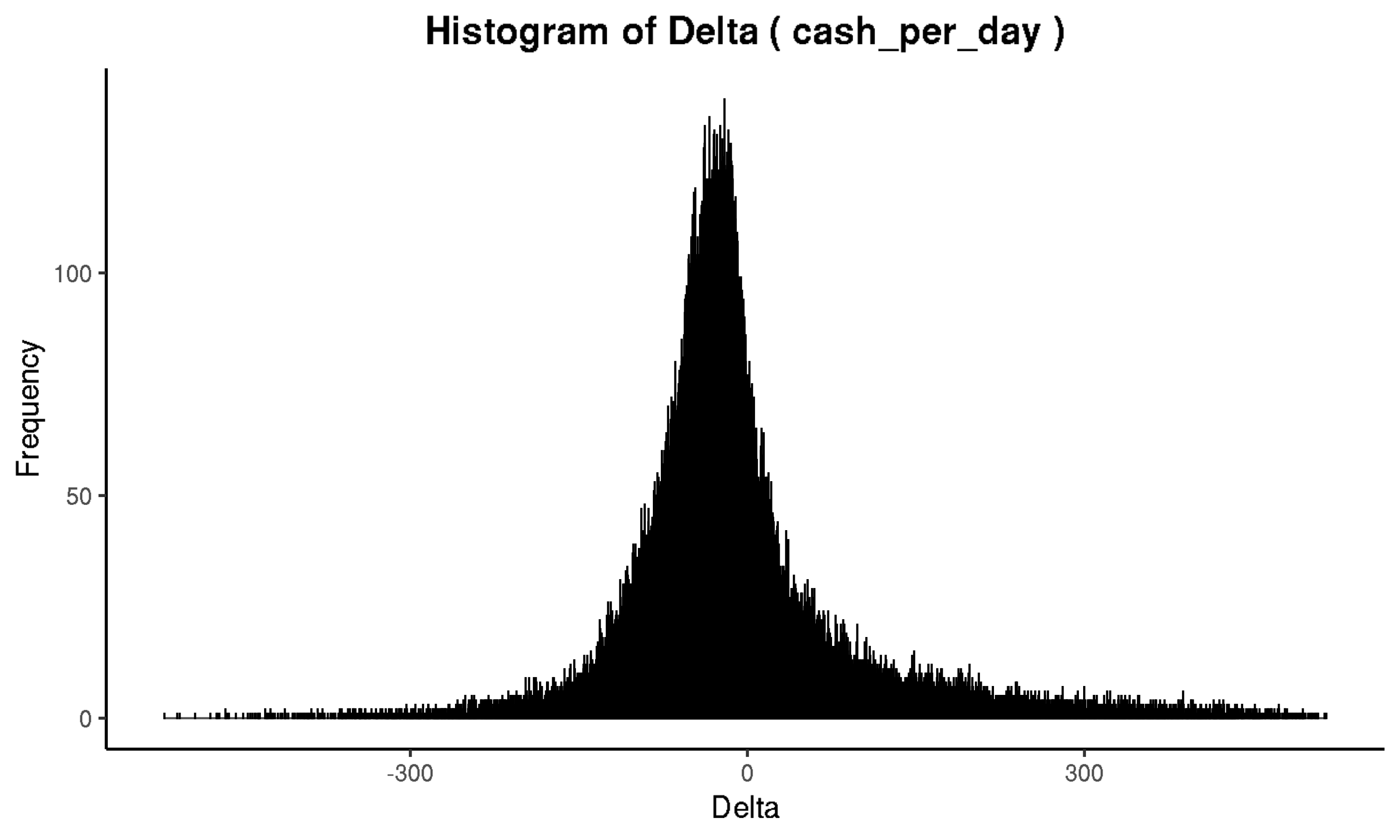}

\vspace{0.5em}

\includegraphics[width=.32\textwidth]{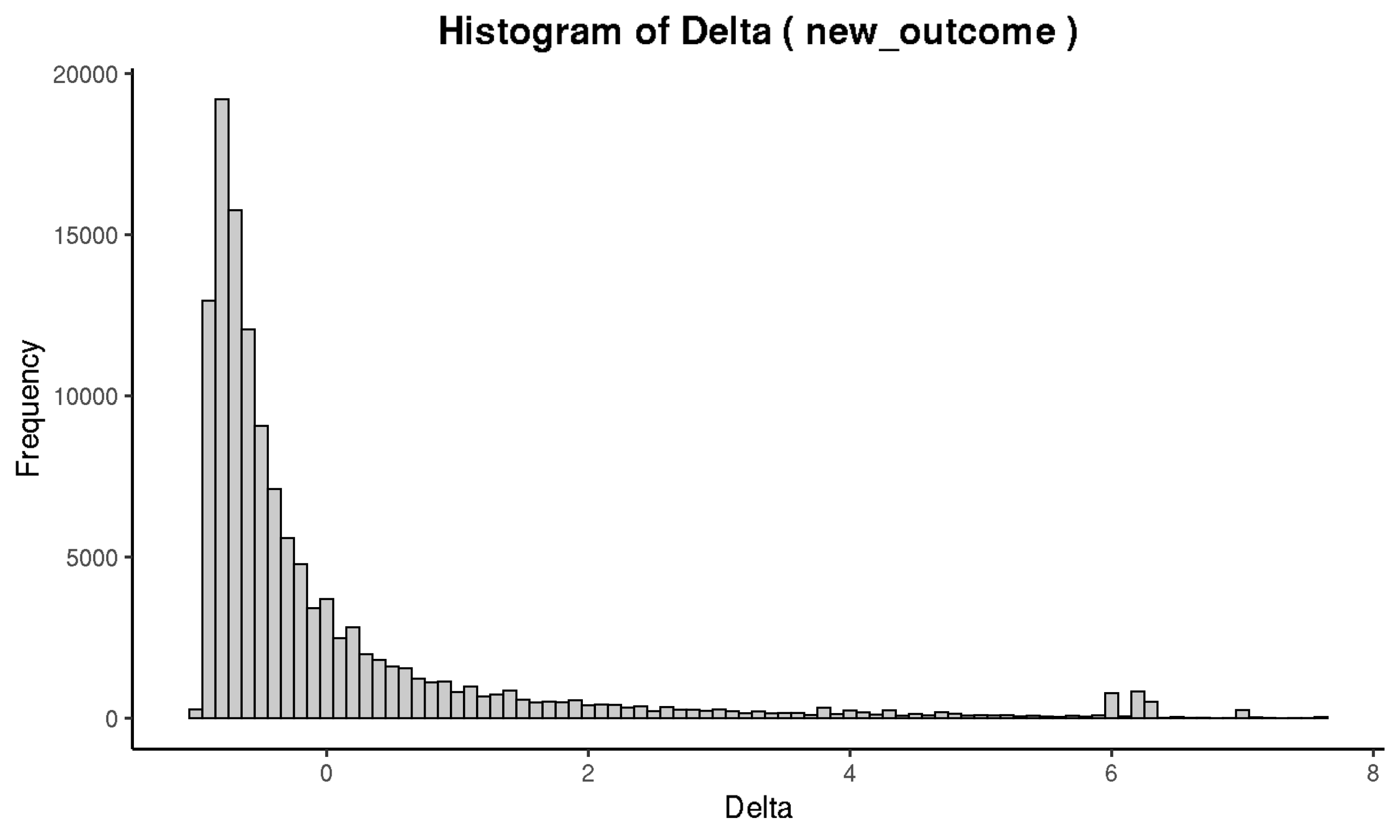}\hfill
\includegraphics[width=.32\textwidth]{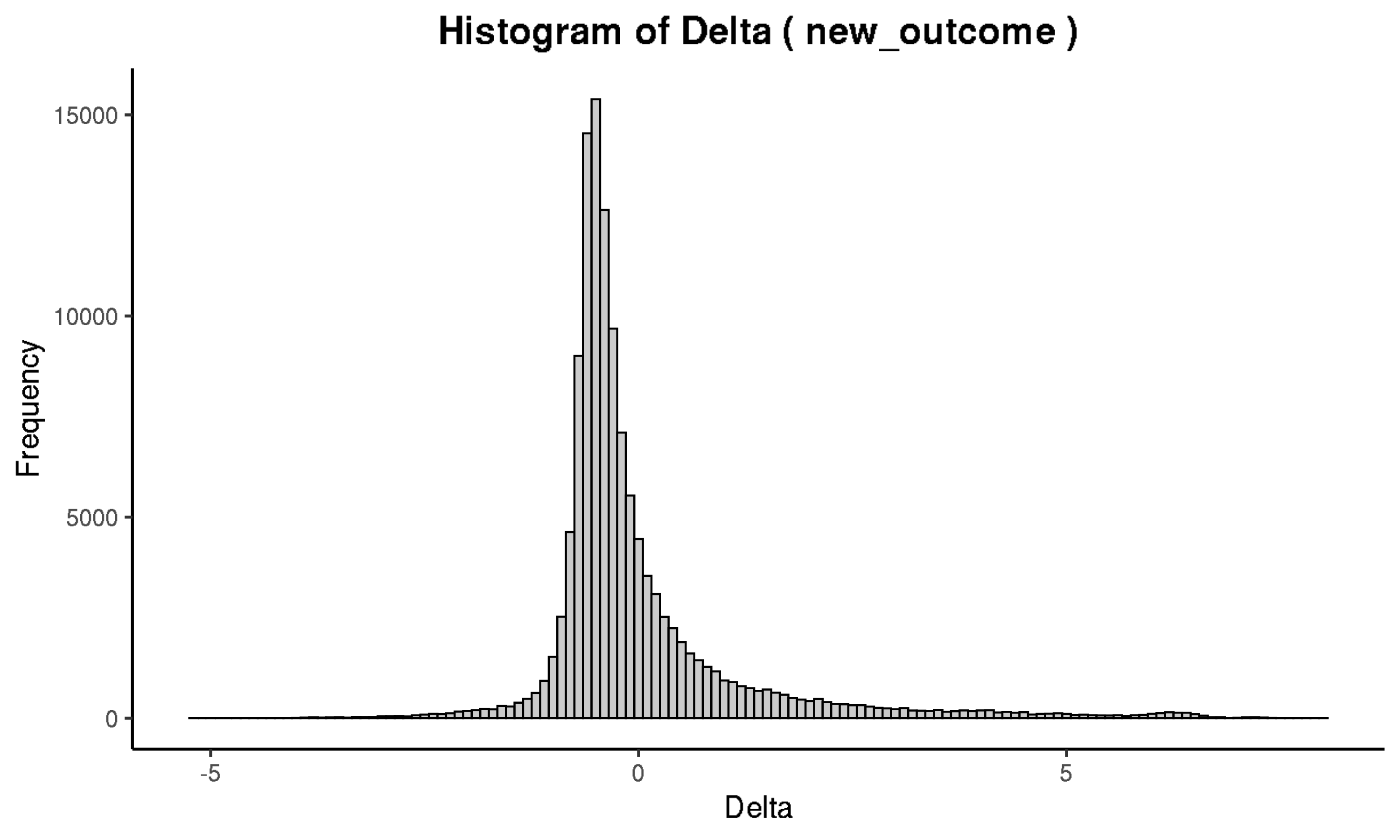}\hfill
\includegraphics[width=.32\textwidth]{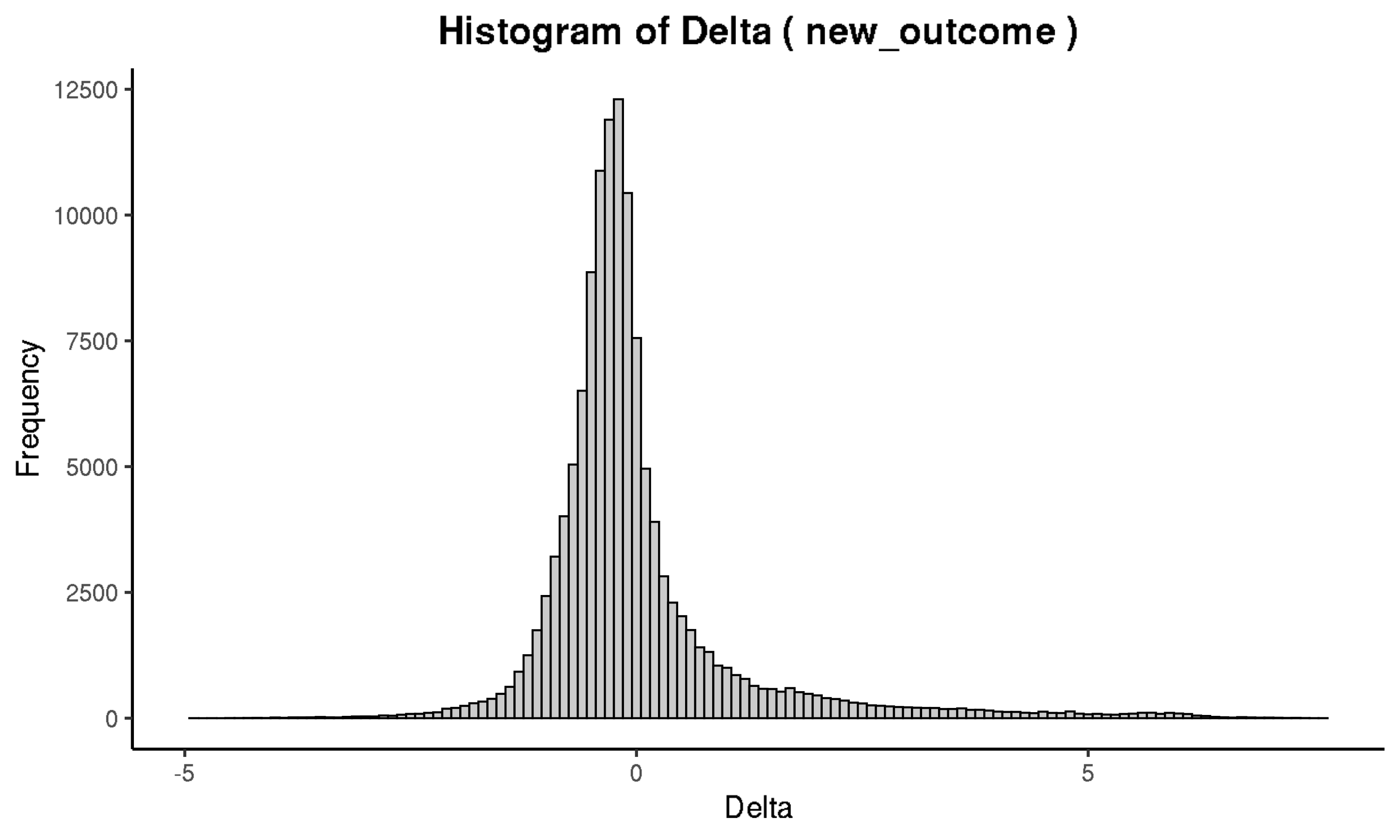}
\caption*{\footnotesize\textit{Note: Histograms of prediction errors from the three GBM specifications. Columns, from left to right: constant-only model, style-features model, and full model. Top row: cash per day. Bottom row: adjusted outcome.}}
\end{figure}

\begin{figure}[ht]
\centering
\caption{Scatter plots of observed versus predicted values across model specifications and outcomes}
\label{fig:diagnostics_scatter}
\includegraphics[width=.32\textwidth]{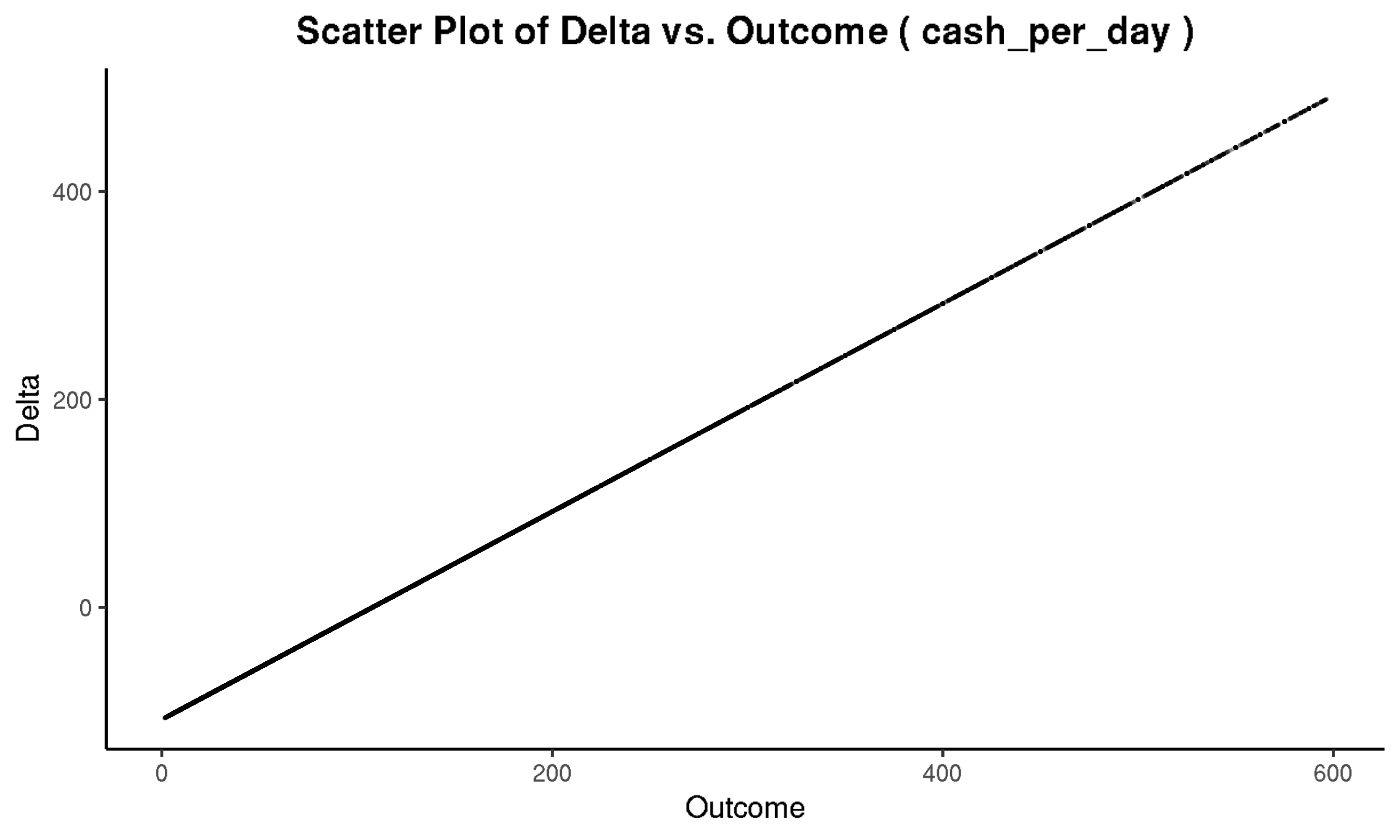}\hfill
\includegraphics[width=.32\textwidth]{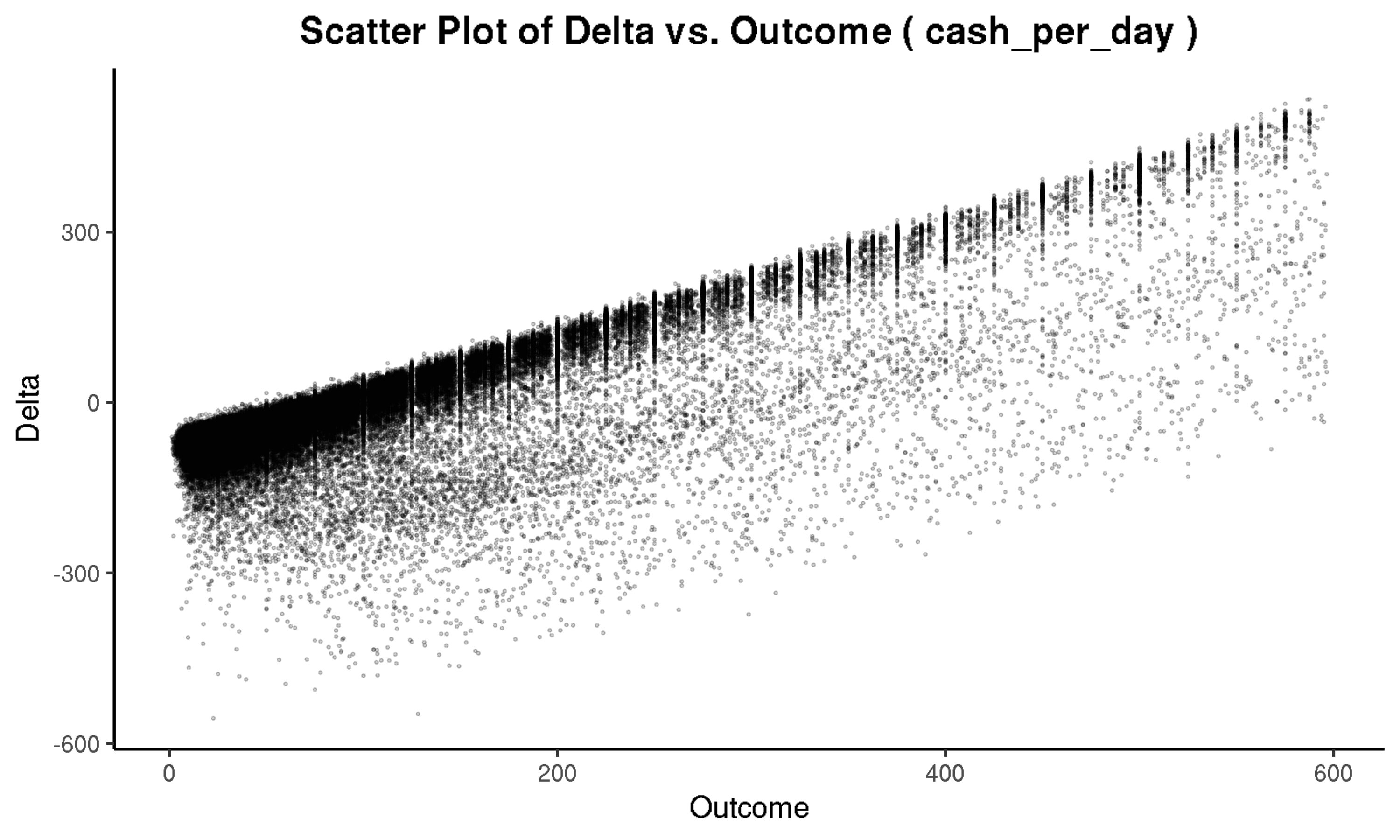}\hfill
\includegraphics[width=.32\textwidth]{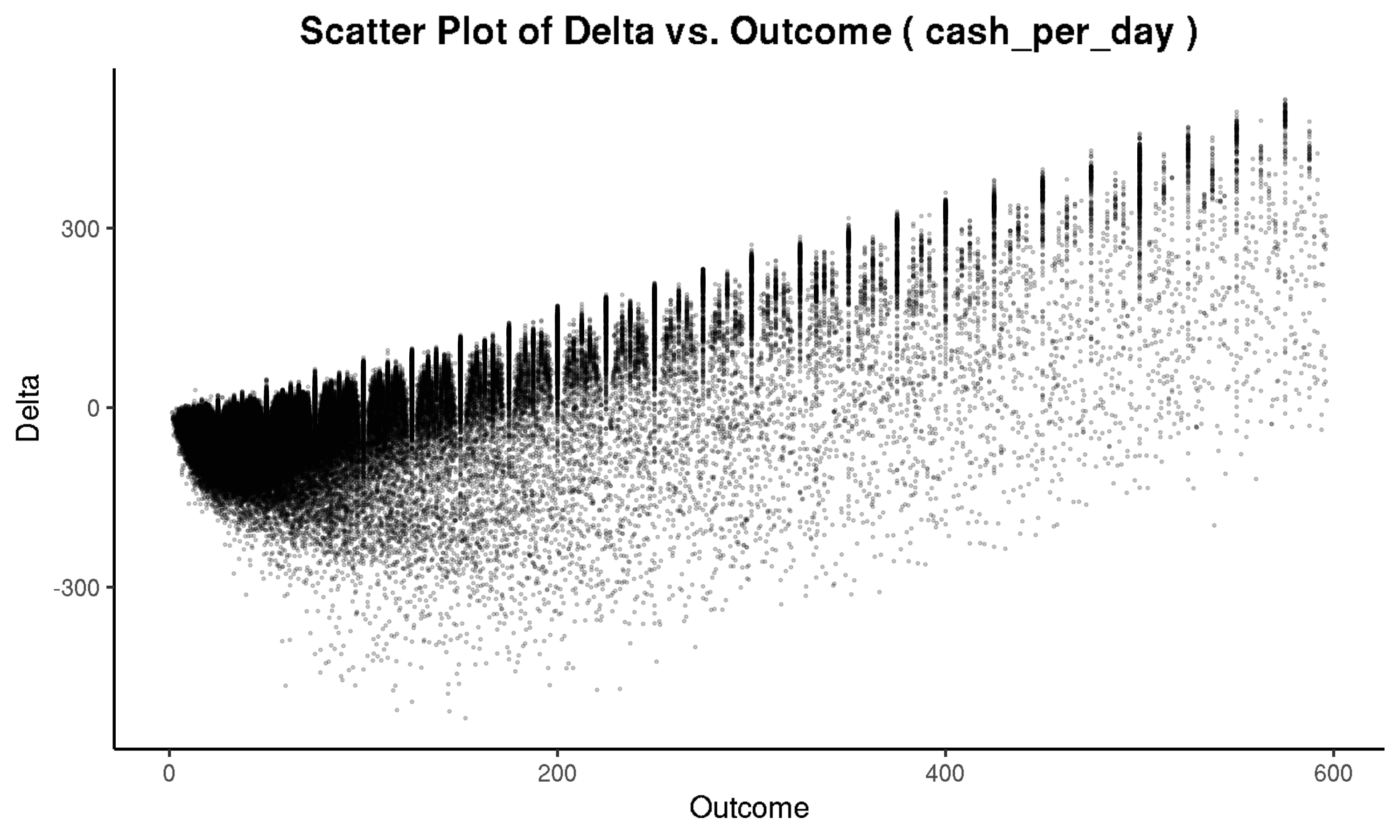}

\vspace{0.5em}

\includegraphics[width=.32\textwidth]{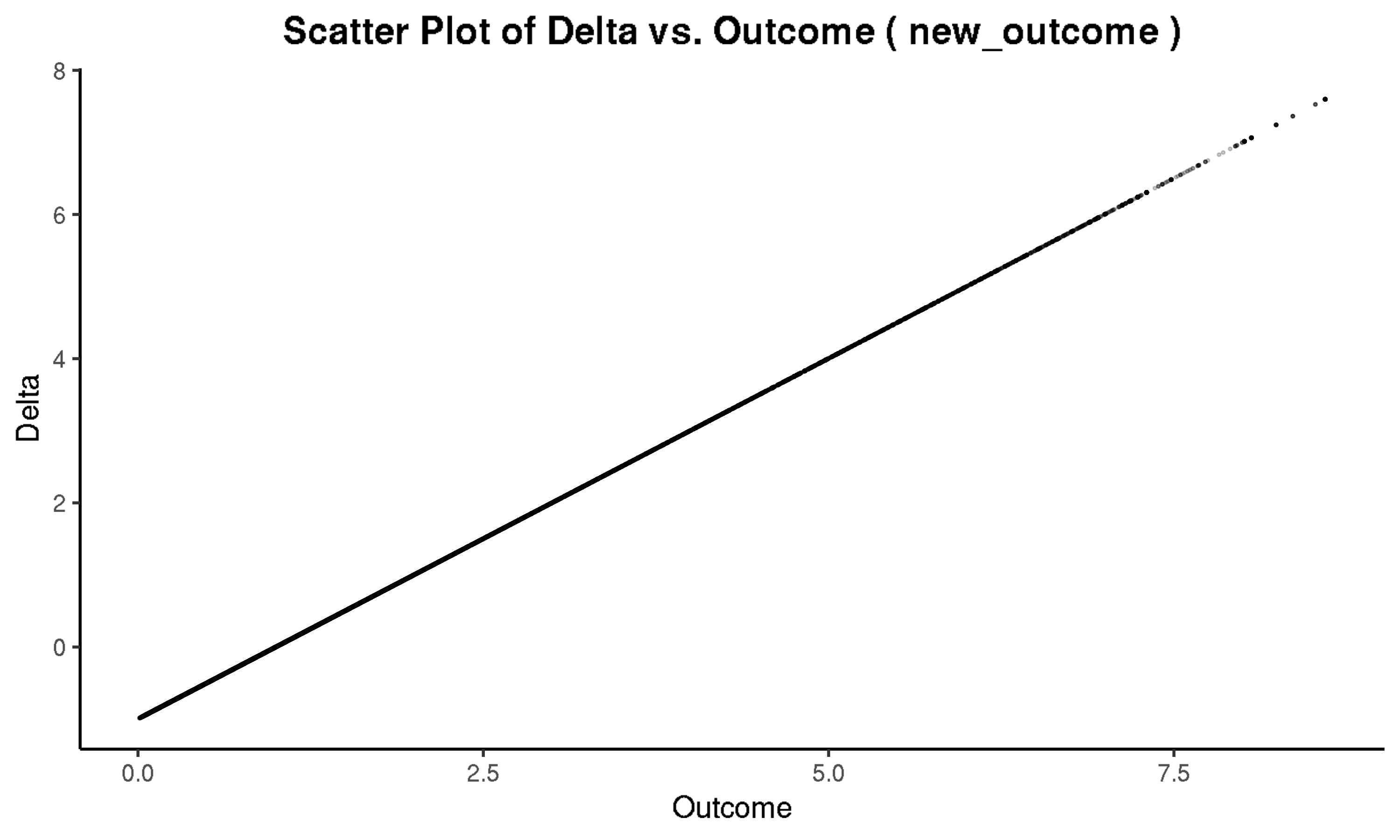}\hfill
\includegraphics[width=.32\textwidth]{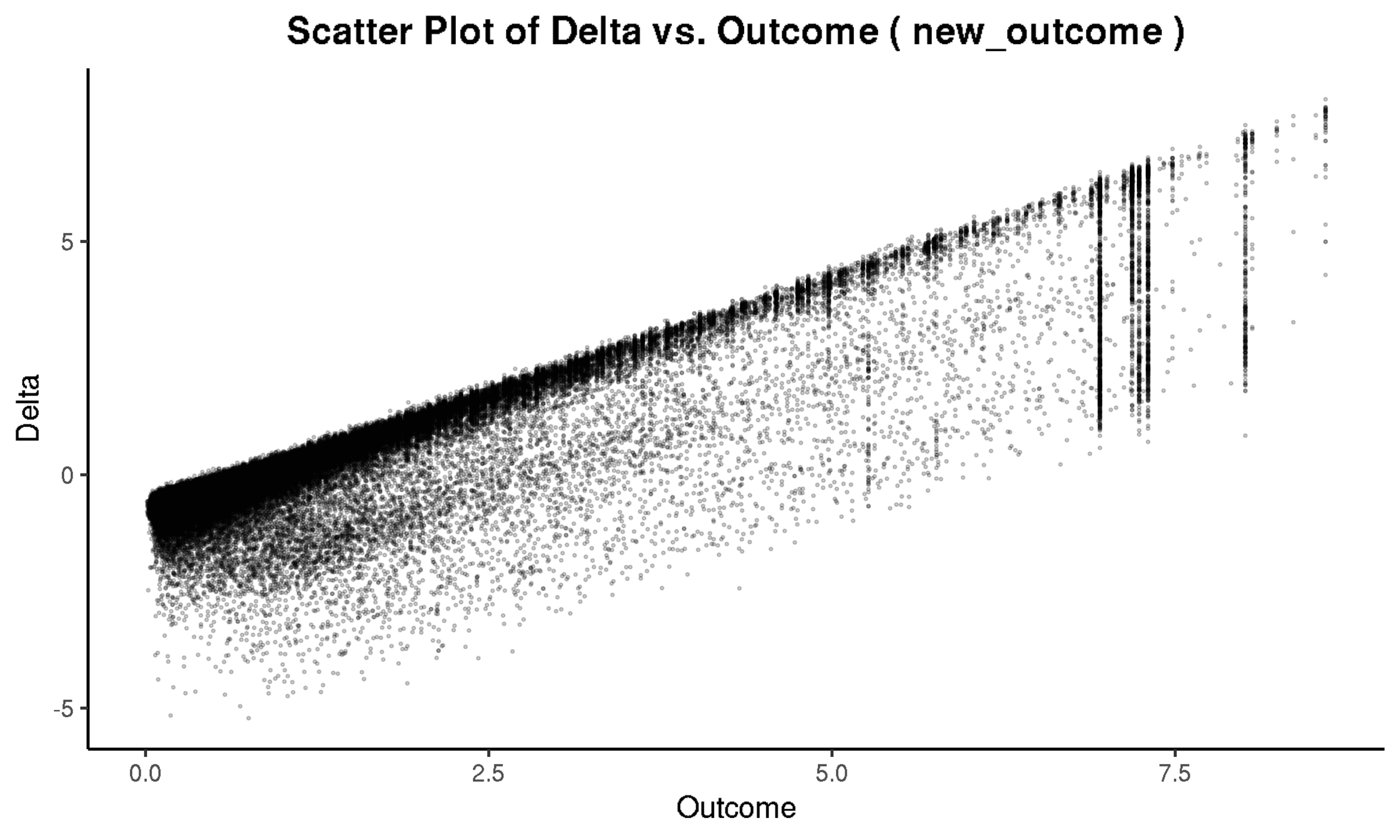}\hfill
\includegraphics[width=.32\textwidth]{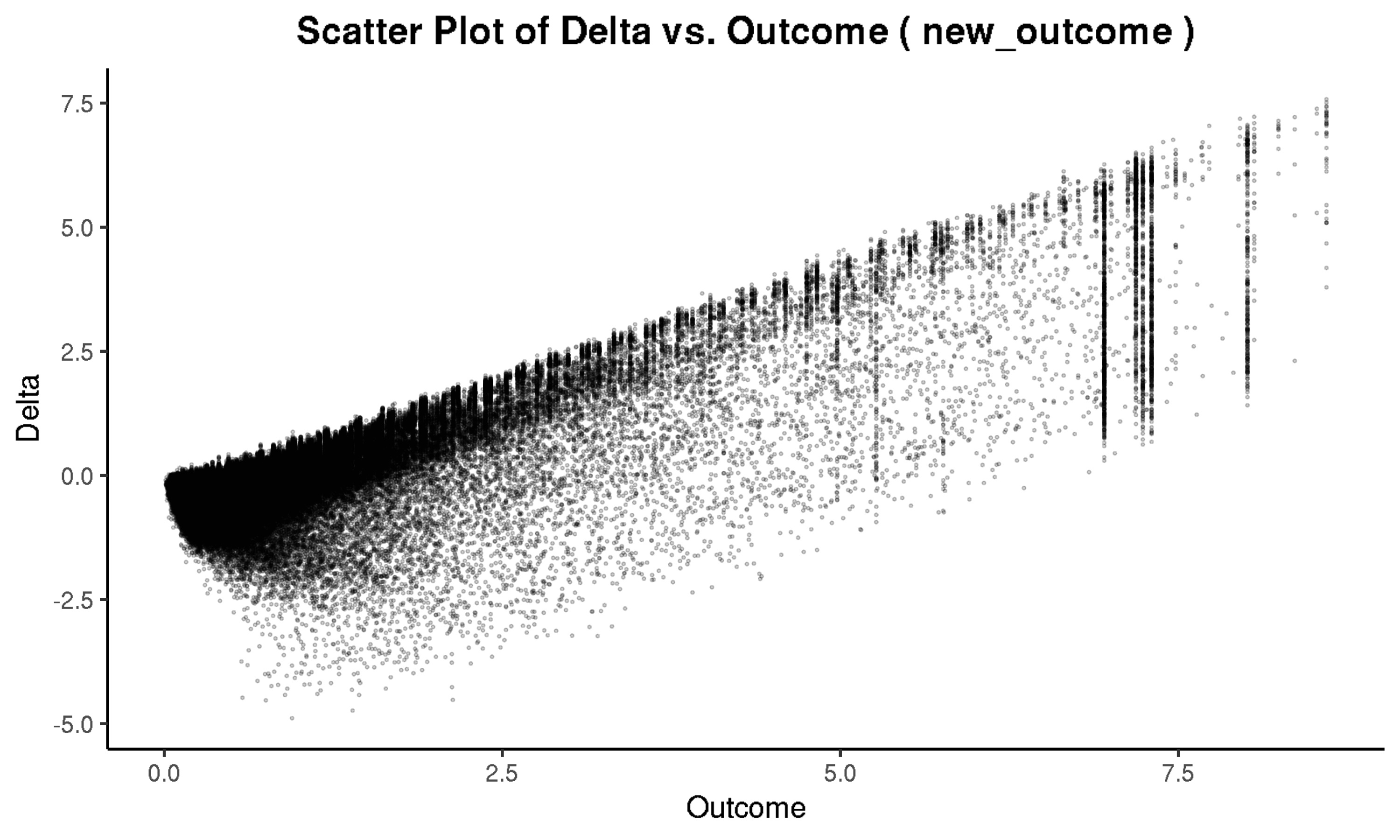}
\caption*{\footnotesize\textit{Note: Scatter plots of predicted (y-axis) against observed (x-axis) values from the three GBM specifications. Columns, from left to right: constant-only model, style-features model, and full model. Top row: cash per day. Bottom row: adjusted outcome.}}
\end{figure}

We now report diagnostic plots for the three models, evaluated on both cash per day (Table \ref{tab:image_pred}) and the adjusted outcome (Table \ref{tab:image_pred_def_new_outcome}). Figure \ref{fig:diagnostics_histograms} shows the distribution of error terms across specifications and outcomes; Figure \ref{fig:diagnostics_scatter} shows the corresponding scatter plots of fitted versus observed values. Across both outcomes, the error distributions from the model with style features and the full model are visibly similar, consistent with the modest gain in MSE from adding the remaining covariates.

\section{Supplementary analysis for AIPW estimates of style features}\label{aipw_robust}
This Appendix \ref{aipw_robust} demonstrates that propensity scores estimated via our machine learning method yield better covariate balance and produce a more closely matched treatment and control group. In Appendix~\ref{aipw_robust}.1, we detail the balance assessment after weighting the treatment and control observations using these propensity scores. Appendices \ref{aipw_robust}.2 and \ref{aipw_robust}.3 illustrate the density plots for the propensity score distributions across different features, provide examples of features removed through 
this procedure, and show those retained for further analysis.

\subsection{Diagnostics for selected style features}
In observational settings, the covariate distributions can differ substantially between treated and untreated individuals, potentially biasing estimates of the average treatment effect (ATE). By adjusting each observation's weight based on its inverse propensity score, we aim to make these distributions more comparable. This section presents the balance check results using the absolute standardized mean difference (ASMD) of covariates in the treatment group (e.g. with \emph{Bodyshot}) versus the control group, before and after propensity-score weighting. We find that applying propensity-score weighting improves the ASMD values, indicating better balance between the treated and untreated groups.

\begin{figure}[!tbp]
\centering
\caption{Covariate balance diagnostics for \emph{Bodyshot} and \emph{Smile}}
\label{fig:diagnostics_balance}
\includegraphics[width=.48\textwidth]{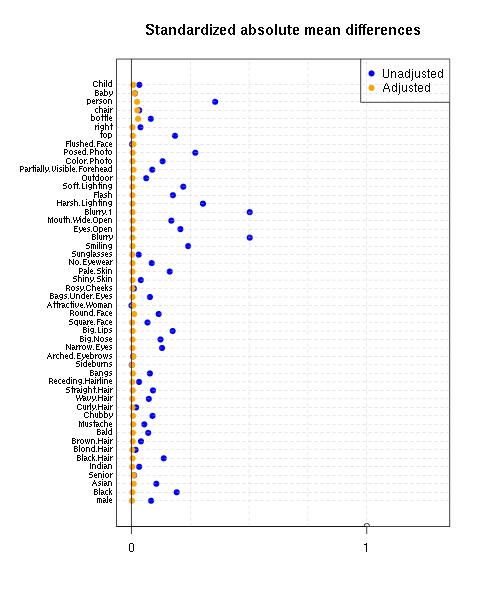}\hfill
\includegraphics[width=.48\textwidth]{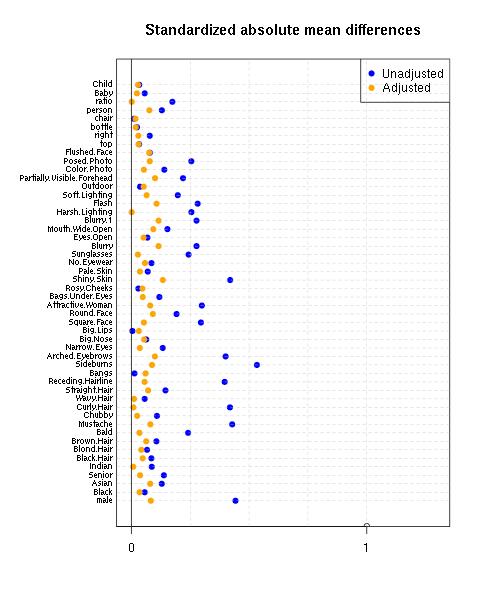}
\caption*{\footnotesize\textit{Note: Standardized absolute mean differences of a selected subset of covariates across profiles with and without the target feature. Left panel: \emph{Bodyshot}. Right panel: \emph{Smile}. Propensity scores used for reweighing were obtained using a GBM model trained on all covariates in Kiva data.}}
\end{figure}

Figure \ref{fig:diagnostics_balance} shows the ASMD of covariates across the treatment and control groups for \emph{Bodyshot} (left) and \emph{Smile} (right). The left-hand side of each panel lists the covariates, revealing that the variables we introduced — along with correlated variables — were far from balanced before weighting (blue dots). After applying propensity-score weighting, the ASMDs (yellow dots) move close to zero in both cases, indicating improved balance between treated and untreated groups.

\subsection{Density Plots of Dropped Features}\label{density_out}
\begin{figure}[htp]
\caption{Propensity Estimates of Dropped Style Features}\label{fig:dropped_style}
\centering
\includegraphics[width=.35\textwidth]{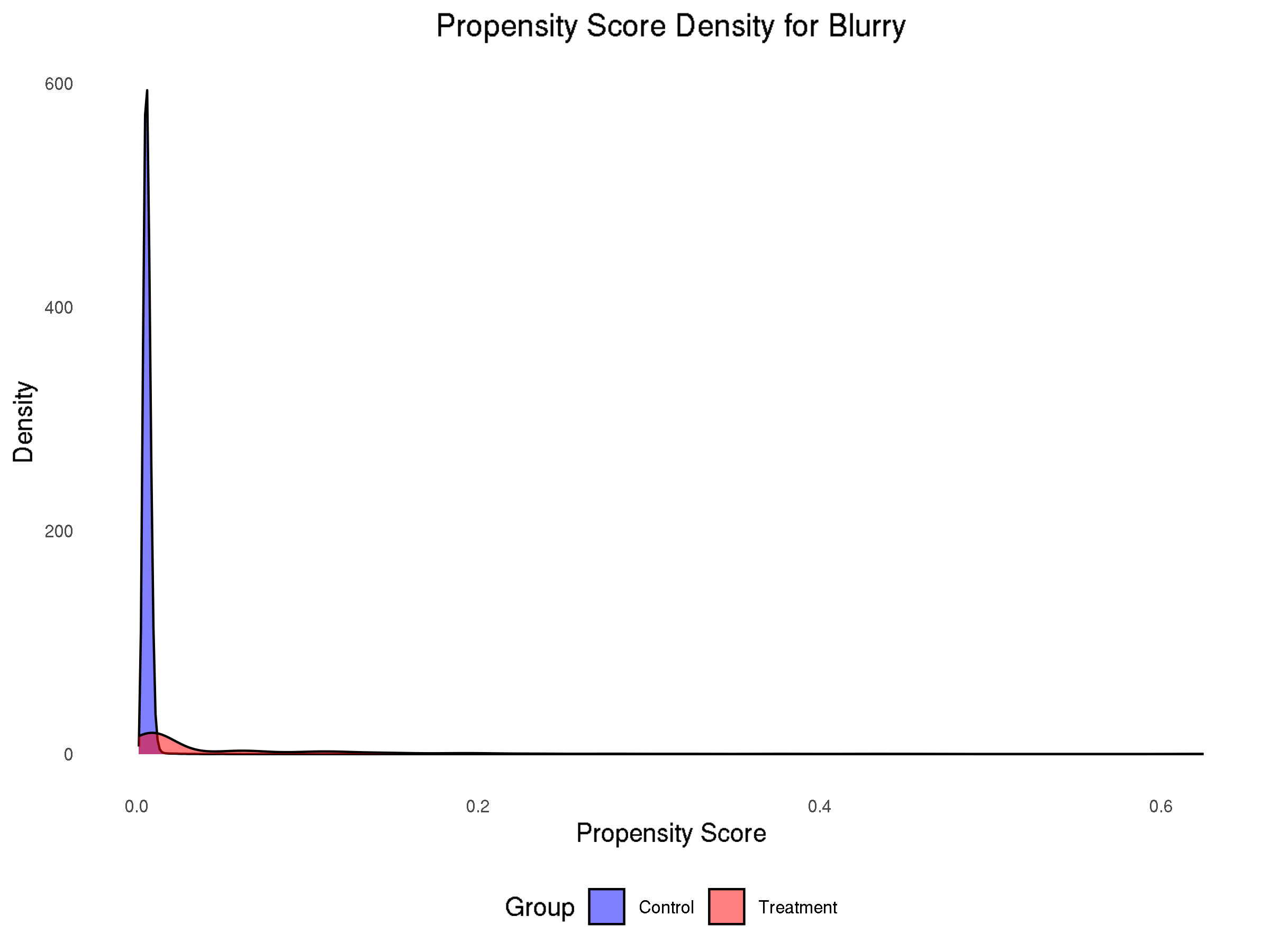}\hfill
\includegraphics[width=.35\textwidth]{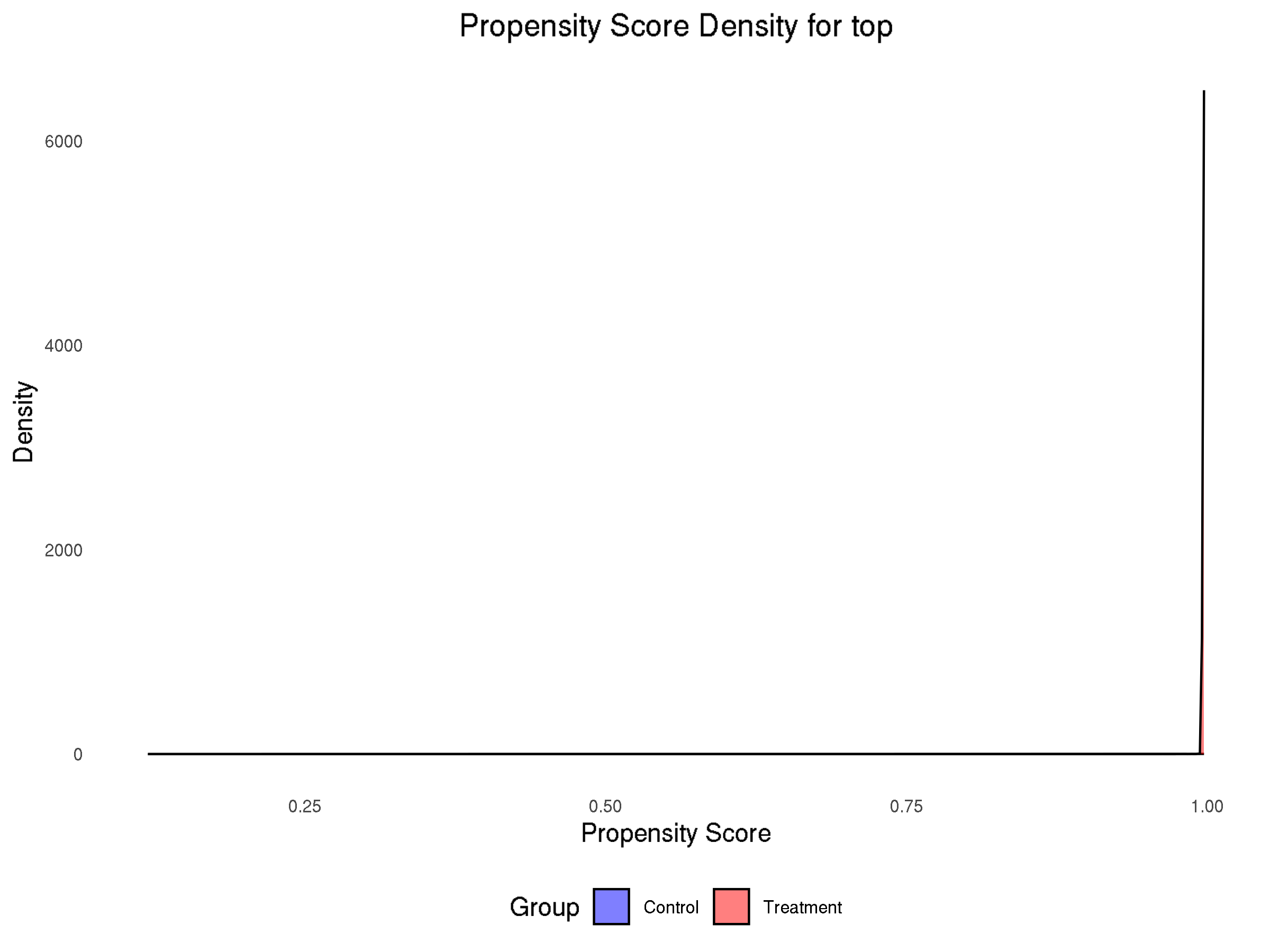}\hfill
\includegraphics[width=.35\textwidth]{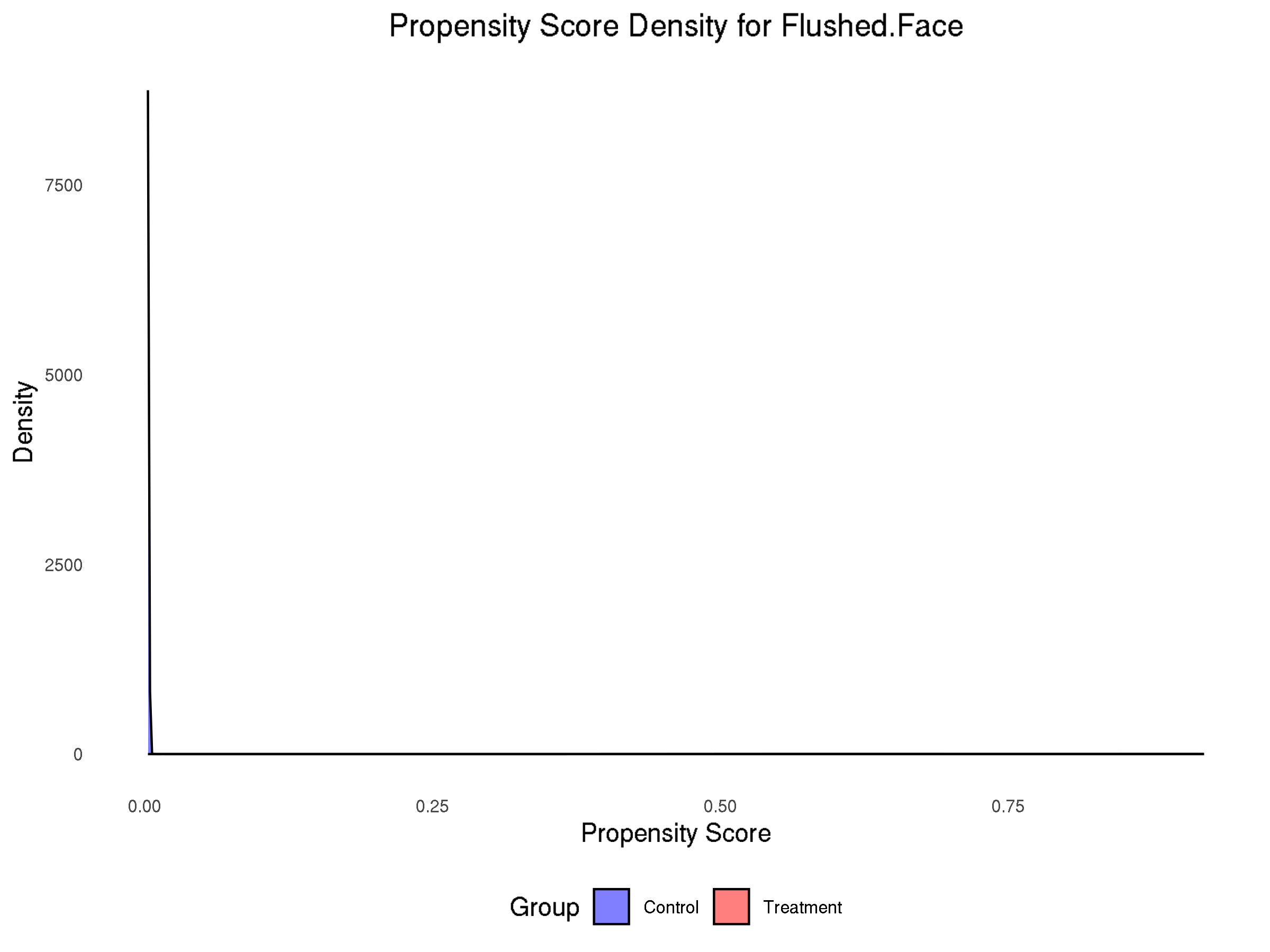}
\includegraphics[width=.35\textwidth]{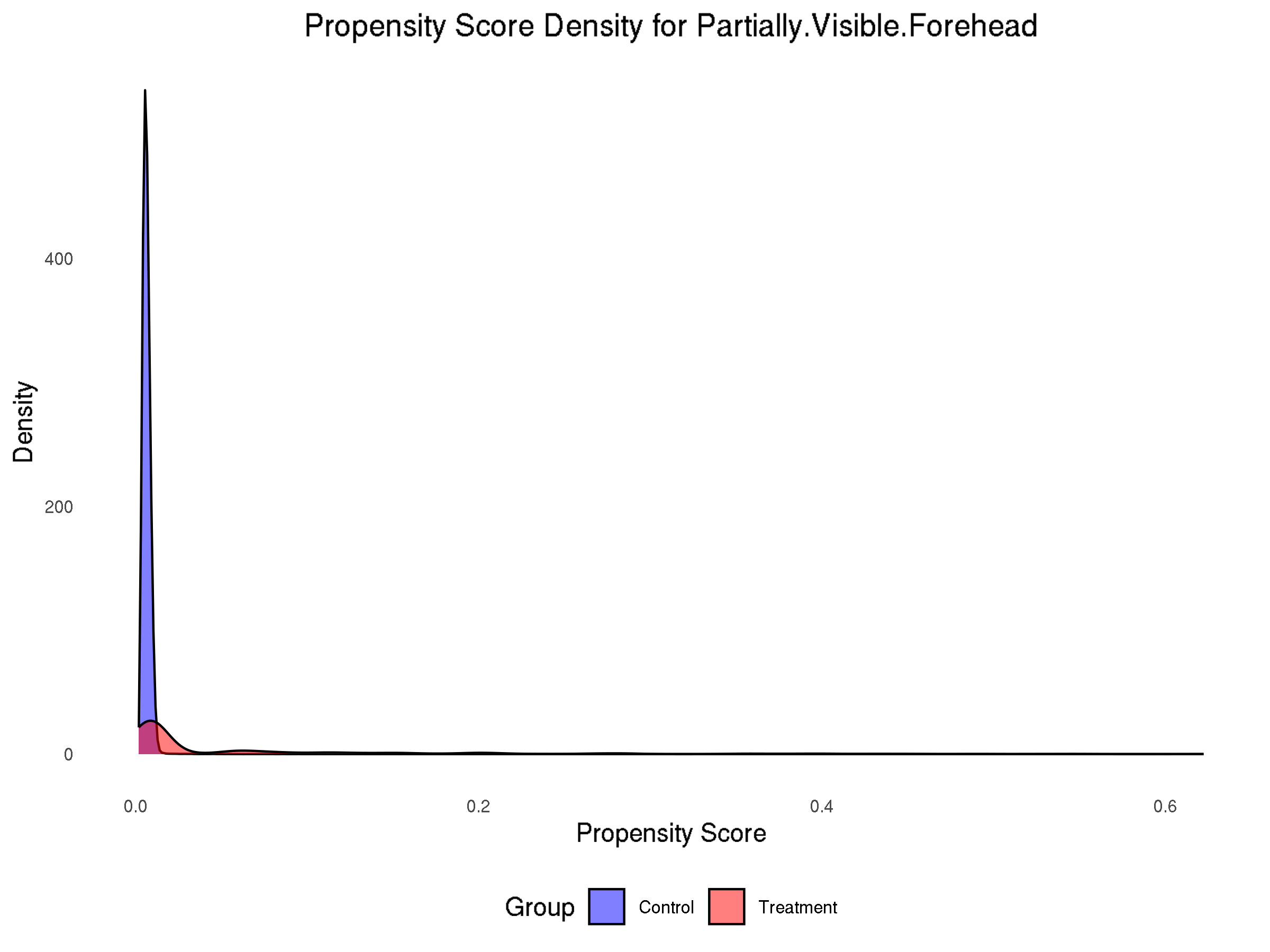}
\caption*{\footnotesize{\textit{Note: These density plots show the gradient boosted machine (GBM) model estimates of the propensity scores for the features \emph{Blurry}, \emph{Top}, \emph{Flushed Face}, and \emph{Partially Visible Forehead}. The lack of overlap between treatment and control groups is evident, justifying their exclusion from the analysis.}}}
\end{figure}

This section provides density plots illustrating the propensity score distributions for features that were excluded from the analysis due to insufficient overlap between the treatment and control groups. As discussed in the main text, features were dropped if either the treatment or control group had propensity score mass below 0.1 or above 0.9, indicating limited comparability between the groups. Figure \ref{fig:dropped_style} highlights the lack of overlap in the propensity score distributions for specific style features.
\subsection{Density Plots of Considered Features}\label{density_in}
\begin{figure}[htp]
\caption{Propensity Estimates of Selected Style Features}\label{fig:nondropped_style}
\centering
\includegraphics[width=.35\textwidth]{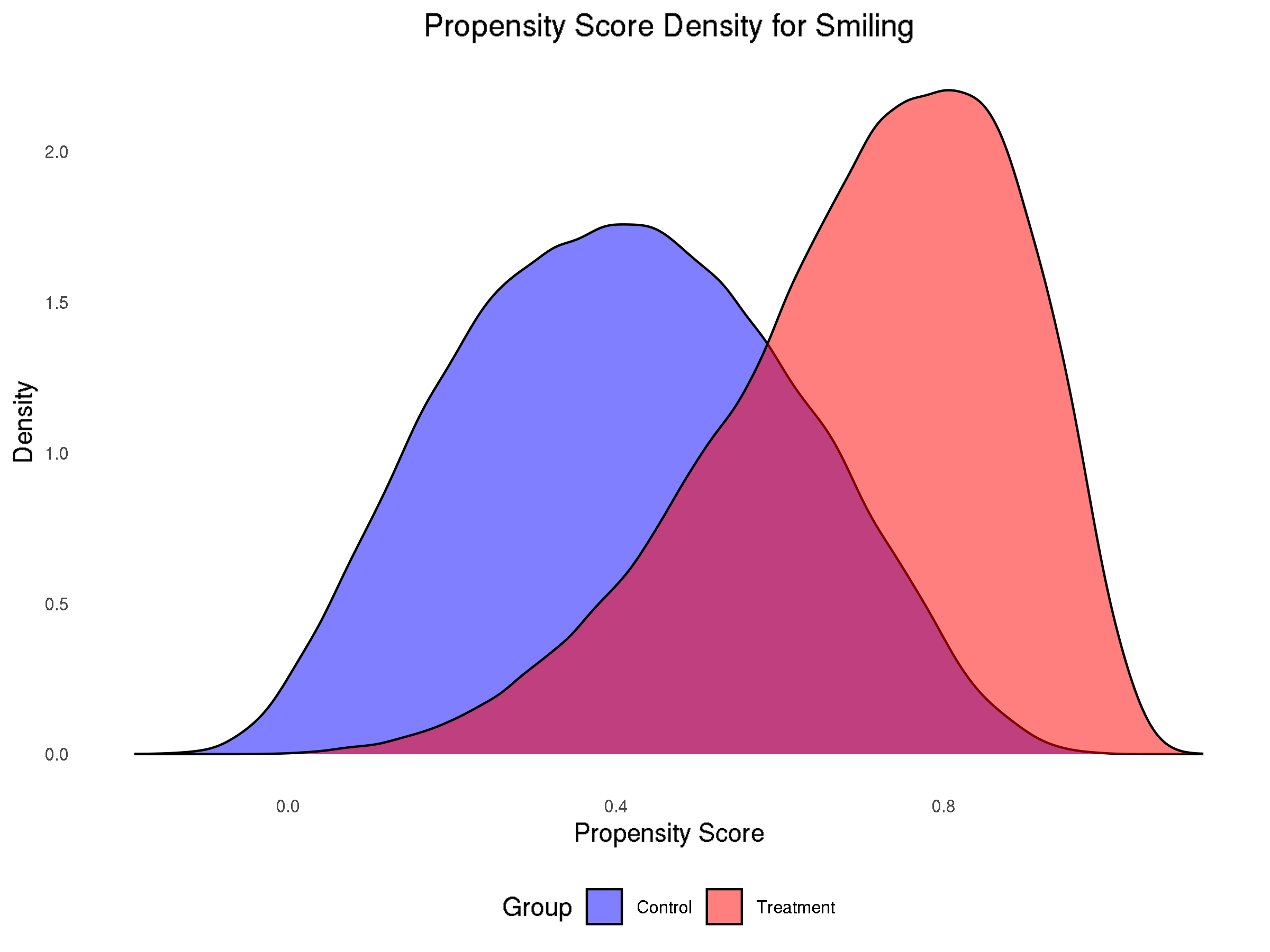}\hfill
\includegraphics[width=.35\textwidth]{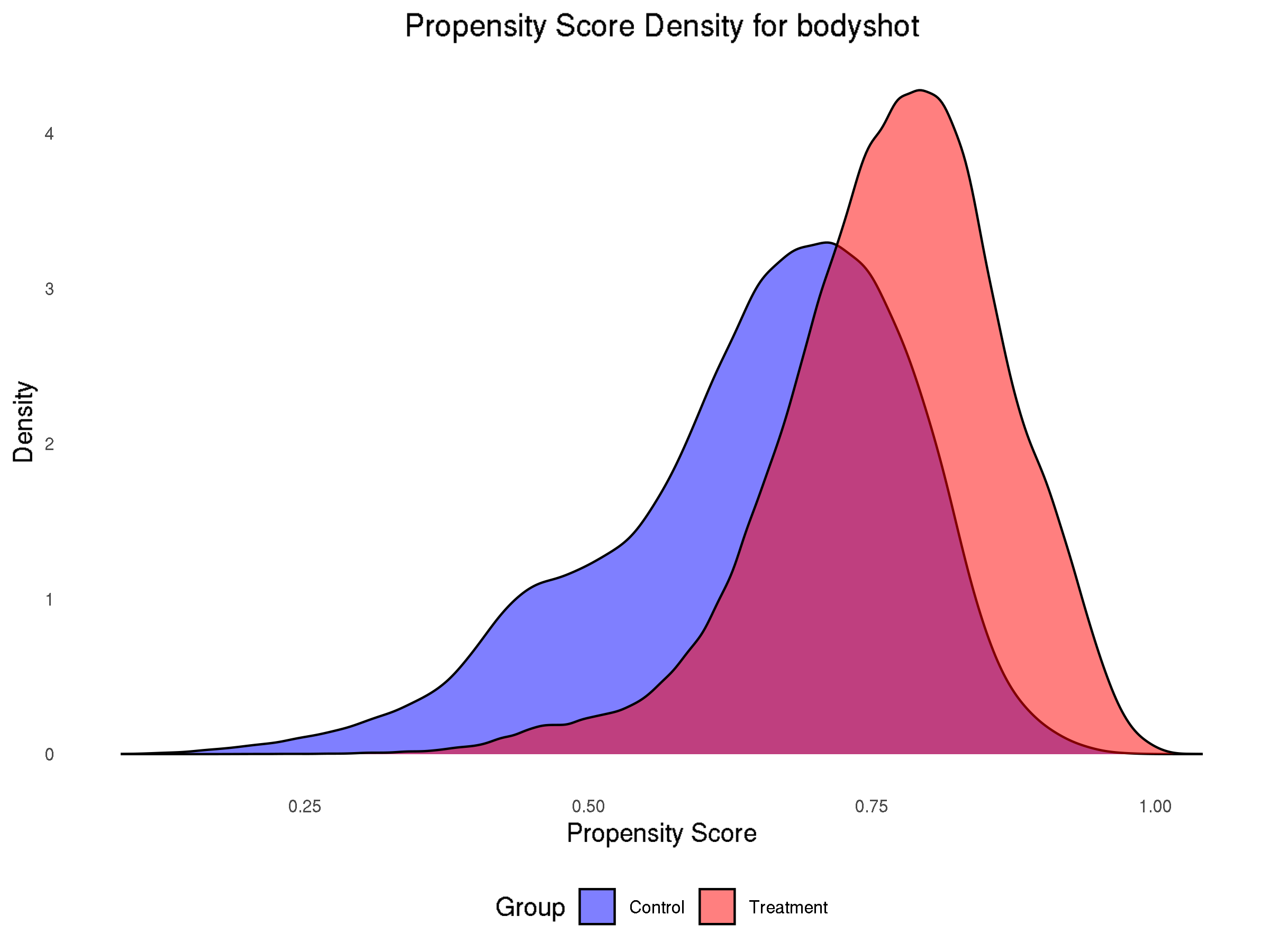}\hfill
\includegraphics[width=.35\textwidth]{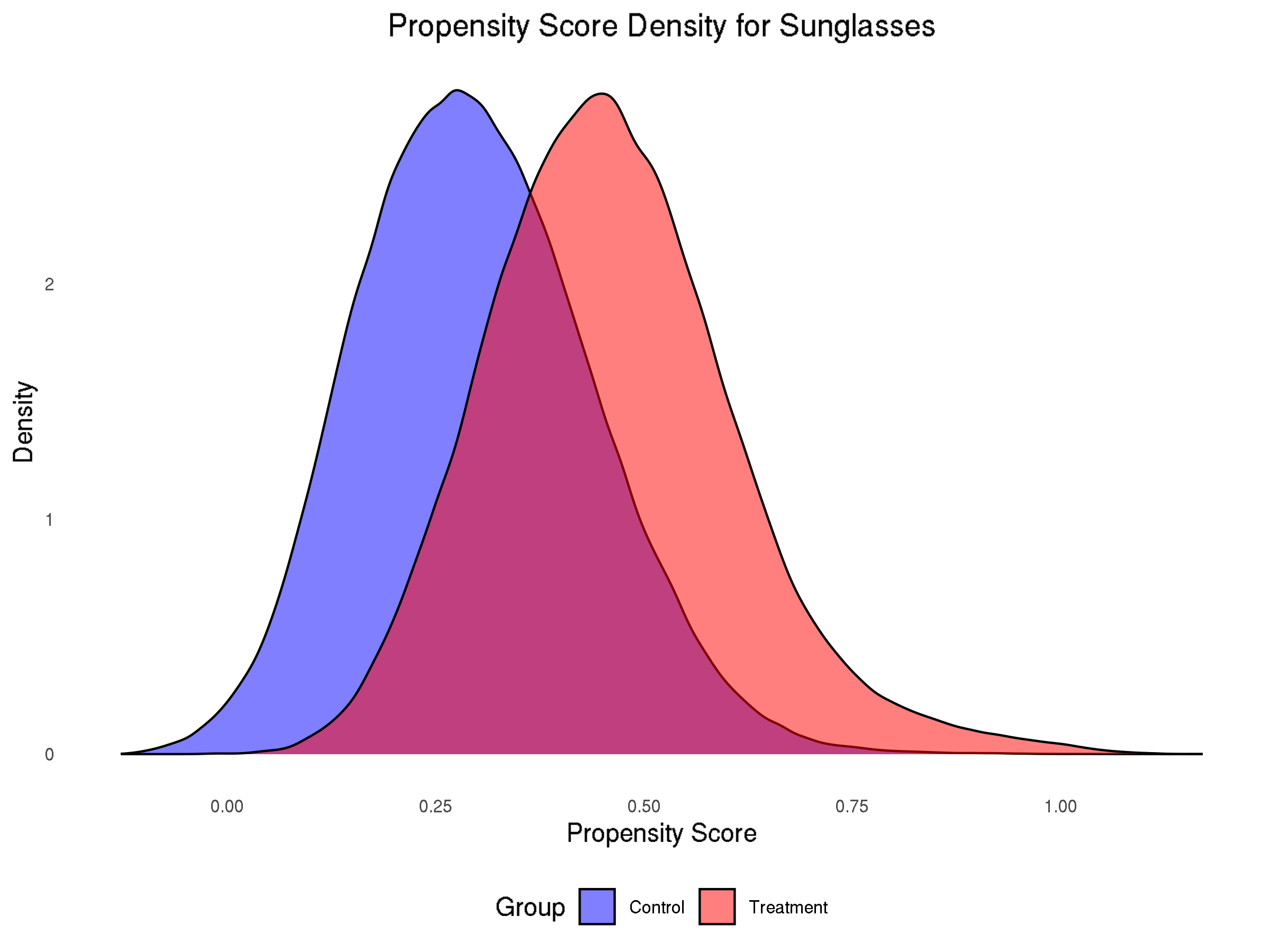}
\includegraphics[width=.35\textwidth]{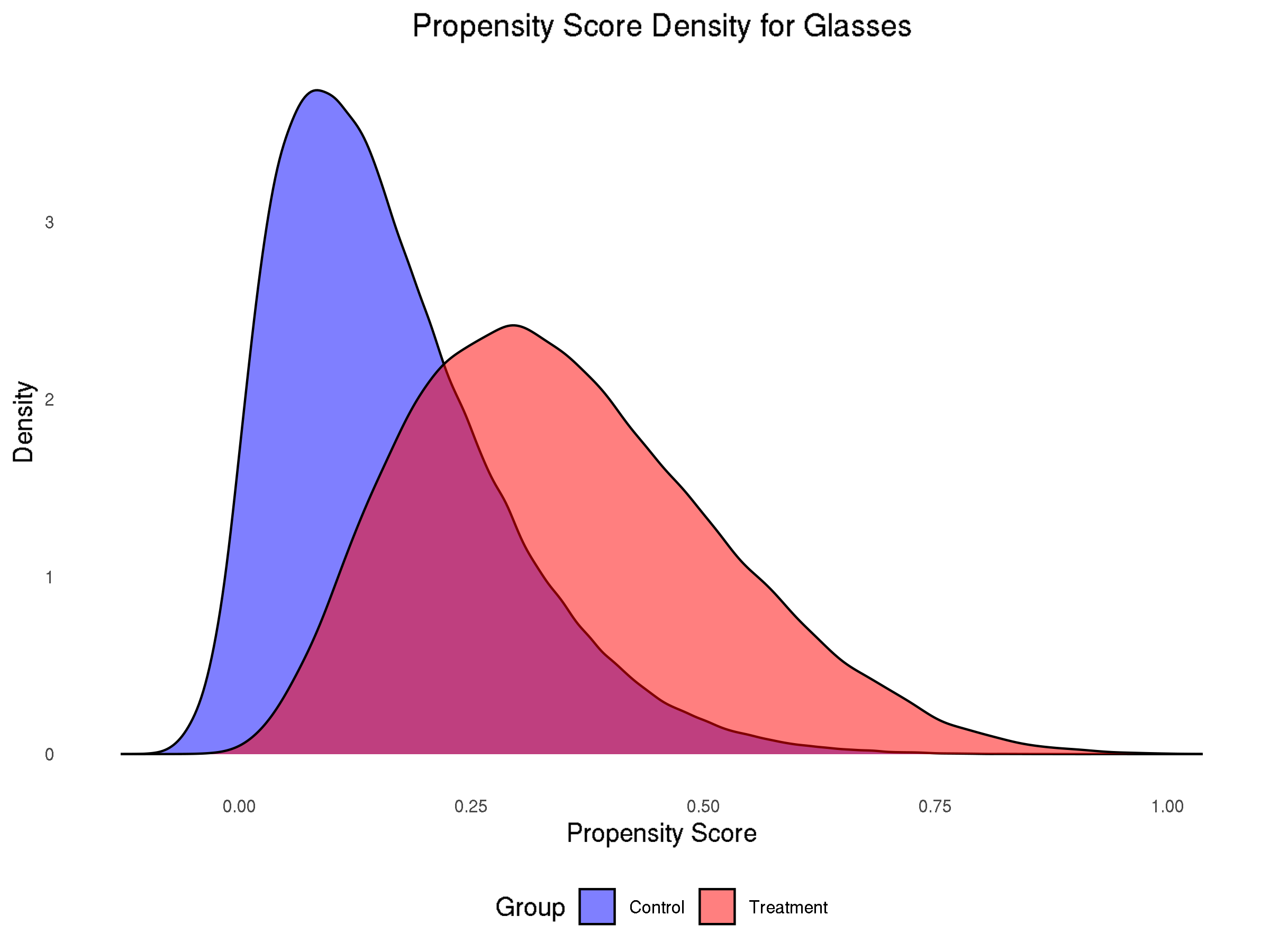}
\caption*{\footnotesize{\textit{Note: These density plots show the gradient boosted machine (GBM) model estimates of the propensity scores for the features \emph{Smiling}, \emph{Bodyshot}, \emph{Sunglasses}, and \emph{Glasses}. The substantial overlap between treatment and control groups indicates common support, justifying their inclusion in the analysis.}}}
\end{figure}

This section shows the propensity density plots for selected style features included in the analysis. Figure \ref{fig:nondropped_style} presents the density plots of propensity scores for the non-dropped \textit{style} features among profiles with and without such features. As the figure illustrates, there is substantial overlap between the two groups, indicating common support in the distribution of these features.

\section{Excluding potential mediators in ATE estimation}\label{ate_robust_mediators}
In this section \label{ate_robust_mediators}, we examine whether certain image-extracted features could, in principle, be “caused” by a given \emph{treatment feature}, making them potential moderators rather than appropriate controls in an AIPW estimator. For example, if the treatment feature is \emph{Smiling}, then a feature like \emph{Crooked Teeth} would only be observed when \emph{Smiling} = 1, implying that it should not be included as a control.  

We focus on three treatment features: \emph{Sunglasses}, \emph{Bodyshot}, and \emph{Smiling}. To assess the potential impact of problematic controls, we estimate two AIPW models: one with the full set of covariates and one in which we exclude potentially endogenous features. Specifically, we remove \emph{Glasses}, \emph{Narrow Eyes}, \emph{Eyes Open}, and \emph{Bags Under Eyes} for \emph{Sunglasses}; \emph{Outdoor}, \emph{Bottle}, \emph{Chair}, \emph{Harsh Lighting}, \emph{Flash}, and \emph{Soft Lighting} for \emph{Bodyshot}; and \emph{Mustache}, \emph{Mouth Closed}, \emph{Mouth Wide Open}, and \emph{Strong Nose-Mouth Line} for \emph{Smiling}. Table \ref{tab:ATE_combined_drop} presents the results. We find that excluding these features does not substantially alter the estimates.

\begin{table}[!htbp] 
    \centering
    \caption{Estimated Average Treatment Effects (ATE)} 
    \label{tab:ATE_combined_drop} 
    \resizebox{0.65\textwidth}{!}{%
    \begin{tabular}{@{\extracolsep{5pt}}lcccccc} 
    \\[-1.8ex]\hline 
    \hline \\[-1.8ex] 
    & \multicolumn{2}{c}{\textbf{ATE Estimates (Model 1)}} & \multicolumn{2}{c}{\textbf{ATE Estimates (Model 2)}} \\ 
    \cline{2-3} \cline{4-5} 
    \\[-1.8ex] 
    & \textbf{Estimate} & \textbf{Std. Err.} & \textbf{Estimate} & \textbf{Std. Err.} \\ 
    \hline \\[-1.8ex] 
    Sunglasses  & -7.42 & 0.64 & -7.54 & 0.68 \\ 
    Bodyshot    & -21.27 & 0.72 & -21.43 & 0.64 \\ 
    Smiling     & 15.54 & 0.74 & 14.16 & 0.67 \\ 
    \hline 
    \hline \\[-1.8ex] 
    \end{tabular} 
    }
    \caption*{\footnotesize{\textit{Note: This table presents the Average Treatment Effects (ATE) estimates for two models. Model 1 has the full set of controls; in Model 2 we are excluding potential mediators from the list of covariates. Model 1 estimates are reported in the second and third columns, while Model 2 estimates appear in the fourth and fifth columns. Standard errors are provided in parentheses.}}}  
\end{table}

\section{Decomposition of the Gender Gap}\label{app:gender_decomp}

This appendix decomposes the gender gap in cash per day into contributions from style features, other type features, sector, geography, time, and market structure. We use an algorithmic prediction of male: the variable indicates that the feature detection algorithm assigns a probability of at least 0.5 that the person in the image is male. As reported in Section~\ref{corr_type_style}, campaigns classified as male raise on average USD 30.2 less per day than those classified as female, and take 5.8 more days to be funded fully.\footnote{In the context of microfinance, the gender gap might be driven by users that aim to correct for discrimination against women in traditional finance. There is a rich literature documenting discrimination against women in traditional entrepreneurial lending. \cite{alesina2013women} shows that women entrepreneurs pay higher rates for access to credit, and \cite{brock2021discriminatory} use a randomized experiment to show that loan officers grant loans to women under less favorable conditions than to men. The phenomenon of over-correcting for discrimination is well documented in experimental psychology \citep{mendes2013brittle, nosek2007pervasiveness}. It is also plausible that Kiva lenders follow broader policy discussions, where the emphasis on developmental policies and aid targeting women is common \citep{kristof2010half}. Furthermore, \cite{ozer2023digital} show that peer-to-peer microlending can effectively advance such social goals.}
We apply a Gelbach decomposition \citep{gelbach2016covariates}, comparing the coefficient on \emph{male} in a univariate regression of cash per day to the coefficient in a full model that includes all variables in the Kiva data, and attributing the change to each group of covariates.
\begin{figure}[htbp]
    \centering
    \caption{Gelbach Decomposition of \emph{Male} Coefficient for cash per day}
    \includegraphics[scale = 0.2]{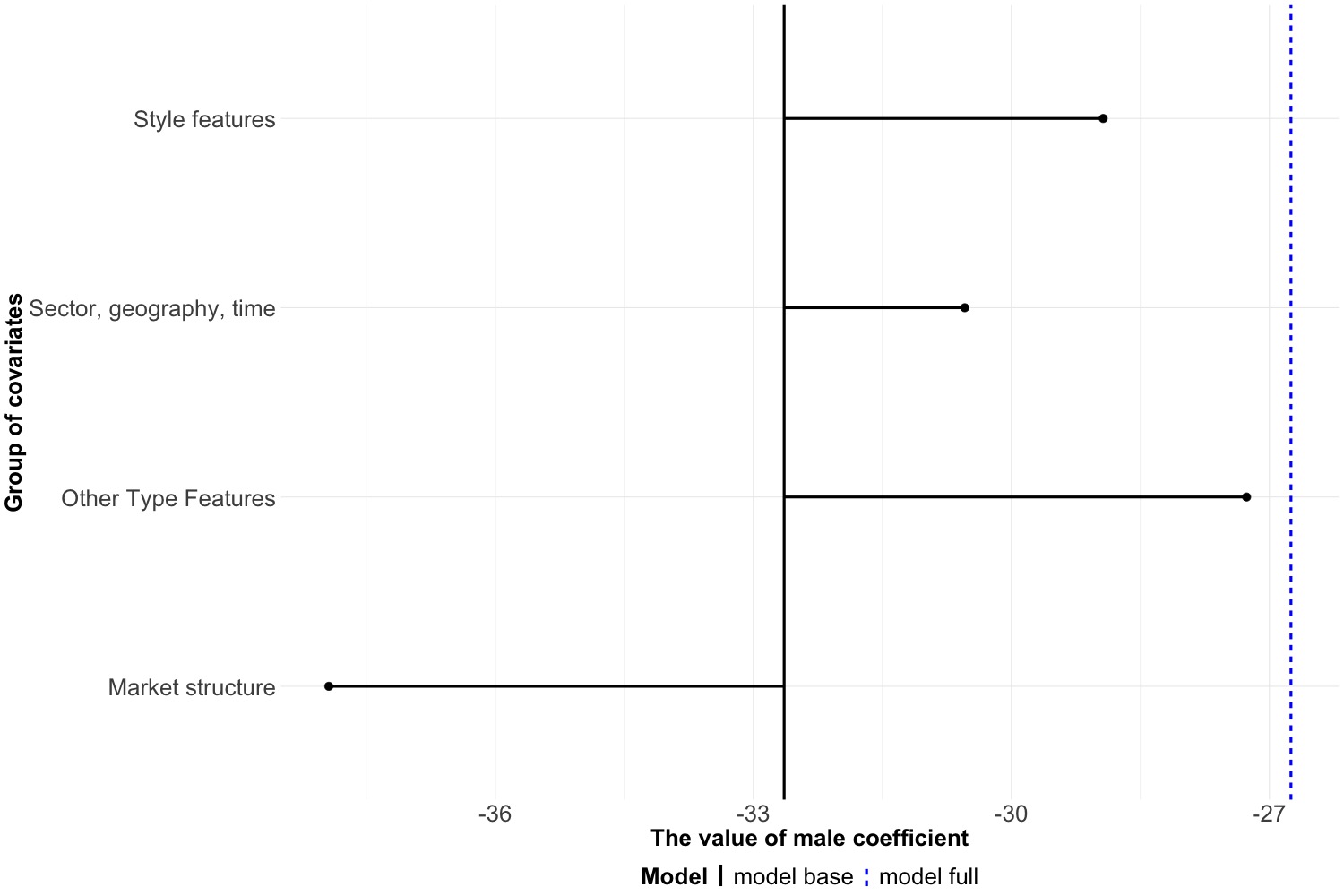}
    \caption*{\footnotesize{\textit{Note: The solid line is the coefficient on \emph{male} from a univariate regression of cash per day; the dashed line is the coefficient adjusted for all variables in Kiva data (OLS). Each horizontal bar reports the position the \emph{male} coefficient would occupy if we adjusted only for the corresponding group of covariates. A bar extending toward zero from the solid line indicates that the group aggravates the gap (controlling for it brings the coefficient toward zero); a bar extending away from zero indicates that the group mitigates it. Type features include all other type features from the image; sector, geography, time includes sector, country, and week fixed effects, loan amount, and repayment details; market structure includes interactions of month and country, month and sector, number of lenders in the week, number of competing campaigns, and share of campaigns of the same race and gender.}}}\label{fig:gelbach}
\end{figure}
Figure~\ref{fig:gelbach} reports the results. Style features aggravate the gender gap: adjusting for them brings the \emph{male} coefficient toward zero, indicating that male borrowers' profile choices contribute to the unadjusted disparity. Consistent with this, prevalence differs sharply by gender: 77\% of female borrowers smile in their profile image, compared with 33\% of male borrowers, while 26\% of male borrowers use a body-shot compared with 22\% of female borrowers (both differences are statistically significant). Other type features and sector, geography, time fixed effects further aggravate the gap, while market structure controls operate in the opposite direction, mitigating it.

\section{Generative Adversarial Networks}\label{appendix_gans}
\subsection{Background}
Generative artificial intelligence (AI) models learn the patterns and probability distributions of training data, then use that understanding to generate new samples. Generative Adversarial Networks (GANs), introduced by \cite{goodfellow2014generative}, are a class of deep generative models designed to produce data that resemble real samples — such as realistic images \citep{ludwig2024machine} and synthetic datasets \citep{athey2024using}. GANs, although do not directly produce estimates of the density or distribution function at a particular point, can be thought of as implicitly estimating the distribution of latent features, and they can be used to generate or output new examples that plausibly could have been drawn from the original dataset.

GANs are composed of two deep models: a generator $G$ and a discriminator $D$. As illustrated by \cite{goodfellow2014generative}, the objective is to learn the training data's distribution $p_{g}$ over data $x$. We define a prior $p_{z}(z)$ on input noise variables, then represent a mapping to data space as $G(z; \theta_{g})$. The Generator receive the input vector sampled from $p_{z}(z)$ and outputs the image. Discriminator $D(x; \theta_{d})$ takes in the image and outputs a single scalar, representing the probability that $x$ the image is real (came from the dataset rather than $p_{g}$ the generation). $D$ and $G$ play the two-player minimax with the objective function: 
\[
\min_{G}\max_{D} V(D, G) = \mathbb{E}_{x \sim p_{data}(x)} [log D(x)] + \mathbb{E}_{z \sim p_{z}(z)} [1 - log D(G(z))]
\]

\subsection{Variation of images in latent space}
\subsubsection{Stimuli generation}

GANs are often employed to modify images and create “Deep Fakes” — fabricated images altered along specific dimensions. In our study, we apply StyleGAN \citep{karras2019stylebased} and StyleGANEX \citep{yang2023styleganex} to generate images that differ by a particular feature of interest. We use a pre-trained GAN generator \(G\) and encoder, which transforms an image \(x_i\) into its latent embedding \(v_i \in \mathcal{Z}\) (in the latent space $\mathcal{Z}$). From there, we obtain a direction vector \(\Delta v_{w}\) (e.g., \(\Delta v_{\text{smile}}\) for a “smile” feature), many of which are available from online open-source codebase. In cases where a direction vector for a specific feature is unavailable off-the-shelf, we propose the procedure in Appendix C.2.2 to obtain the vector.

After encoding each image $x$ into its latent space, we adjust its embedding along \(\Delta v_{w}\) and feed the result into the generator to generate it back to the image:
\[
\tilde{x}_i = G\bigl(v_i \pm \alpha \,\Delta v_{w}\bigr),
\]
where \(\alpha\) is a continuous scale parameter. By increasing or decreasing \(\alpha\), we audit the output images $\tilde{x}$ at both extremes and select those reflecting the desired feature alterations. 

\begin{figure}[!htbp]
\caption{Experiment 1: Example of facial attributes alternations (Gender and Smile) via the corresponding gradient}\label{ganstudy1}
\centering
\includegraphics[width=1.1\linewidth]{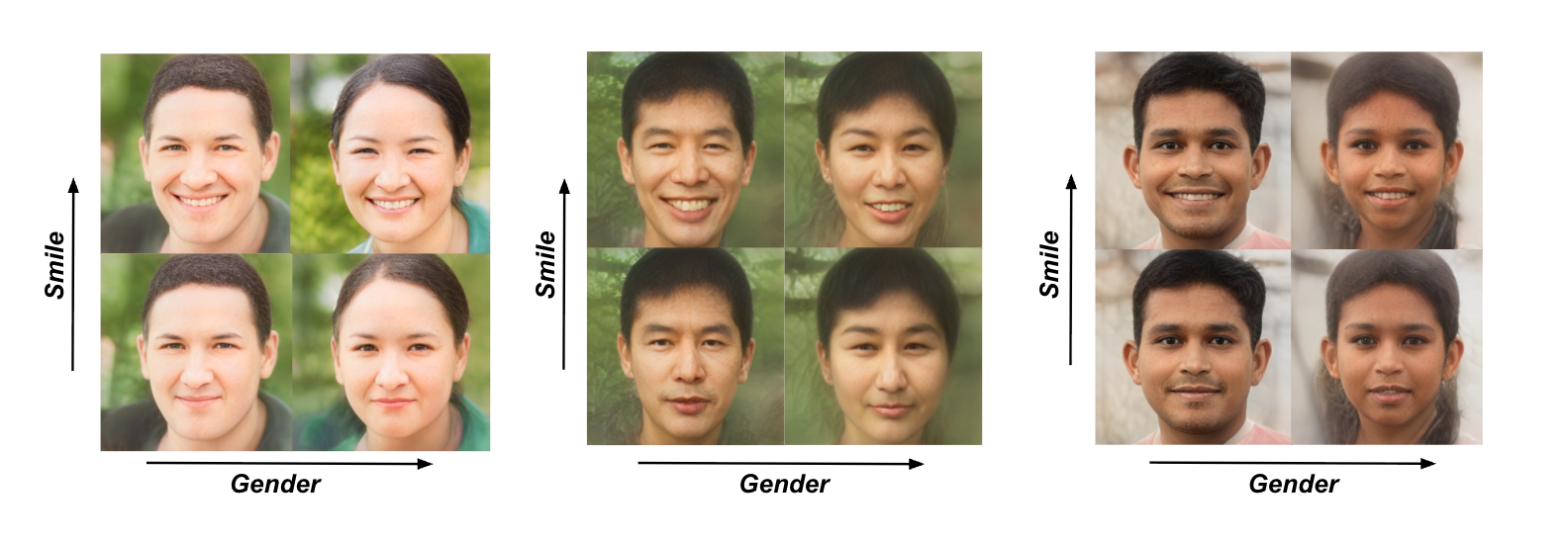}\\
\end{figure}

\begin{figure}[!htbp]
\caption{Experiment 2: Example of facial attributes alternations (\textit{Age}, \textit{Hair Color}, and \textit{Glasses}) via the corresponding gradient}\label{ganstudy2}
\centering
\includegraphics[width=1.1\linewidth]{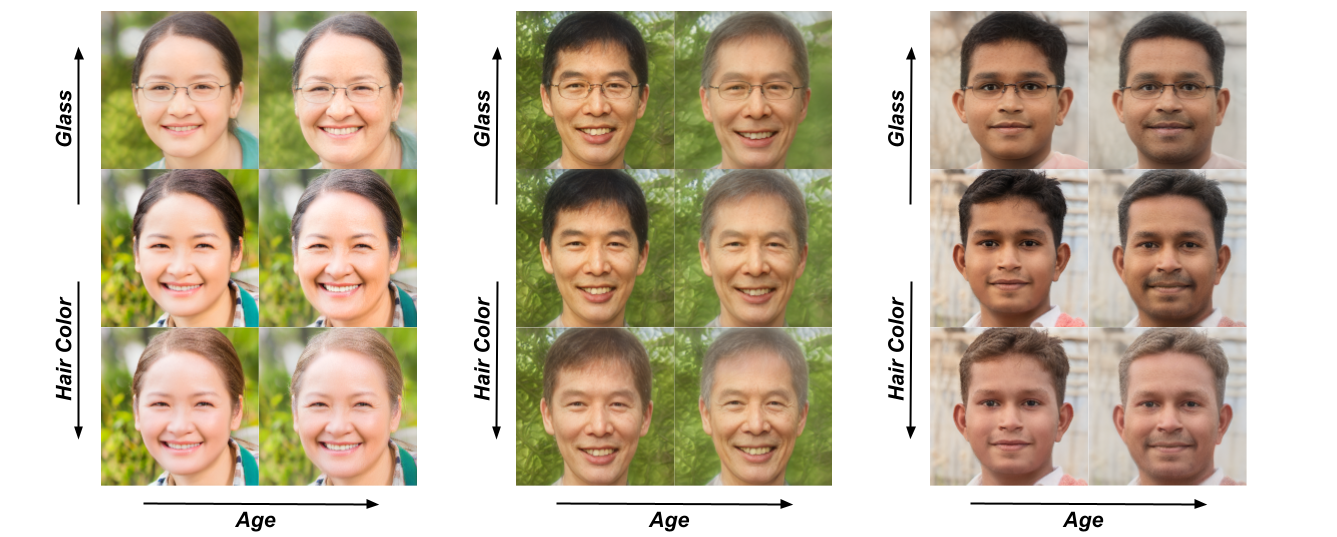}\\
\end{figure}

In Experiments~1 and~2, we apply this technique to manipulate features such as \textit{gender} and \textit{smile}, as well as \textit{age}, \textit{hair color}, and \textit{glasses}. Figures~\ref{ganstudy1} and~\ref{ganstudy2} illustrate examples of these modifications using this approach.

Image features usually have correlations, and this method works even in the presence of correlation between the visual features in the images. We introduce hyper-parameters to control the degree of alteration in the desired direction, and we fine-tune these parameters on a per-image basis. Once the attribute is modified, it is integrated back into the original (unmodified) image. To ensure the result appears realistic and seamless, we apply a series of post-processing steps, including deblurring, inpainting, and auto-blending. After post-processing, the margins around the human headshot are inpainted and blended with the background. Figure 10 in our manuscript shows the full images after post-processing.

\subsubsection{Identify direction vector}
Since not all direction vectors are readily available off-the-shelf, we adapt recent work and propose a method to derive the direction vector e.g. $\Delta v_{gender}$ for image manipulation in latent space.

We assume there exists a bijection mapping between the feature of interest in the image and the 
relevant dimensions of its corresponding latent embedding vector. Consequently, we can use 
the direction vector to obtain the “potential outcome” of the image with that feature 
altered, while holding other attributes constant. This parallels the concept of Average 
Treatment Effects in causal inference, and by using it in the opposite way, we can find 
the relevant latent dimensions by computing the difference in mean latent embeddings 
between images that exhibit the feature and those that do not, with other features 
randomly selected.

Consider the \textit{race} feature as an example (though it is ultimately 
outside this paper’s scope). No published, open-source pre-trained models provide 
an off-the-shelf direction vector for \textit{race} partially due to its complexity, 
especially when multiple racial categories (\textit{Asian}, \textit{Black}, 
\textit{White}) must be toggled. To manipulate between \textit{Asian} 
and \textit{Black} in the latent space, for instance, we randomly sample 
about 200 images identified as \textit{Asian} and 200 identified as 
\textit{Black}, with other characteristics randomly chosen\footnote{We used images labeled by CNN and verified by human audit}. Let 
\( w_i = 1 \) indicate \textit{Black} and \( w_i = 0 \) otherwise, with 
\( s_i \) indicating other features. Each image \( x_i \) yields a latent embedding 
\( v_i \) depending on \( (w_i, s_i) \) in the image, and we assume any change in some relevant 
latent dimension arises solely from the feature of interest, without other unmeasured confounders. We then identify the relevant dimension in the latent embedding that corresponds to the change of the interested feature via the following 
approach, under our assumption:

Let \(w_i=1\) indicate \textit{Black} and \(w_i=0\) indicate \textit{Asian}. 
Let \(s_i\) denote an observed image attribute used to construct the comparison
sample. For example, \(s_i=1\) if the person in the image is smiling and 
\(s_i=0\) otherwise. Before encoding the images into the latent space, we construct
the two comparison groups so that the distribution of \(s_i\) is balanced across
the \(w_i=1\) and \(w_i=0\) groups. In the binary smiling example, this requires

\begin{align}
    \frac{1}{n_1}\sum_{i:w_i=1} s_i
    =
    \frac{1}{n_0}\sum_{i:w_i=0} s_i,
\end{align}

where \(n_1=|\{i:w_i=1\}|\) and \(n_0=|\{i:w_i=0\}|\). More generally, when
\(s_i\) contains multiple observed image attributes, we construct the comparison
groups so that the empirical distribution of these attributes is comparable across
the two groups.

Each image \(x_i\) is then mapped into a latent embedding \(v_i\). Under this
balanced sampling procedure, the direction vector associated with changing the
feature of interest \(w_i\) is estimated by the difference in mean latent embeddings:

\begin{align}
    \Delta \hat v_w
    =
    \frac{1}{n_1}\sum_{i:w_i=1} v_i
    -
    \frac{1}{n_0}\sum_{i:w_i=0} v_i .
\end{align}

The resulting vector \(\Delta \hat v_w\) is intended to capture the latent-space
direction associated with the feature of interest, net of the observed image
attributes used to construct the balanced comparison groups. In the example above,
the comparison is balanced on smiling; when additional attributes are used in the
sampling step, the same logic applies to those attributes as well. This interpretation
relies on the quality of the balance and on the assumption that remaining unobserved
image attributes are not systematically correlated with \(w_i\) within the balanced
sample.

\begin{figure}[!htbp]
\caption{Example of facial-attribute alterations (\textit{Race}) using the direction vector constructed by the method above.}\label{race}
\centering
\includegraphics[width=0.6\linewidth]{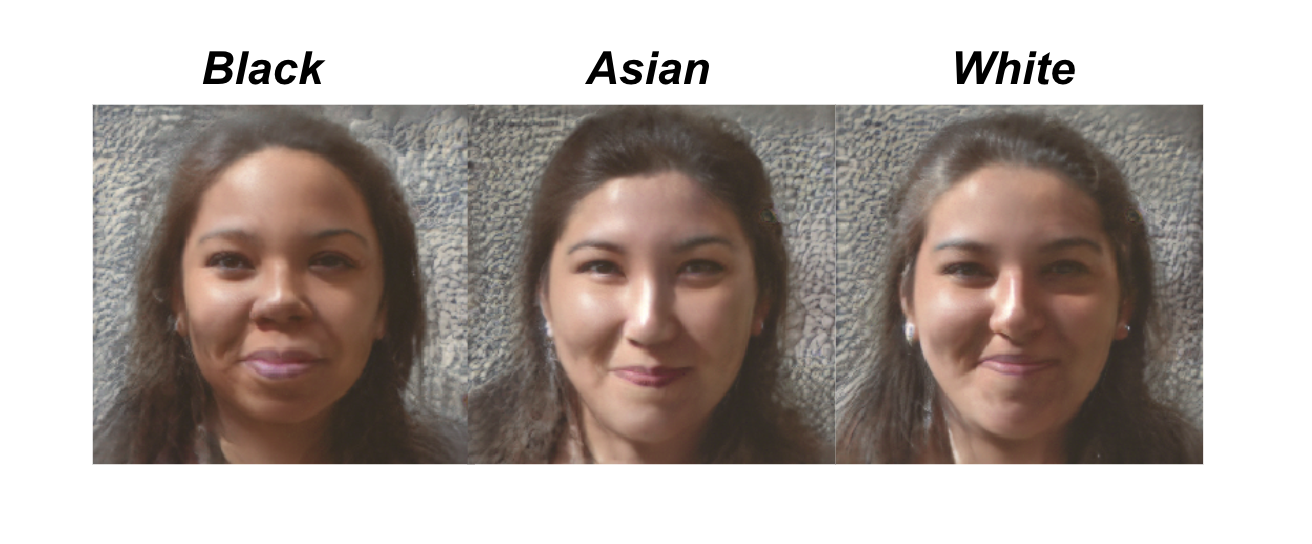}\\
\end{figure}

\section{Comparison statistics of GANs generated images and Kiva images}\label{GAN_emo_check}

This appendix compares summary statistics for our GAN-generated images and a sample of Kiva profile images, to check that the generated images plausibly reflect the variation in real borrower profiles. To evaluate this alignment, again we use an external API provided by \cite{peterson2022deep} to estimate psychological traits in each set of images.

We estimated psychological traits in both GAN-generated images and sub-sampled Kiva images. We include nearly 200 images from each set, obtain the scores of the psychological traits, and then  plot out the score distribution. When comparing the trait distributions, we find that they substantially overlap, with similar means and standard deviations for the majority of traits.

\begin{figure}[!htbp]
\caption{Traits score density comparison between sub-sampled GAN (red) generated images and Kiva images.}\label{densitytraits}
\centering
\includegraphics[width=0.9\linewidth]{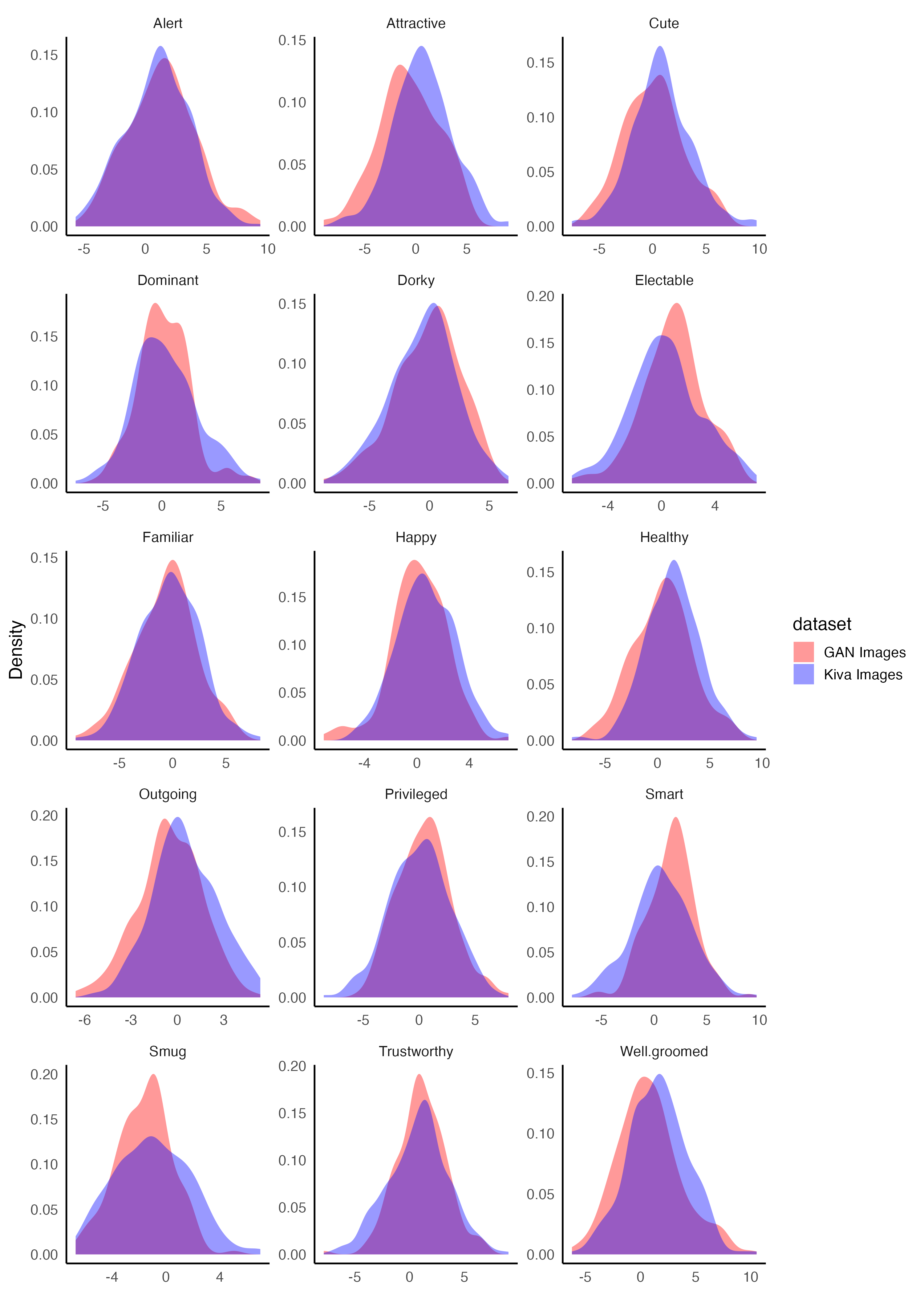}\\
\end{figure}
These checks are not conclusive, but they support the claim that perceived psychological attributes in our GAN images track those in real Kiva borrower profiles. It also shows that modifications to the generated images produce trait shifts comparable in size to the trait differences observed across Kiva images.

\section{Attention checks in the experiment}\label{attention_checks}
To check the quality of experimental data, we included attention checks in the survey. Attention checks are questions designed explicitly to detect inattentive responses through direct queries of attention or through questions designed to catch inattentive respondents (\cite{abbey2017attention}). There are three purposes of the attention checks in our experimental setting First, attention checks ensure that the recruited subjects are fully informed of their roles in the correct context before subjects make decisions. Second, attention checks prevent the subjects from careless decision-making and help the recruited subjects make rational decisions. In addition, attention checks also give us the flexibility to filter the data in order to have high-quality ones, depending on whether we would like to tighten or loosen our criteria.

In order to avoid the attention checks themselves inducing a deliberative mindset and becoming a threat to the validity, we try to ask the subjects to recall detail in a previous image after they make the choice and the correct answer to that gives us the reason to believe that people have been paying rational attention to their choices.\footnote{\cite{kung2018attention} encourage researchers to justify the use of attention checks without compromising scale validity} 

The Attention check in Figure \ref{Attention_1} asks \emph{What is the objective of a lender on a micro-lending platform?}. This question clarifies the lenders' role by differentiating the role between profit-making investors and non-profit investors.  By answering this question correctly, the recruited subject understands that, as a donor in a non-profit micro-lending platform dedicated to expanding equal and reachable loan access, their goal should be supporting the poor borrowers and communities in need, instead of investing for profit (a prompt with this information was provided earlier in the survey).

\begin{figure}[!htb]
    \centering
    \caption{Attention check 1}\label{Attention_1}
    \includegraphics[scale = 0.4]{fig/attention0.pdf}
    
\end{figure}

\begin{figure}[!htbp]
    \centering
    \begin{subfigure}[t]{0.49\textwidth}
        \centering
        \includegraphics[width=\linewidth]{fig/attention2.pdf}
        \caption{Attention check 2}\label{fig:att2}
    \end{subfigure}\hfill
    \begin{subfigure}[t]{0.49\textwidth}
        \centering
        \includegraphics[width=\linewidth]{fig/attention3.pdf}
        \caption{Attention check 3}\label{fig:att3}
    \end{subfigure}
    \caption{Attention checks}\label{fig:att_combined}
\end{figure}

Attention checks in Figures \ref{fig:att2} and \ref{fig:att3} are conducted in the format of a quiz. Attention check in Figure \ref{fig:att2} is an open-ended query asking the subject for the reason of their decisions.\footnote{\cite{abbey2017attention} uses this type of attention checks and manipulation validations to detect inattentive respondents in primary empirical data collection.} The last check is a multiple choice query asking about the occupation of the borrower on the previous slide.

\begin{figure}[!htbp]
    \centering
    \caption{Responses to Attention check 1 (lender's objective)} \label{attention_obj}
    \includegraphics[width = 0.5\textwidth]{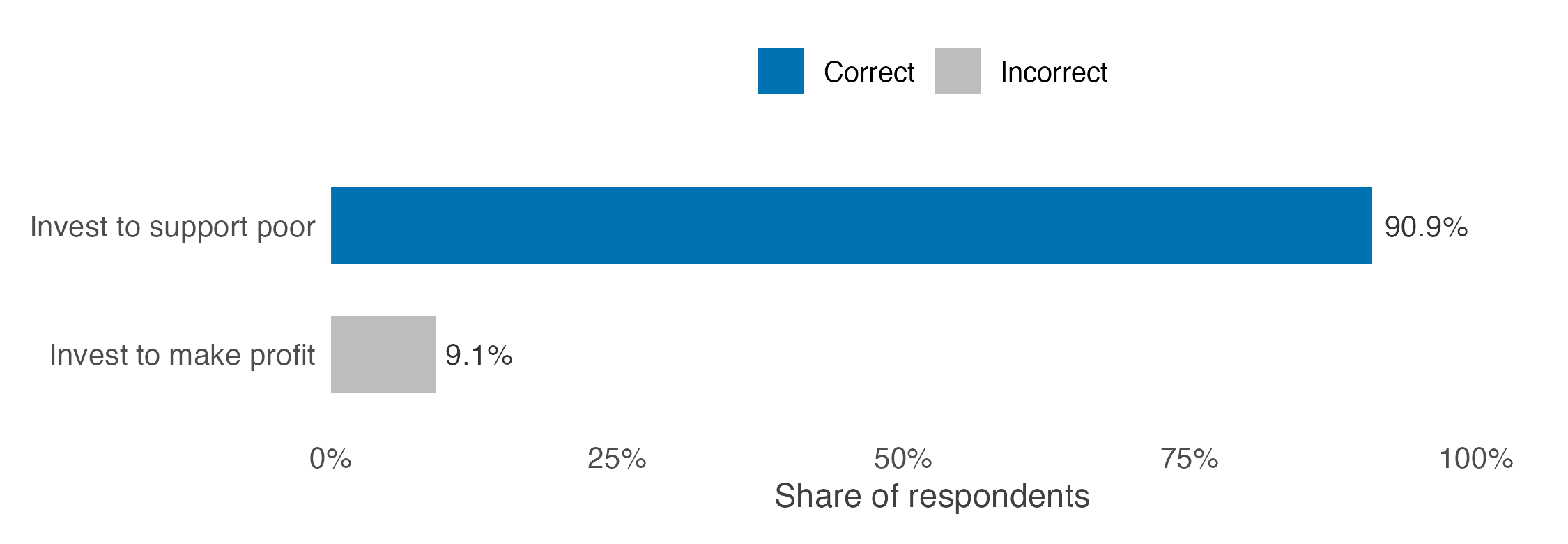}
    \caption*{\footnotesize{\textit{Note: Share of subjects choosing each answer to ``What is the objective of a lender on a micro-lending platform?'' The correct answer (\emph{Invest to support poor}, blue) was chosen by 90.9 percent of subjects.}}}
\end{figure}

\begin{figure}[!htbp]
    \centering
    \caption{Responses to Attention check 3 (borrower's occupation)} \label{attention_occupation}
    \includegraphics[width = 0.5\textwidth]{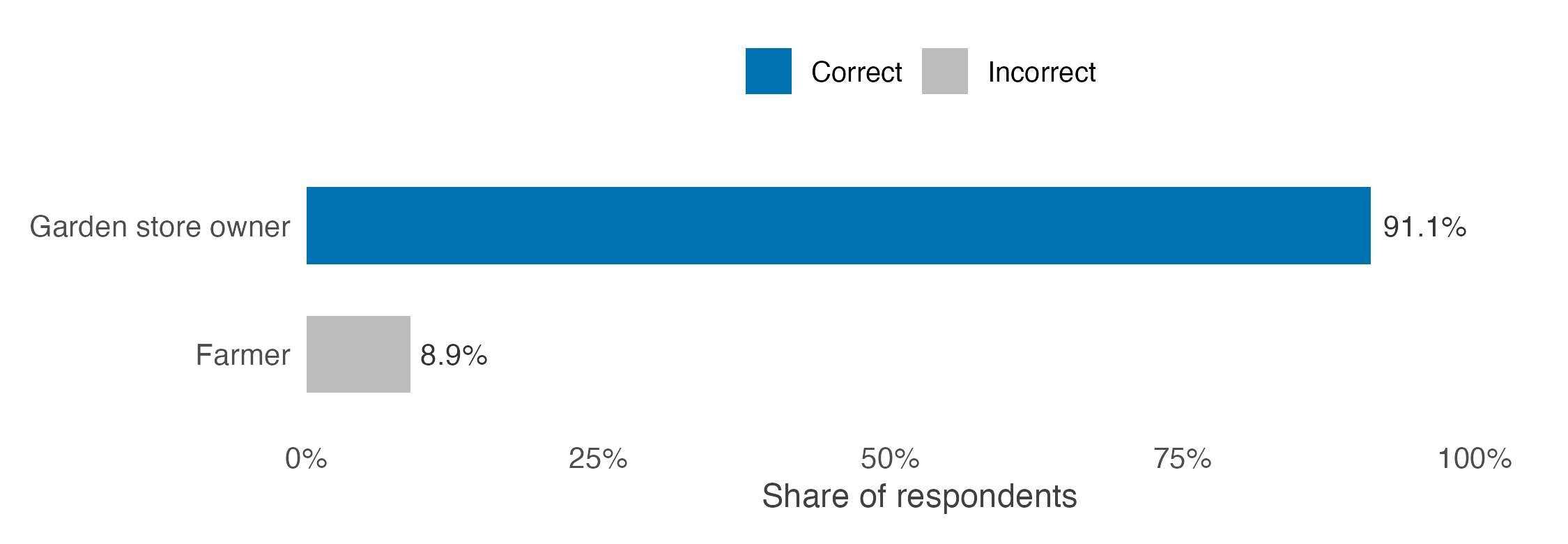}
    \caption*{\footnotesize{\textit{Note: Share of subjects choosing each answer to the multiple-choice query about the occupation of the borrower on the preceding slide. The correct answer (\emph{Garden store owner}, blue) was chosen by 91.1 percent of subjects.}}}
\end{figure}

Figures \ref{attention_obj} and \ref{attention_occupation} show how subjects answered attention checks. The correct answer was chosen by 90.9 percent of subjects on the lender-objective question and by 91.1 percent on the occupation question. We take these high accuracy rates as evidence that subjects were paying attention to their choices.

\section{Summary statistics from the experiment}\label{sum_stats_recruited}

\begin{figure}[!htbp]
    \centering
    \begin{subfigure}[t]{0.32\textwidth}
        \centering
        \includegraphics[width=\linewidth]{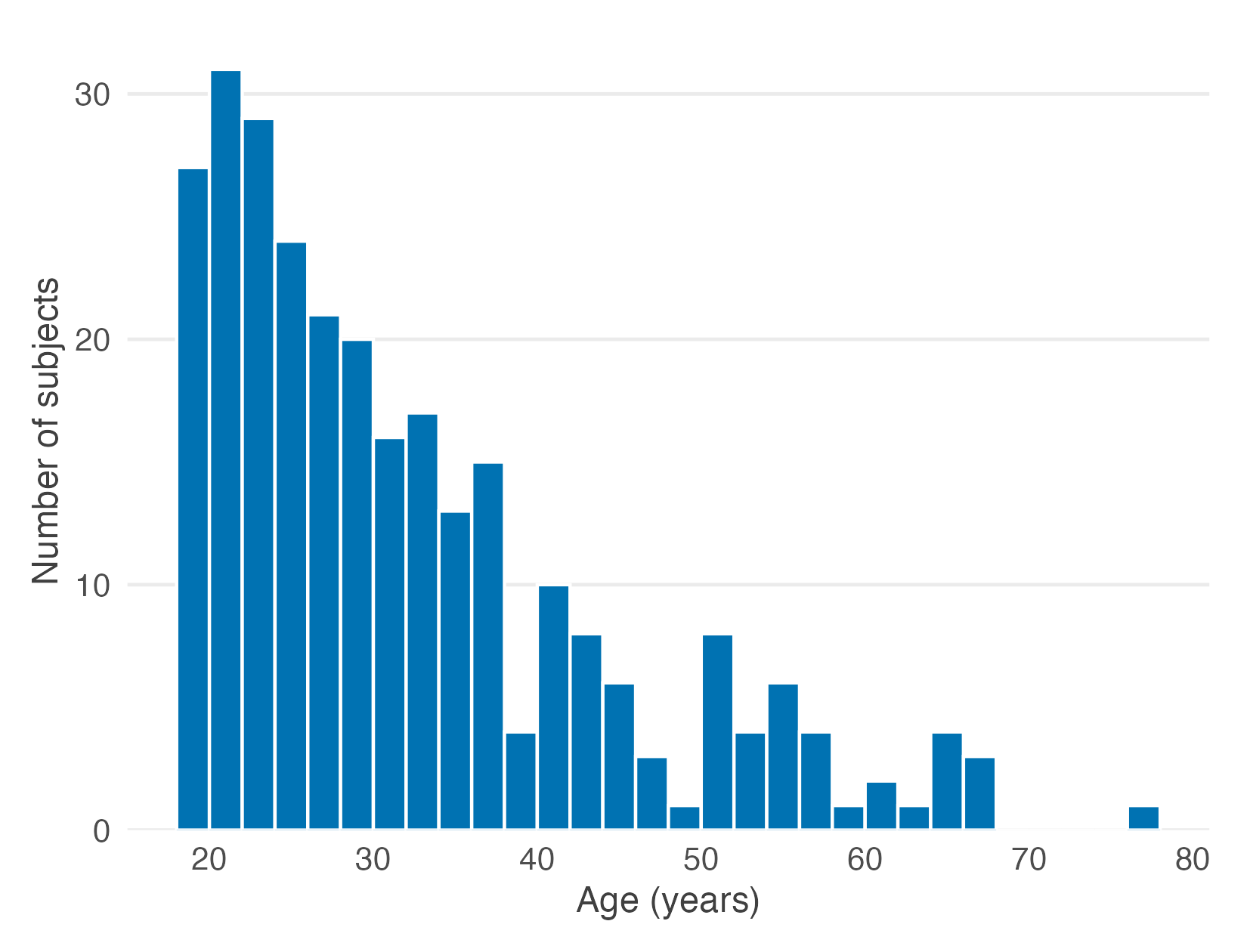}
        \caption{Age}\label{fig:demoexp1age}
    \end{subfigure}\hfill
    \begin{subfigure}[t]{0.66\textwidth}
        \centering
        \includegraphics[width=\linewidth]{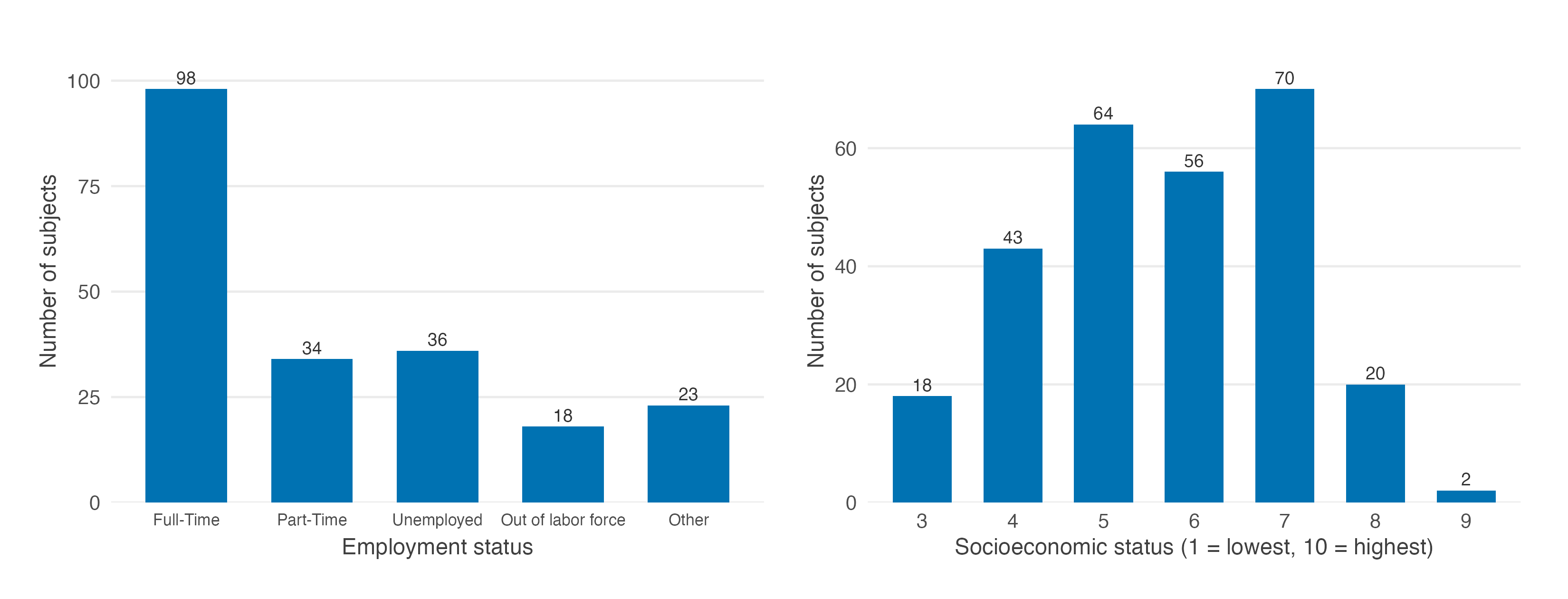}
        \caption{Employment and socio-economic status}\label{fig:demoexp1}
    \end{subfigure}
    \caption{Experiment 1 subject demographics}\label{fig:demoexp1_combined}
    \caption*{\footnotesize\textit{Note: All fields are self-reported. Panel (a) shows the distribution of age (bin width: 2 years). Panel (b), left: employment status; we group full-time and ``due to start a new job within the next month'' responses, and drop observations where the Prolific-side employment field had expired (29 percent of subjects). Panel (b), right: self-assessed socio-economic status on a 1--10 scale; we screened participants to be of at least status 3.}}
\end{figure}

Figure \ref{fig:demoexp1_combined} describes the subjects who took part in Experiment 1. The age distribution skews young: subjects in their twenties account for the largest share, and the median age is below thirty. The modal subject is full-time employed, and the modal self-assessed socio-economic status is between 5 and 7 on a 1--10 scale (we required participants to score at least 3). Roughly 29 percent of the sample is missing employment information, mostly because the Prolific-side employment field had expired at the time of the experiment.

\section{Diagnostics of the Recruited Experiments}\label{exp_diagnostics}
Table \ref{tab:covariate_balance} and Table \ref{tab:balance_summary} show tests for covariate differences in the two experiments. We compare characteristics of subjects across all the treatments. We report standard errors not-adjusted for multiple hypotheses testing. We find that subjects' characteristics are balanced across treatments in the first experiment. In the second experiment, there is one statistically significant difference -- there is a difference in the probability of being full time employed between subjects who saw borrowing campaigns with glasses and without glasses. Note, that each subject saw all potential combinations of features, however, some subjects saw specific features multiple times.

\begin{table}[ht]
\centering
\caption{Covariate Balance Across Treatments}
\label{tab:covariate_balance}
\begin{tabular}{lccc}
\hline
\textbf{Covariate} & \textbf{Smile (1 vs 0)} & \textbf{Bodyshot (1 vs 0)} & \textbf{Male (1 vs 0)} \\
\hline
Full-Time Employed & -0.001 (0.014) & -0.022 (0.014) &  0.002 (0.013) \\
High Charity       &  0.002 (0.007) & -0.004 (0.007) & -0.003 (0.006) \\
High Status        &  0.010 (0.014) &  0.020 (0.014) & -0.001 (0.013) \\
Male Subject       &  0.007 (0.015) &  0.005 (0.015) & -0.002 (0.014) \\
Student            & -0.005 (0.013) &  0.012 (0.013) &  0.002 (0.012) \\
\hline
\end{tabular}
\begin{flushleft}
\footnotesize{
\textit{Notes:} Each cell reports the difference in means of the covariates for the specified treatment comparison. Standard errors are shown in parentheses. Significance levels: * $p < 0.1$, ** $p < 0.05$, *** $p < 0.01$.
}
\end{flushleft}
\end{table}

\begin{table}[ht]
\centering
\caption{Covariate Balance Across Groups}
\label{tab:balance_summary}
    \resizebox{0.95\textwidth}{!}{%
\begin{tabular}{lcccccc}
\hline
\textbf{Covariate} & \textbf{Old (1 vs 0)} & \textbf{SE (Old)} & \textbf{p-value (Old)} & \textbf{Dark Hair (1 vs 0)} & \textbf{SE (Dark Hair)} & \textbf{p-value (Dark Hair)} \\
\hline
Full-Time Employed &  0.005 & 0.012 & 0.678 & -0.002 & 0.013 & 0.868 \\
High Charity       &  0.000 & 0.006 & 0.957 & -0.005 & 0.006 & 0.378 \\
Male Subject       &  0.012 & 0.013 & 0.365 & -0.023 & 0.013 & 0.064 \\
Student            & -0.001 & 0.009 & 0.928 & -0.006 & 0.008 & 0.502 \\
\hline
\textbf{Covariate} & \textbf{Glasses (1 vs 0)} & \textbf{SE (Glasses)} & \textbf{p-value (Glasses)} & \textbf{Sunglasses (1 vs 0)} & \textbf{SE (Sunglasses)} & \textbf{p-value (Sunglasses)} \\
\hline
Full-Time Employed & -0.031 & 0.013 & 0.020 &  0.015 & 0.013 & 0.261 \\
High Charity       & -0.005 & 0.006 & 0.453 &  0.003 & 0.006 & 0.623 \\
Male Subject       & -0.010 & 0.013 & 0.446 & -0.000 & 0.013 & 0.976 \\
Student            & -0.001 & 0.009 & 0.868 &  0.002 & 0.009 & 0.819 \\
\hline
\end{tabular}
}
\begin{flushleft}
\footnotesize{
\textit{Notes:} The table reports the differences in covariates across groups. Columns show the difference in means, standard errors (SE), and p-values for each covariate and treatment comparison. Significance levels: * $p < 0.1$, ** $p < 0.05$, *** $p < 0.01$.
}
\end{flushleft}
\end{table}

\begin{figure}[!htbp]
    \centering
    \caption{Mean outcomes by borrower profile}\label{fig:mean_outcomes}
    \begin{subfigure}[b]{0.4\textwidth}
        \centering
        \includegraphics[width=\linewidth]{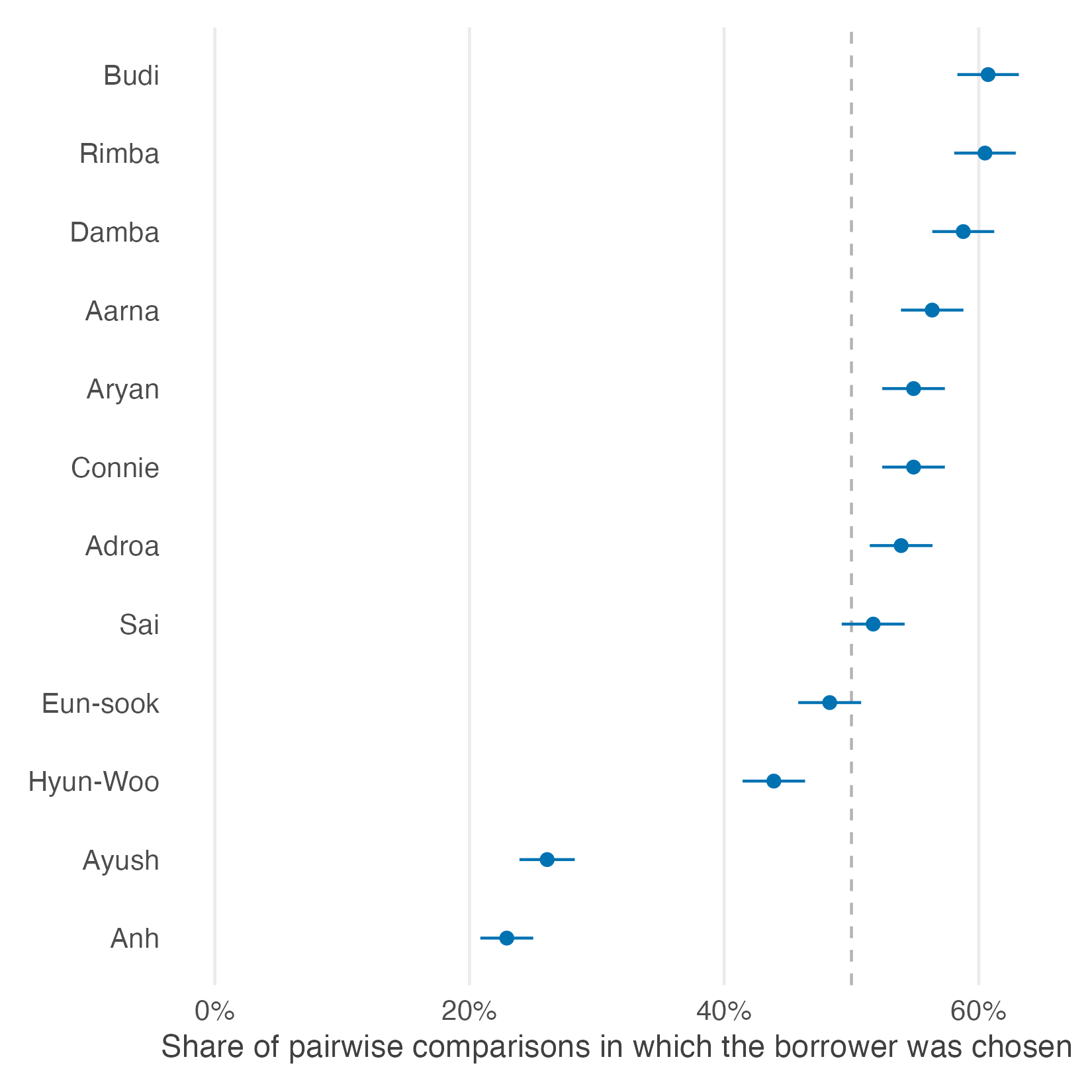}
        \caption{Experiment 1}\label{fe_1}
    \end{subfigure}\hfill
    \begin{subfigure}[b]{0.4\textwidth}
        \centering
        \includegraphics[width=\linewidth]{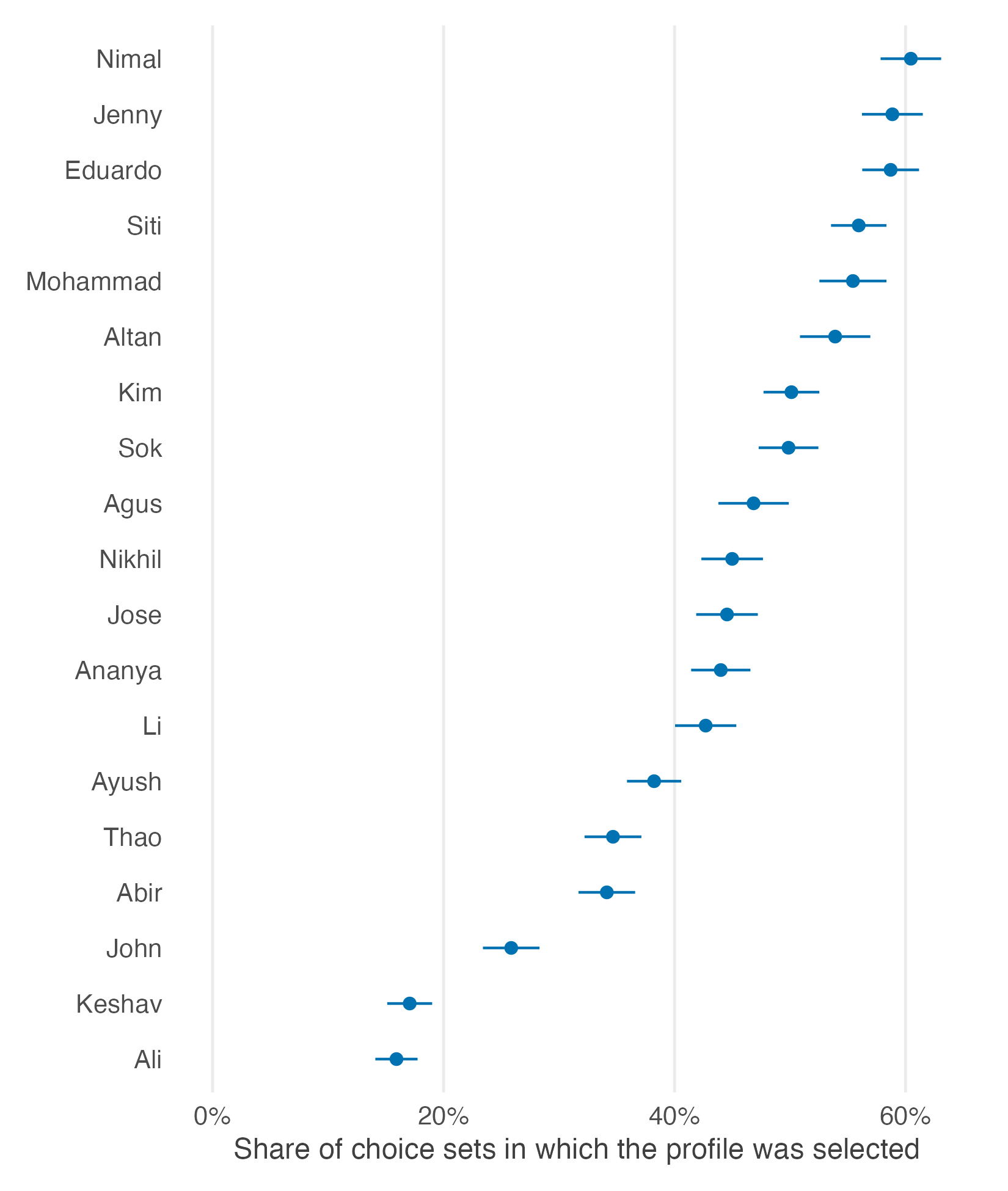}
        \caption{Experiment 2}\label{fe_2}
    \end{subfigure}
    \caption*{\footnotesize{\textit{Note: For each borrower profile, the share of choice sets (Experiment 1: pairwise comparisons; Experiment 2: full menus) in which the profile was selected, with $\pm 1$ standard error of the mean. In Panel (a), the dashed reference line marks the 50 percent rate that would obtain under random choice in a binary comparison. Profiles are ordered by their mean selection rate.}}}
\end{figure}

Figure \ref{fig:mean_outcomes} reports the average choice rate for each borrower profile in the two experiments. In Experiment 1 (Panel a), the highest-rated profiles (Budi, Rimba, Damba) are chosen in roughly 60 percent of pairwise comparisons, while the lowest-rated profiles (Anh, Ayush) are chosen in fewer than 25 percent. In Experiment 2 (Panel b), the spread is wider: profiles range from a 17 percent selection rate (Ali) to 60 percent (Nimal). These persistent gaps motivate the inclusion of profile fixed effects in the choice models reported in the main text.

\section{Comparison of Observational, Experimental, and SIMEX-Adjusted Estimates}
\label{app:comparison_estimates}

\begin{figure}[!htbp]
\centering
\caption{Comparison of observational, experimental, and SIMEX-adjusted estimates}
\includegraphics[width=0.65\textwidth]{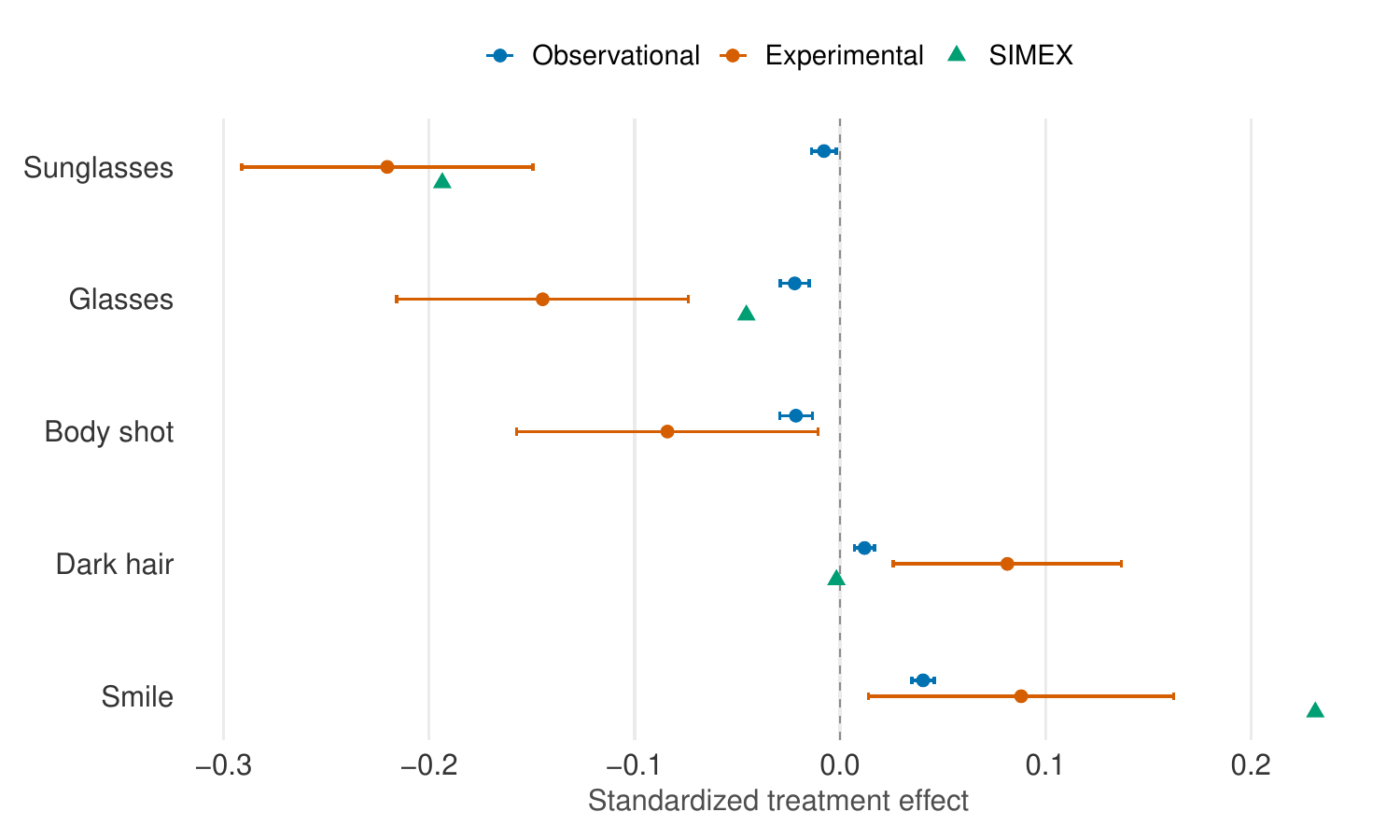}
\label{fig:obs_vs_exp_vs_simex}
\caption*{\footnotesize{\textit{Note:} The figure reports standardized treatment effect estimates across three approaches: observational estimates (blue), experimental estimates (red), and SIMEX-adjusted observational estimates (green). Error bars represent 95\% confidence intervals for observational and experimental estimates. SIMEX estimates are shown without confidence intervals because extrapolation-based standard errors are not available in the consolidated summary.}}
\end{figure}

Figure~\ref{fig:obs_vs_exp_vs_simex} compares standardized treatment-effect estimates from the observational analysis, the randomized experiments, and the SIMEX-corrected observational approach. The three sets of estimates are broadly aligned, but they need not coincide for several reasons.

First, the observational and experimental estimates are drawn from different populations. The experimental data consist of choices made by recruited participants in a controlled setting, whereas the observational estimates reflect behavior by lenders on the Kiva platform. Differences in user composition may therefore generate systematic differences in estimated treatment effects.

Second, the observational and SIMEX-adjusted estimates rely on an unconfoundedness assumption. Although we control for a rich set of covariates, remaining unobserved heterogeneity correlated with both visual features and outcomes could bias the estimates. By contrast, the experimental estimates are identified by random assignment and do not rely on this assumption.

Third, the observational estimates use ML-generated features that are subject to non-classical measurement error. The SIMEX procedure adjusts for this bias using misclassification rates estimated from the audit study. The correction assumes that extrapolating from observed noise levels to the no-noise counterfactual recovers the underlying parameters. This is a standard approach, but it introduces a modeling layer that is absent from both the raw observational and experimental estimates.

Despite these differences, the estimates agree in sign and differ primarily in magnitude. With the exception of \textit{Dark hair}, the SIMEX-adjusted estimates are larger in absolute value than the corresponding observational estimates, consistent with attenuation from measurement error. The baseline observational estimates are therefore conservative for most features.

\section{Analysis of defaults across default types}\label{default_types}

In this section, we analyze alternative outcomes related to loan repayment. The non-repayment of a Kiva loan is typically attributable to a borrower's default. However, in certain instances, a microfinance organization's default might precipitate the non-repayment. It is plausible that the borrower's image features might be indicative of the borrower's propensity for repayment but not necessarily predictive of a default by the microfinance organization. To investigate this, we define a new outcome variable, \emph{default by borrower}. This variable takes the value of 0 in scenarios where the loan has been repaid or the microfinance organization has defaulted; otherwise, it takes the value of 1.

In some cases, a borrower defaults not on the entirety of the loan but only on a part of it. To capture this we introduce an outcome \emph{share not repaid} that takes values between 0 and 1 and represents the proportion of funds left unpaid by the borrower.

We analyze whether style features are predictive of these new outcomes. We train a Boosted Forrest (GBM) on 70\% of data and report the predictive performance on the 30\% test set. We consider three model specifications: a constant model, a model incorporating all style features, and a full model which includes all covariates. The results are presented in Table \ref{tab:def_borrower}. We find that style features do not improve the predictive performance of models of either of the outcomes. 

\begin{table}[!ht]
    \centering
          \caption{Comparison of the test-set predictive performance of default models with and without image features.}
\begin{tabular}{lccc}
\\[-1.8ex]\hline 
\hline \\[-1.8ex] 
Outcome variable & Model specification & MSE & Standard error \\ 
  \hline
Default borrower & Constant & 0.0058 & 0.0004 \\ 
  Default borrower & Style & 0.0057 & 0.0003 \\ 
Default borrower & Full & 0.0039 & 0.0002 \\  \hline
Share not repaid & Constant & 0.0030 & 0.0002 \\ 
Share not repaid & Style & 0.0033 & 0.0002 \\ 
Share not repaid & Full & 0.0031 & 0.0002 \\ 
\\[-1.8ex]\hline 
\hline \\[-1.8ex] 
\end{tabular}
            \caption*{\footnotesize{\textit{Note: Test set performance of selected predictive models with different sets of covariates.}}}\label{tab:def_borrower}
\end{table}

\section{Mediation Analysis}
\label{sec:mediation_appendix}

Section~\ref{psych_traits} established that style features causally shift predicted psychological traits. This appendix takes up a more demanding question: how much of the effect of a style feature on funding operates through the trait channel? This decomposition is not identified by the assumptions of Section~\ref{psych_traits}; it requires two further conditional-independence assumptions, which we state and apply below.

\paragraph{Data.} We use the observational sample throughout this appendix. The GAN sample is too small to support the bootstrap-based mediation procedure with stable inference. We note that the use of observational data implies that style features will be systematically related to other characteristics of images (in contrast to our GAN image dataset, where each base image has a fixed number of variants corresponding to the styles of interest, implying that other image characteristics are independent of style in the dataset).  We score a random sample of $1{,}496$ observational borrower images using \citet{peterson2022deep}'s model, following the same procedure as for the GAN images.

\begin{figure}[htbp]
\centering
\caption{Effects of Style Features on Predicted Psychological Traits: Observational vs.\ Experimental}
\includegraphics[width=0.9\textwidth]{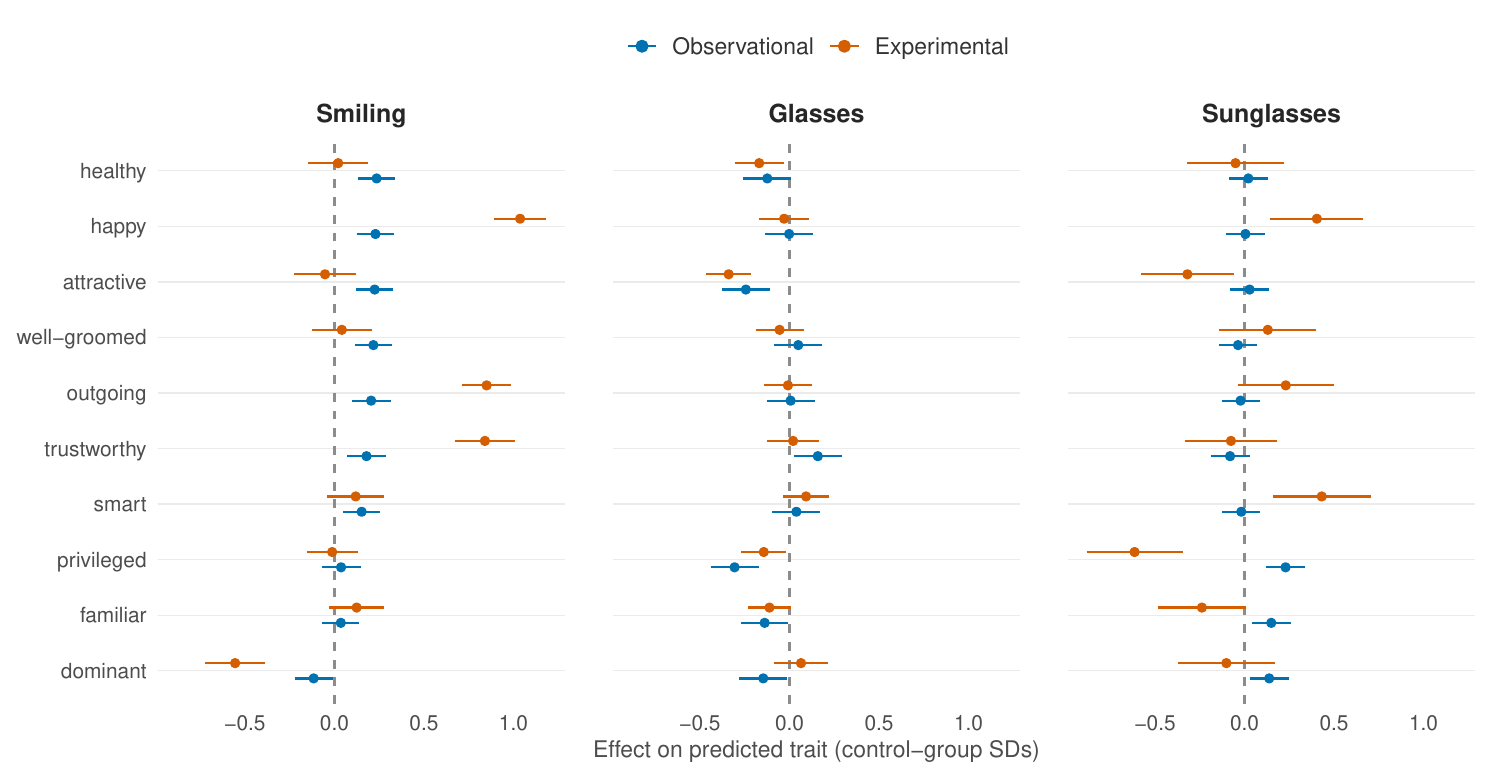}
\caption*{\footnotesize{\textit{Notes:} Each point is the estimated effect of a style feature on a predicted psychological trait, in control-group standard deviations of the trait. Horizontal bars are 95\% confidence intervals. Blue points: observational sample, with standard errors clustered at the loan level. Orange points: GAN-based experimental sample.}}
\label{fig:apath_obs_vs_exp}
\end{figure}

Figure~\ref{fig:apath_obs_vs_exp} compares the estimated effects of style features on predicted traits across the two samples; both sets of estimates are standardized by the control-group standard deviation of the trait. The two sets agree in sign for the majority of (style, trait) pairs, and the observational estimates are more precise, resolving additional channels as statistically significant. \emph{Smiling} is the clearest case: the experimental analysis already identifies the headline channels (\emph{trustworthy}, \emph{happy}, \emph{outgoing}, \emph{dominant}), and the observational analysis adds \emph{healthy}, \emph{attractive}, \emph{well-groomed}, and \emph{smart}. \emph{Glasses} behave similarly across the two samples, with one exception: the estimate for \emph{dominant} flips from a positive and not statistically significant experimental estimate to a negative and statistically significant observational estimate. \emph{Sunglasses} are the least consistent of the three: although most observational estimates are not statistically significant, those for \emph{privileged}, \emph{familiar}, and \emph{dominant} are statistically significant with small magnitudes, and no trait reaches statistical significance with a consistent sign in both samples.

In what follows, the treatment $T_i \in \{0,1\}$ is one of the three style indicators, the mediator $M_i$ is one of the ten predicted traits, the outcome $Y_i$ is daily funding (cash per day), and the covariates $X_i$ are the loan-level controls used elsewhere in the paper.
 
\paragraph{Estimand.} The estimand of interest is the share of a style feature's effect on funding that is mediated by the predicted trait. Following \citet{imai2010general}, we define this quantity using potential outcomes. For each unit $i$, let $M_i(t')$ denote the potential value of the trait under style feature value $t' \in \{0,1\}$, and let $Y_i(t,m)$ denote the potential value of funding under style feature value $t \in \{0,1\}$ and trait value $m \in \mathcal{M}$, where $\mathcal{M} \subseteq \mathbb{R}$ is the support of the predicted trait. The composite potential outcomes

\begin{equation}\label{pot_out_mediation}
Y_i\!\left(t,\, M_i(t')\right), \qquad (t, t') \in \{0,1\}^2,
\end{equation}
permit the style feature value that enters the outcome ($t$) to differ from the value that determines the trait ($t'$). The natural indirect effect at treatment level $t$ fixes $t$ in the outcome and varies $t'$ in the trait:
\begin{equation}
\overline{\mathrm{NIE}}(t)
\;=\;
\mathbb{E}\!\left[\, Y_i\!\left(t, M_i(1)\right) - Y_i\!\left(t, M_i(0)\right) \right].
\label{eq:nie}
\end{equation}
That is, $\overline{\mathrm{NIE}}(t)$ measures the change in expected funding induced by shifting the trait from $M_i(0)$ to $M_i(1)$ while holding the style feature itself at $t$. The natural direct effect is defined symmetrically, $\overline{\mathrm{NDE}}(t') = \mathbb{E}[Y_i(1, M_i(t')) - Y_i(0, M_i(t'))]$, and together the two effects decompose the average treatment effect $\tau \equiv \mathbb{E}[Y_i(1, M_i(1)) - Y_i(0, M_i(0))]$:
\begin{equation}
\tau \;=\; \overline{\mathrm{NIE}}(t) \;+\; \overline{\mathrm{NDE}}(1-t), \qquad t \in \{0,1\}.
\label{eq:mediation_decomp}
\end{equation}

\paragraph{Identification.} The composite potential outcomes in Equation~\eqref{pot_out_mediation} with $t \ne t'$ are never observed: they describe funding under one style-feature regime evaluated at the trait value the unit would display under a different regime. Identifying the natural direct and indirect effects therefore requires assumptions beyond the unconfoundedness condition invoked in Section~\ref{offline_data} (which was satisfied by construction in our experimental data with GAN images). We follow \citet{imai2010identification} and impose two such assumptions.

\begin{ass}[Sequential ignorability of treatment]
\label{ass:a1}
For all $(t, t', m) \in \{0,1\}^2 \times \mathcal{M}$,
\begin{equation}
\{Y_i(t', m),\, M_i(t)\} \;\perp\!\!\!\perp\; T_i \;\big|\; X_i.
\label{eq:a1}
\end{equation}
\end{ass}

Assumption~\ref{ass:a1} strengthens the unconfoundedness condition used in Section~\ref{offline_data} along two dimensions. First, it asserts conditional independence of $T_i$ from the funding potential outcome at every treatment–mediator pair $(t', m)$, not only at the realized pair. Second, it requires the same conditional independence to hold for the mediator: $M_i(t) \perp\!\!\!\perp T_i \mid X_i$. The covariates $X_i$ must therefore be rich enough to render the style feature as good as randomly assigned with respect to both the predicted trait and the funding outcome. Assumption~\ref{ass:a1} would fail if, for example, borrowers working with a more skilled photographer have photographs that are more likely to include a smile but are also more likely to be scored as trustworthy based on factors not captured by $X_i$.

\begin{ass}[Cross-world independence of the mediator]
\label{ass:a2}
For all $(t, t', m) \in \{0,1\}^2 \times \mathcal{M}$,
\begin{equation}
Y_i(t', m) \;\perp\!\!\!\perp\; M_i(t) \;\big|\; T_i = t,\, X_i.
\label{eq:a2}
\end{equation}
\end{ass}

Assumption~\ref{ass:a2} states that, conditional on covariates and on the realized style feature $T_i = t$, the mediator $M_i(t)$ is independent of the funding potential outcome $Y_i(t', m)$ at every $(t', m)$. When $t \ne t'$, the assumption equates conditional distributions across counterfactual regimes that are never jointly observed; it is therefore inherently untestable, and it has no analog elsewhere in the paper. To interpret the assumption, note that for a borrower who did not smile, the trait she would have displayed had she smiled is a counterfactual we never observe. Assumption \ref{ass:a2} requires that, among non-smiling borrowers with the same $X_i$, this counterfactual trait value tells us nothing about the funding the borrower would receive under any style and trait combination. It would fail if, for example, borrowers with more symmetric faces gain more perceived trustworthiness from smiling and also attract more funding for reasons unrelated to the trait, such as a general lender preference for symmetric faces. In this scenario, the counterfactual trait would signal something about funding, namely facial symmetry.

The estimates reported below are identified under, and only under, the conjunction of Assumptions~\ref{ass:a1} and~\ref{ass:a2}.

\citet{imai2010identification} show that, under Assumptions~\ref{ass:a1} and~\ref{ass:a2}, the natural indirect and direct effects are identified from the joint distribution of $(Y_i, T_i, M_i, X_i)$ as
\begin{equation}
\overline{\mathrm{NIE}}(t)
=
\int\!\!\int \mathbb{E}\!\left[Y_i \mid T_i = t,\, M_i = m,\, X_i = x\right]
\Big[dF_{M \mid T=1, X=x}(m) - dF_{M \mid T=0, X=x}(m)\Big] dF_X(x),
\label{eq:nie_id}
\end{equation}
and
\begin{equation}
\overline{\mathrm{NDE}}(t)
=
\int\!\!\int \Big[\mathbb{E}\!\left[Y_i \mid T_i = 1,\, M_i = m,\, X_i = x\right] - \mathbb{E}\!\left[Y_i \mid T_i = 0,\, M_i = m,\, X_i = x\right]\Big] dF_{M \mid T=t, X=x}(m)\, dF_X(x).
\label{eq:nde_id}
\end{equation}
The two expressions have symmetric structures. Equation~\eqref{eq:nie_id} fixes the conditional outcome regression at treatment level $t$ and integrates it against the difference between the mediator distributions under treatment and control, isolating the change in funding attributable to shifting the trait distribution while the outcome response is held fixed. Equation~\eqref{eq:nde_id} reverses the roles, fixing the mediator distribution at the $T_i = t$ regime and integrating against the difference between the outcome regressions under treatment and control.

\paragraph{Estimation.}

We estimate $\overline{\mathrm{NIE}}(t)$ and $\overline{\mathrm{NDE}}(t)$ by the simulation-based plug-in estimator of \citet{imai2010general}: we replace each population object in Equations~\eqref{eq:nie_id}--\eqref{eq:nde_id} with its sample analog, simulate counterfactual mediator draws from the fitted mediator distribution, and average over units.

We model both nuisance components with generalized additive models, following the specification of \citet{imai2010general}. The outcome regression is additive in penalized smoothed functions of the continuous covariates and admits treatment-arm-specific smoothed functions of the trait, and the mediator regression is a Gaussian GAM in the same covariates. Allowing the smoothing in the trait to vary by treatment arm admits nonlinearity in the trait-funding relationship.

For each (style feature, trait) pair, the procedure consists of four steps: (i) fit the outcome and the mediator models on the full sample; (ii) simulate counterfactual mediators $\widehat{M}_i(0)$ and $\widehat{M}_i(1)$ from the fitted mediator model; (iii) predict the four composite counterfactual outcomes $\widehat{Y}_i\!\left(t, \widehat{M}_i(t')\right)$ for $(t, t') \in \{0,1\}^2$; and (iv) average over $i$ to form $\widehat{\overline{\mathrm{NIE}}}(t)$ and $\widehat{\overline{\mathrm{NDE}}}(t)$. We construct $95\%$ confidence intervals by the nonparametric bootstrap, resampling loans with replacement, refitting both regressions and repeating steps (i)--(iv) within each of $B$ replications.

We assess sensitivity to the functional form of the trait--funding relationship by re-estimating the entire procedure with linear specifications for both nuisance components (OLS in place of GAM) and report the two sets of estimates side by side. This comparison explores robustness to functional-form misspecification of the conditional outcome regression.

We estimate the decomposition separately for each of the ten predicted traits, with that trait as the sole mediator and the other nine traits excluded from both the outcome and mediator regressions. The predicted traits are themselves plausibly causally linked: a smile may raise perceived trustworthiness in part by raising perceived attractiveness, so conditioning on a second trait when estimating the indirect effect through a focal trait would block part of the causal pathway of interest. As a consequence of this design, each single-mediator NIE captures the total indirect effect through its mediator and any downstream traits causally linked to it. The ten NIEs within a given style feature therefore overlap on shared causal segments \citep{vanderweele2014mediation}.

\begin{figure}[htbp]
\centering
\caption{Standardized Natural Indirect Effects of Style Features, by Trait}
\includegraphics[width=\textwidth]{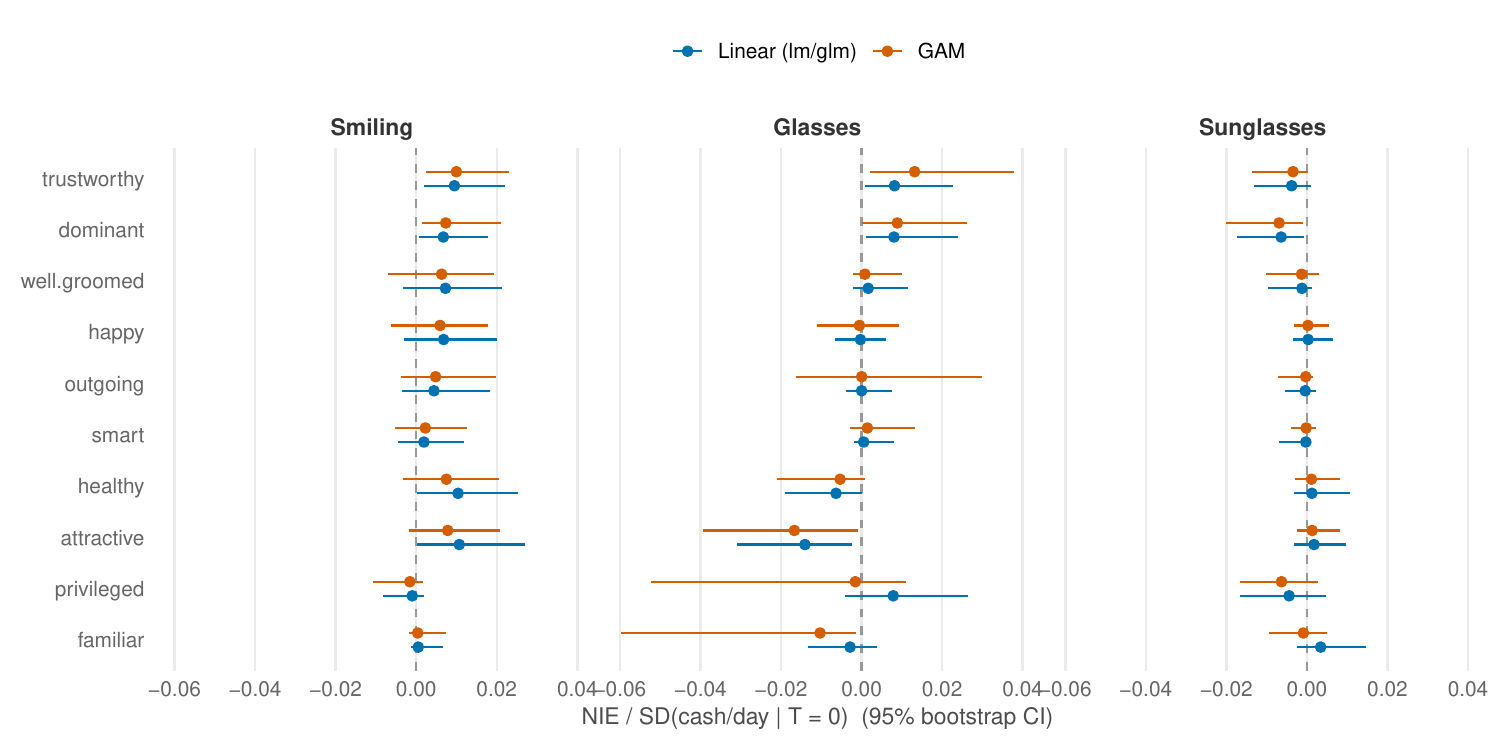}
\caption*{\footnotesize{\textit{Notes:} Each point is $\widehat{\mathrm{NIE}}/\widehat{\mathrm{SD}}(Y \mid T = 0)$ for one (style feature, trait) pair, observational sample. Horizontal bars are 95\% bootstrap percentile intervals over $1{,}000$ replicates. Blue points: linear specification. Orange points: GAM specification.}}
\label{fig:appendix_nie_std}
\end{figure}

\paragraph{Results.} Figure~\ref{fig:appendix_nie_std} reports $\widehat{\overline{\mathrm{NIE}}}$ for every (style feature, trait) pair, scaled by the control-group standard deviation of daily funding under the corresponding treatment. The figure overlays estimates from the GAM and OLS specifications; the two lead to similar conclusions across all $30$ (style feature, trait) pairs, so the results below do not depend on the GAM's flexibility in the trait--funding relationship.

The three style features differ substantially in how much of their funding effect operates through measured trait channels. \emph{Smiling} raises daily funding by approximately $0.15$ control-group standard deviations in total. Two trait channels are statistically distinguishable from zero: perceived trustworthiness contributes $\widehat{\overline{\mathrm{NIE}}} = 0.010$ ($7\%$ of the total) and perceived dominance contributes $\widehat{\overline{\mathrm{NIE}}} = 0.007$ ($5\%$). Confidence intervals for the remaining eight (smile, trait) pairs include zero.

\emph{Glasses} raise daily funding by approximately $0.12$ control-group standard deviations in total, but the trait decomposition reveals four offsetting channels of comparable magnitude. Two are positive---trustworthiness ($+0.013$, or $+11\%$ of the total) and dominance ($+0.009$, $+7\%$)---and two are negative---attractiveness ($-0.017$, $-14\%$) and familiarity ($-0.010$, $-8\%$). The positive total effect of glasses on funding thus reflects substantial cancellation across measured trait channels rather than a single dominant pathway.

\emph{Sunglasses} reduce daily funding by approximately $0.20$ control-group standard deviations in total. Only perceived dominance is borderline statistically distinguishable from zero, at $\widehat{\overline{\mathrm{NIE}}} = -0.007$ ($3\%$ of the total); the remaining nine intervals include zero. The sunglasses penalty therefore operates almost entirely outside the ten trait channels we measure.

The contrast across style features is the main finding of this appendix. The results suggest that the smile and glasses effects on funding can be decomposed as the joint product of how the style feature shifts perceived traits (Section~\ref{psych_traits}) and how those traits in turn shift funding, where the estimated natural indirect effects through the traits are substantial. The sunglasses penalty cannot be understood in the same way. Although Section \ref{psych_traits} establishes that sunglasses causally shift several perceived traits, the estimated natural indirect effects through these traits are close to zero. 

These conclusions are made under Assumptions~\ref{ass:a1} and~\ref{ass:a2}, which we do not test empirically. Without Assumption~\ref{ass:a2}, the same estimates have a descriptive interpretation, providing summaries of how style features, predicted traits, and funding co-vary in the data.

The decomposition has one practical implication for the policy counterfactuals. Because part of the smile effect on funding operates through perceived trustworthiness, a platform that recommends profile-photo changes can verify, image by image, that a recommended photograph in fact shifts the trait through which the policy is meant to operate, using the same predicted-trait scores reported here. Compliance (did the borrower follow the recommendation?) and mechanism (did the recommended photograph shift perceived trustworthiness?) are distinct questions, and our results suggest that policies should include verification that compliance is leading to shifts in traits that positively affect outcomes.

\section{Stability of coefficient estimates}\label{coef_stability}
In the analysis of counterfactual platform policies, we assume that the lenders' preferences for image features are stable across different market structures. However, the impact of specific image features might vary across different ways of organizing the marketplace, particularly when they're used as quality signals. The extent to which image features affect beliefs about quality depends on lenders' beliefs about how common it is for high-quality borrowers to have images with those features, and how frequent these features are in general. If a change in the market design alters the set of borrowers lenders consistently see, it's plausible that it will consequently shift both their prior beliefs about the borrowers' quality and their perception of how image features impact their posterior beliefs about quality. In the extreme case, when all borrowers a lender sees on the platform have a certain image feature, that image feature will not affect the beliefs about quality. In contrast, when image features impact lenders' utility from selecting a borrower, they will impact the outcomes irrespective of lenders' beliefs about how informative image features are of the borrowers' quality.

\begin{figure}[!htb]
    \centering
    \caption{Kiva website prior to 28th of May 2016}\label{old_kiva}
    \includegraphics[scale = 0.4]{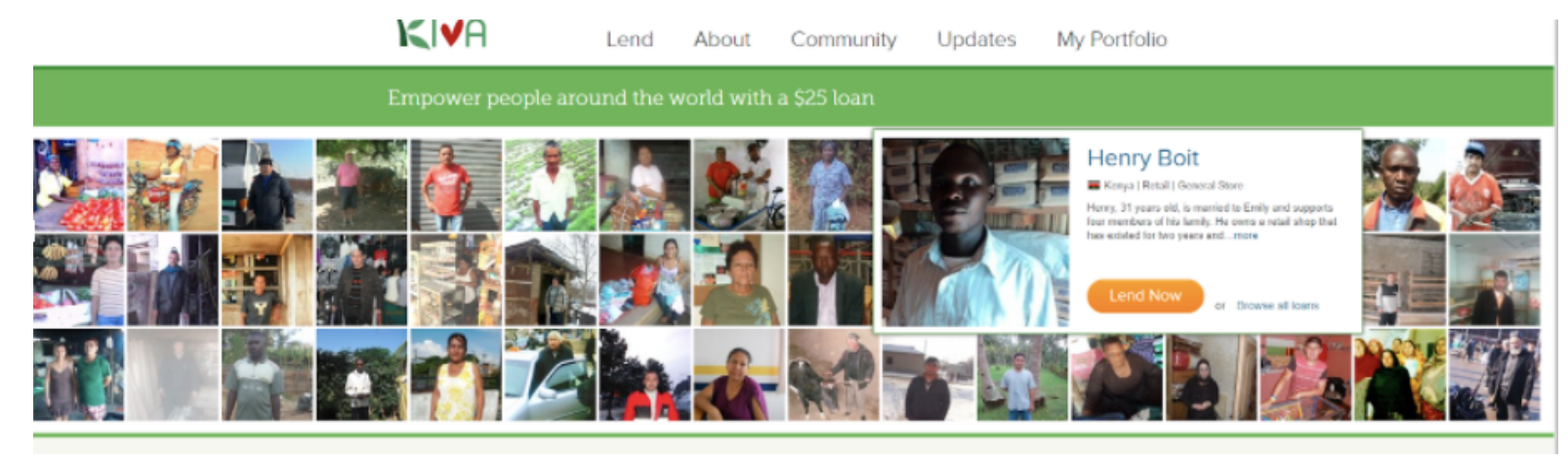}
    \caption*{\footnotesize{\textit{Note: Screenshot from https://kiva.org landing page. Source - https://archive.org/}}}
\end{figure}

\begin{figure}[!htb]
    \centering
    \caption{Borrowers' categories introduced after 28th of May 2016}\label{new_kiva}
    \includegraphics[scale = 0.25]{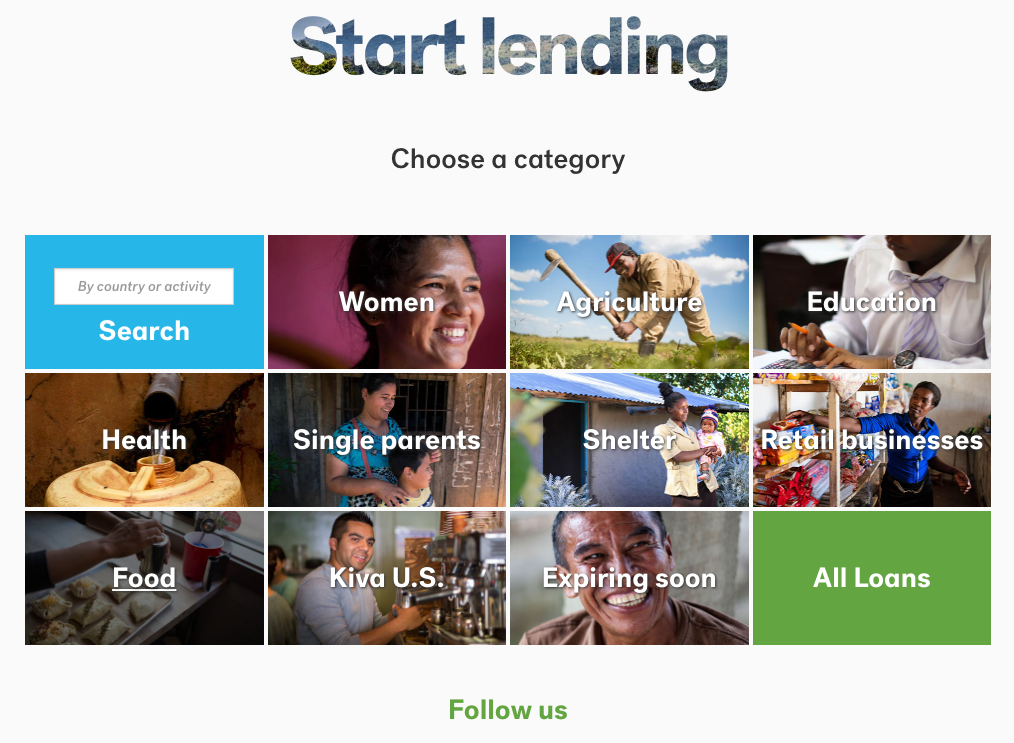}
    \caption*{\footnotesize{\textit{Note: Screenshot from https://kiva.org landing page. Source - https://archive.org/}}}
\end{figure}

To test this, we exploit a natural experiment in the form of the Kiva landing page redesign. On the 28th of May 2016, Kiva carried out a major website change; before that, all borrowers were displayed on the same page (see Figure \ref{old_kiva}). In the updated design, borrowers were sorted into categories. Figure \ref{new_kiva} shows the available categories as displayed on Kiva's new landing page. After the change, lenders can quickly select categories, and in doing so, they'll see a different pool of borrowers than before the website update. If style features are used to compare the available borrowers and mostly act as signals of underlying quality, then it's plausible that the change in the way borrowers are displayed should change the impact of style on lenders' choices.

To evaluate the stability of style coefficients, we consider two periods: period \emph{before}, which starts on 5/28/2015 and ends on 4/28/2016, and the period \emph{after}, which starts on 5/28/2016 and ends on 8/28/2017. Within these periods, the website was organized following the same logic, but across the periods, the display of borrowers differed. We end the \emph{before} period one month before the change so that most of the borrowing campaigns posted in the \emph{before} period would have ended before the introduction of the new system. 

We estimate the average treatment effect of smile on cash per day. We use the AIPW estimator, where the propensity and outcome models are estimated using Gradient Boosted Trees. Table \ref{tab:before_after} presents the estimated average treatment effects and their difference. We find that the impact of smile on cash per day was statistically significant and positive in each of these two periods. Their difference is not statistically significant. However, we also note that either of these estimates is lower than the estimate from Figure \ref{fig:ATE_cash}. This can, for example, reflect the change in the user base of Kiva over time. In the main analysis, we restricted attention to borrowing campaigns posted between 2006 and 2016, as we only had information on loan repayment for this period.

\begin{table}[!ht]
    \centering
          \caption{Comparison of the impact of \emph{Smile} before and after website redesign.}
\begin{tabular}{lcc}
  \hline  \hline
Period & Estimate & Std.err  \\ 
  \hline
Before period & 2.842 & 0.339 \\ 
After period & 2.781 & 0.506 \\ 
Difference & 0.061 & 0.609 \\ 
   \hline   \hline
\end{tabular}
            \caption*{\footnotesize{\textit{Note: Estimates of the impact of smile on cash-per-day before and after the website redesign. Estimates from the AIPW estimator. Last row estimate before less estimate after.}}}\label{tab:before_after}
\end{table}

\subsection{Coefficient stability across markets with different feature prevalence}\label{coef_stability_markets}

As an additional check on the stability of style coefficients, we examine whether the estimated impact of selected image features changes systematically across markets in which these features are more or less prevalent. The concern is that if lenders use a style feature primarily as a signal, then its informational value may decline in markets where the feature is very common. In contrast, if lenders derive utility from what the feature represents, its impact on outcomes need not disappear when it becomes more prevalent.

To study this, we define markets as \emph{sector} $\times$ \emph{month} cells. Within each market, we compute two estimates for each feature: (i) the unadjusted difference in means in cash per day between borrowers with and without the feature, and (ii) a simplified AIPW estimate of the average treatment effect, where the propensity score and outcome model adjust for market-level competition and borrower image characteristics. We then organize markets from low to high prevalence of a given feature and plot the corresponding estimates against feature prevalence. We focus on four style features that play a central role in the paper: \emph{Smile}, \emph{Glasses}, \emph{Sunglasses}, and \emph{Dark Hair}.

\begin{figure}[!htb]
    \centering
    \caption{Market-level coefficient estimates by feature prevalence}\label{fig:market_prevalence_stability}

    \begin{minipage}{0.49\textwidth}
        \centering
        \includegraphics[width=\textwidth]{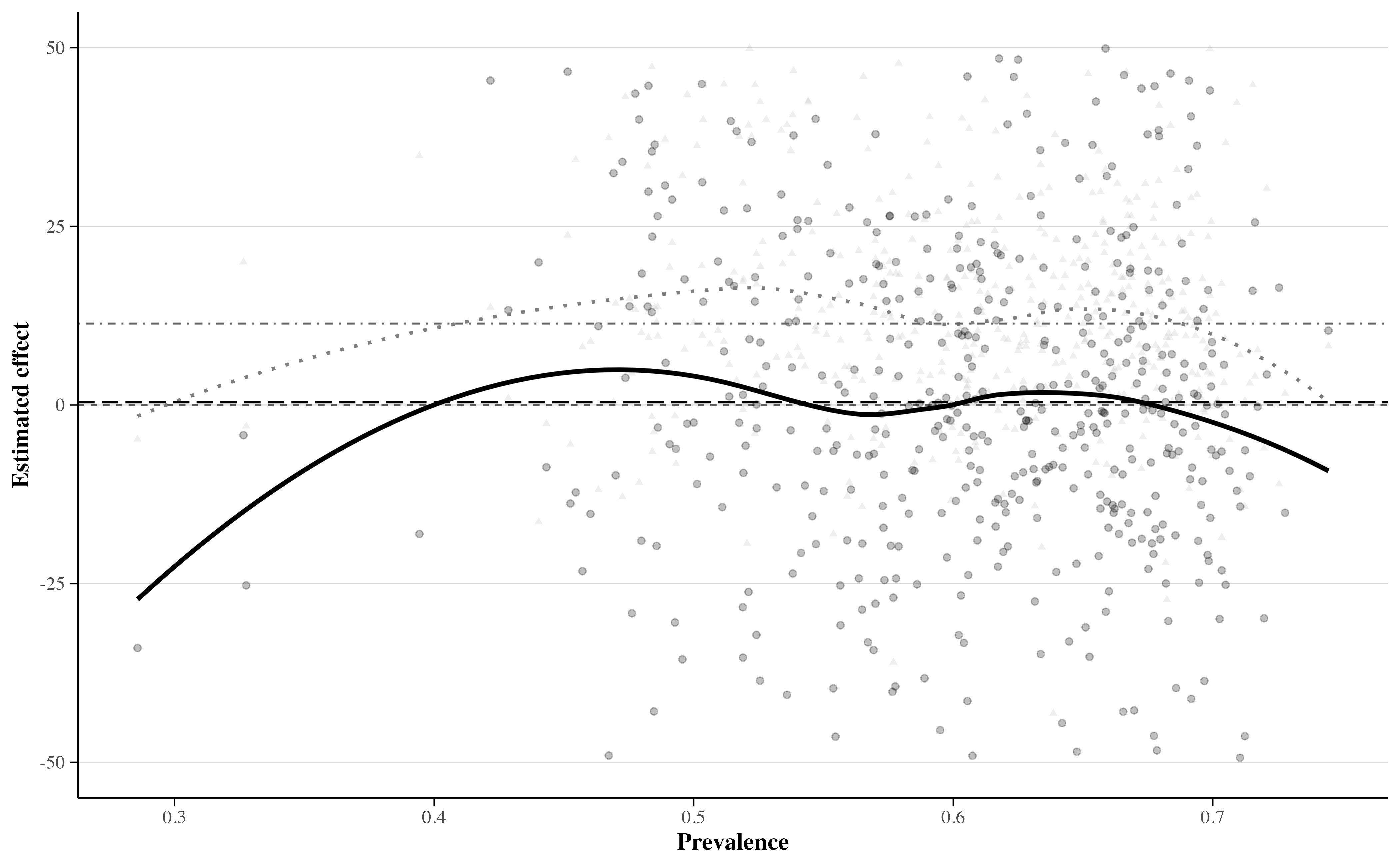}
        \vspace{0.2em}
        
        {\footnotesize (a) \emph{Smile}}
    \end{minipage}
    \hfill
    \begin{minipage}{0.49\textwidth}
        \centering
        \includegraphics[width=\textwidth]{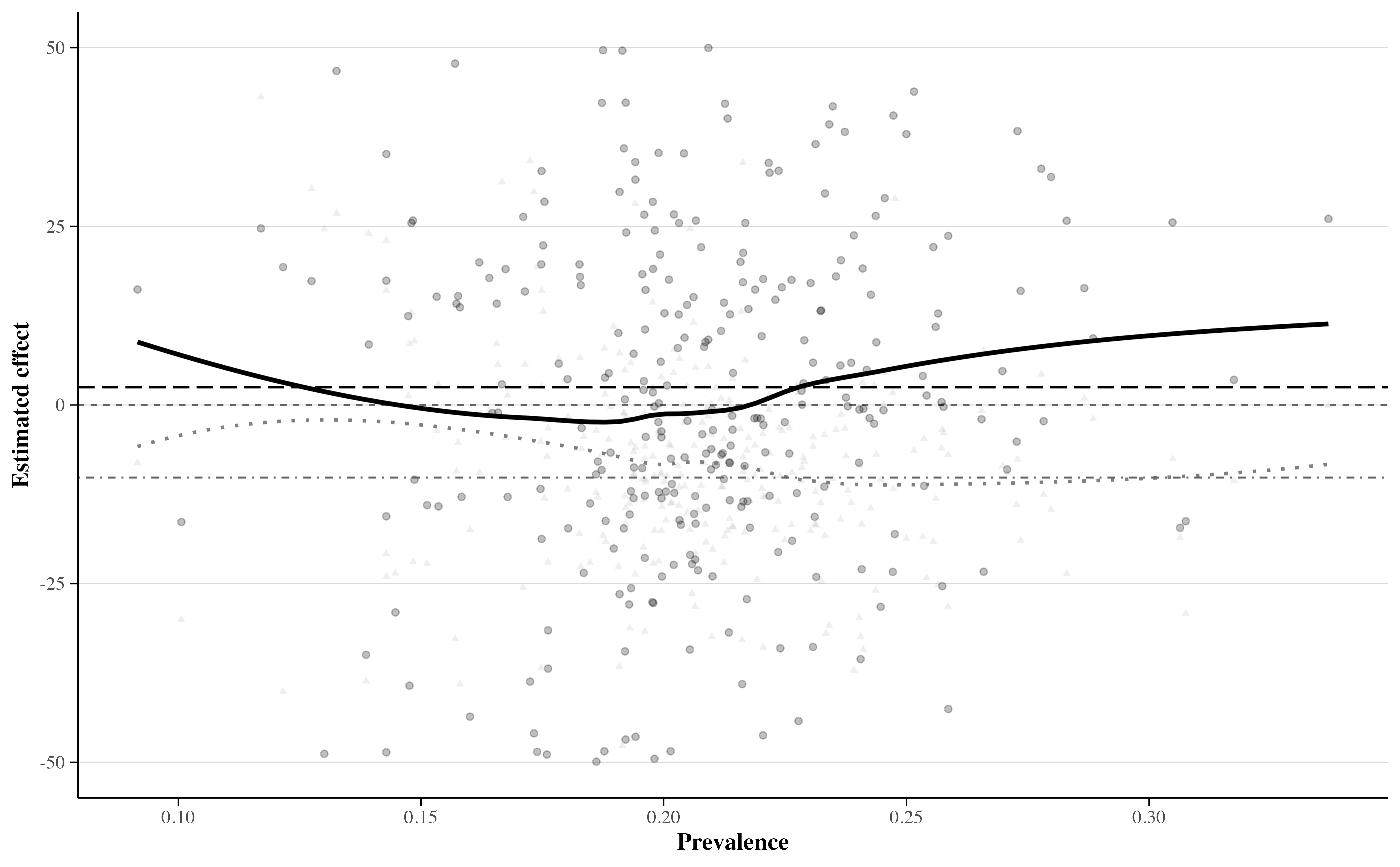}
        \vspace{0.2em}
        
        {\footnotesize (b) \emph{Glasses}}
    \end{minipage}

    \vspace{0.8em}

    \begin{minipage}{0.49\textwidth}
        \centering
        \includegraphics[width=\textwidth]{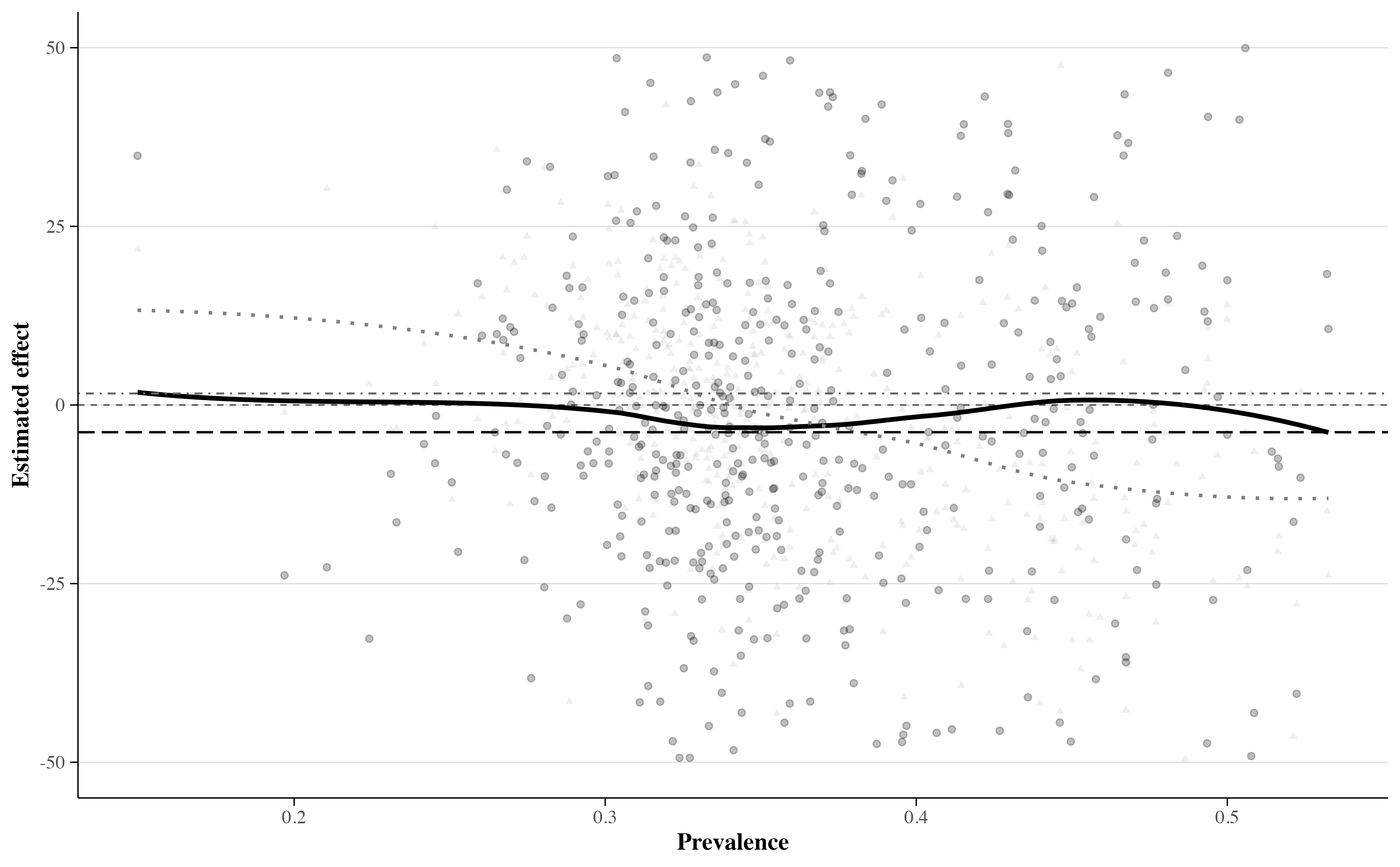}
        \vspace{0.2em}
        
        {\footnotesize (c) \emph{Sunglasses}}
    \end{minipage}
    \hfill
    \begin{minipage}{0.49\textwidth}
        \centering
        \includegraphics[width=\textwidth]{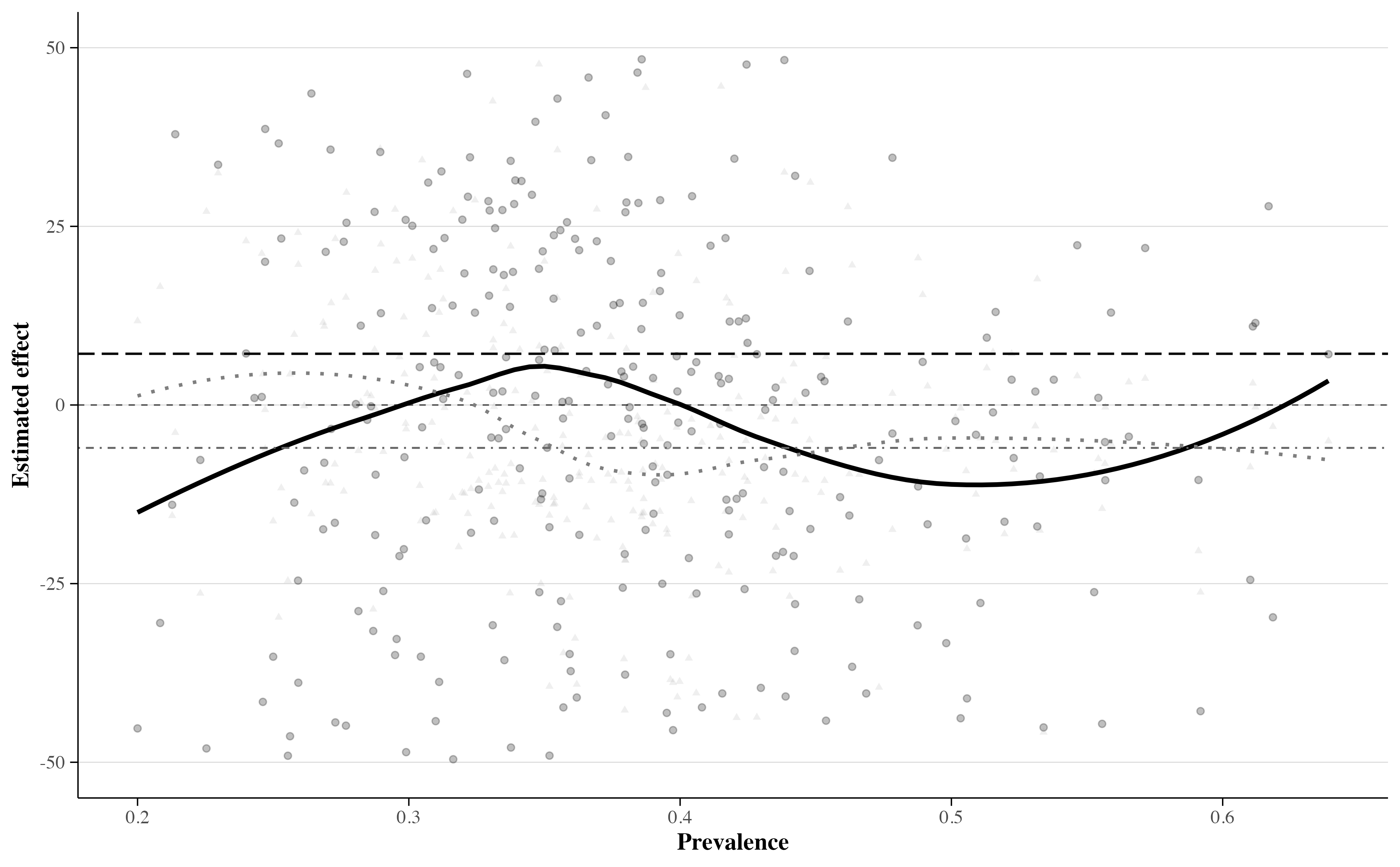}
        \vspace{0.2em}
        
        {\footnotesize (d) \emph{Dark Hair}}
    \end{minipage}

    \caption*{\footnotesize \textit{Note: Markets are defined as sector $\times$ month cells. In each panel, the horizontal axis shows the prevalence of the indicated feature within the market. Black points show market-level simplified AIPW estimates of the effect of the feature on cash per day; light triangles show unadjusted differences in means. The solid line is a smoothed trend in the AIPW estimates and the dotted line is a smoothed trend in the unadjusted differences in means. The dash-dotted horizontal line shows the overall AIPW estimate in the full sample, and the long-dashed horizontal line shows the overall unadjusted difference in means. The vertical axis is restricted to the interval $[-50,50]$ for readability.}}
\end{figure}

Figure \ref{fig:market_prevalence_stability} presents the results. The market-level estimates are naturally noisy, as many sector-month cells are relatively small. This is visible in the substantial dispersion of both the unadjusted and AIPW estimates. Nevertheless, the figures do not reveal a clear pattern in which higher prevalence systematically reduces the estimated impact of the feature. In particular, the smoothed AIPW trends do not show the kind of monotone decline that would be expected if these image features mainly operated through a signal-extraction channel that becomes uninformative once the feature is common in the market. Instead, the estimated effects remain broadly stable across markets with different feature prevalence.

Taken together, these results provide additional support for the assumption that the coefficients on style features are reasonably stable across different market environments. While the market-level evidence is necessarily noisy, we do not find clear evidence that a feature loses its impact because it becomes more common in the local pool of borrowers observed by lenders.

 \section{Algorithm for counterfactual simulations}\label{algo_cf}

In this section \ref{algo_cf}, we describe the algorithm for generating outcomes under counterfactual policies in more detail. We divide the algorithm into two parts: (i) simulation of a market, and (ii) simulation of lenders' choices. We focus on a simplified case in which we consider only \emph{male}, \emph{bodyshot}, and \emph{smile}.

\begin{algorithm}
\caption{Simulation of a market}\label{alg:market_sim}
\begin{algorithmic}
\State $\tilde{\eta} \gets U(\mathcal{N};22)$ \Comment Draw 22 fixed effects uniformly from the set of estimated fixed effects
\State $\tilde{male} \gets \mathbf{E}_{G}[male|D(\tilde{\eta});22] $ \Comment Draw 22 gender realizations
\State $\tilde{bodyshot} \gets \mathbf{E}_{G}[bodyshot|D(\tilde{\eta}), \tilde{male}; 22] $
\State $\tilde{smile} \gets \mathbf{E}_{G}[smile|D(\tilde{\eta}), \tilde{male}; 22] $

\If{$H \in \left\{Partial compliance\right\}$}
    \If{$\tilde{bodyshot} == 1$}
        \State $\tilde{bodyshot} = B_{0.25}$ \Comment Bernoulli trial with $p =0.25$ \EndIf
    \If{$\tilde{smile} == 0$}
        \State $\tilde{smile} = B_{0.75}$ \EndIf
        \EndIf
\State $x \gets \left(\tilde{\eta}, \tilde{male}, \tilde{bodyshot}, \tilde{smile} \right)$
\\
\If{$H \in \left\{ Restrict Competition  \right\}$} 
    \State $\mathcal{M} \gets h(x;5)$
\Else   \State $\mathcal{M} \gets h(x;11)$ \Comment Draw borrowers from the pool following the probability function $h$
\EndIf
\State $\mathcal{M} \gets (\mathcal{M}, \omega)$ \Comment add outside option\\
\Return{} $\mathcal{M}$
\end{algorithmic}
\end{algorithm}

Algorithm \ref{alg:market_sim} proceeds in two steps, first, simulates the pool of borrowers and, second, samples from the pool to construct the market. Policies impact the distribution of the features in the pool (\emph{partial compliance}), the size of the market (Restrict Competition), and the probability of being sampled into the market (through the function $h$). 

Once a market is simulated we determined lenders' choices with Algorithm \ref{alg:lend_choices}. We first simulate the preferences of a lender, then compute the utility associates from different borrowers, and, finally, determined which borrower is selected.

\begin{algorithm}
\caption{Simulation of a lender choice}\label{alg:lend_choices}
\begin{algorithmic}
\State $(\tilde{\alpha}, \tilde{\beta}, \tilde{\gamma}) \gets (N(\alpha,sd_{\alpha}), N(\beta, sd_{\beta}), N(\gamma, sd_{\gamma}))$ \Comment draw preference parameters
\State $\tilde{\epsilon} \gets GEV$ \Comment draw random utility parameters for each borrowing campaign

\State $u \gets U(\mathcal{M} ;\tilde{\alpha},\tilde{\beta},\tilde{\gamma}, \tilde{\epsilon})$ \Comment compute utilities from choosing any of the borrowers

\State $choice \gets max(u)$\\
\Return{} $choice$
\end{algorithmic}
\end{algorithm}

\section{Fake and genuine smiles distinction}\label{fake_smiles_section}

We develop an algorithm to distinguish fake from genuine smiles and apply it to the Kiva observational data. We show that only genuine smiles increase funding outcomes. A policy that asks borrowers to smile may therefore be less effective if the new smiles are perceived as forced, and platforms should pair the recommendation with guidance that helps borrowers produce genuine smiles.

The effectiveness of policies that encourage borrowers to change facial expressions, specifically to smile, relies on the premise that the previously non-compliant borrowers can create images with desired features and that these newly added features impact lenders' choices; for example, the platform policy might be ineffective if lenders perceive the facial expressions in new images as not genuine. This section argues that this concern is legitimate by showing that non-genuine smiles do not increase funding rates.

To introduce a distinction between genuine and fake (forced) smiles, we train an algorithm that classifies the type of smile. We develop this algorithm using a dataset of 6442 images classified by human annotators as fake or genuine smiles.\footnote{The dataset and the original model structure are referred here: \url{https://github.com/vviveks/FakeSmileDetection}; we modified the original algorithm to the task of binary prediction --- genuine or fake. Our algorithm predicts the fakeness of smiles using three different detected components of each face: whole face, eyes, and mouth. We train three deep neural networks (ResNet, DenseNet, and AlexNet) jointly and concatenate the learned latent vectors to make a joint prediction of whether the smiling is fake in the last layer. The model achieves a cross-entropy loss of 0.67 on the test set (0.66 on the train set) and an F1-score of 0.70 on the test set (0.71 on the train set).}

We use the algorithm in a random sample of 45 thousand profiles from the Kiva observational dataset. First, we predict whether the person in the image smiles and whether the smile is genuine or fake. Next, we group the borrowers by the predicted type of smile and compute the average cash collected per day. Finally, we estimate the average impact of each type of smile on cash collected per day; to do that we use the AIPW estimator (we follow the same methodology as in Section \ref{offline_data}). Table \ref{tab:fake_smiles_tab} shows the results.

\begin{table}[!ht]
    \centering
          \caption{The impact of different types of smile on cash per day.}
    \begin{tabular}{lllll}
\\[-1.8ex]\hline 
\hline \\[-1.8ex] 
        Estimand & Not smiling & Any smile & Genuine smile & Fake smile \\ \hline
        Mean outcome in group & 115.0 (0.7) & 131.8 (0.7)	 &	136.5 (0.8)	& 116.4 (0.7) \\ 
        Average treatment effect & - & 	7.3 (1.0) &	15.3 (1.2)	& 0.8 (1.2) \\ \hline

        \\[-1.8ex]\hline 
\hline \\[-1.8ex] 
    \end{tabular}
            \caption*{\footnotesize{\textit{Note: The first row shows the mean cash per day across four groups of borrowers: not smiling, having any type of smile, having a genuine smile, and a fake smile. The second row shows the average effect of having a smile on cash collected per day. We estimate the effect using the AIPW estimator, which adjusts for all other observable characteristics. The comparison group includes borrowers that do not have images with a smile. Standard errors are in parentheses.}}}\label{tab:fake_smiles_tab}
\end{table}

Results presented in Table \ref{tab:fake_smiles_tab} indicate that only smiles that our algorithm predicted to be genuine lead to higher outcomes. Specifically, we estimate that a genuine smile increases the cash collected per day by \$15, while a fake smile has no statistically significant impact. 

This analysis points to an important limitation of any policy based on facial expressions. If lenders perceive some of the smiles created in response to the new policy as not genuine, they might not increase funding rates. In the simulation exercise, we assumed that 75\% of the previously non-compliant borrowers become compliant under the new policy. The policy becomes less effective when the share of borrowers that create images with genuine smiles decreases.

To reduce the risk of an ineffective policy, the platform could give borrowers immediate feedback on their uploaded photos so they can adjust before submitting. An algorithm similar to the one developed in this section can be a part of such a policy.


\begin{thebibliography}{}

\bibitem[Abbey and Meloy, 2017]{abbey2017attention}
Abbey, J.~D. and Meloy, M.~G. (2017).
\newblock {Attention by Design: Using Attention Checks to Detect Inattentive Respondents and Improve Data Quality}.
\newblock {\em {Journal of Operations Management}}, 53:63--70.

\bibitem[Abeler et~al., 2014]{abeler2014representative}
Abeler, J., Becker, A., and Falk, A. (2014).
\newblock {Representative Evidence on Lying Costs}.
\newblock {\em {Journal of Public Economics}}, 113:96--104.

\bibitem[Aggarwal et~al., 2015]{AGGARWAL201555}
Aggarwal, R., Goodell, J.~W., and Selleck, L.~J. (2015).
\newblock {Lending to Women in Microfinance: Role of Social Trust}.
\newblock {\em {International Business Review}}, 24(1):55--65.

\bibitem[Alesina et~al., 2013]{alesina2013women}
Alesina, A.~F., Lotti, F., and Mistrulli, P.~E. (2013).
\newblock {Do Women Pay More for Credit? Evidence from Italy}.
\newblock {\em {Journal of the European Economic Association}}, 11:45--66.

\bibitem[Andreoni and Petrie, 2008]{andreoni2008beauty}
Andreoni, J. and Petrie, R. (2008).
\newblock {Beauty, Gender and Stereotypes: Evidence from Laboratory Experiments}.
\newblock {\em {Journal of Economic Psychology}}, 29(1):73--93.

\bibitem[Asplund et~al., 2020]{asplund2020auditing}
Asplund, J., Eslami, M., Sundaram, H., Sandvig, C., and Karahalios, K. (2020).
\newblock {Auditing Race and Gender Discrimination in Online Housing Markets}.
\newblock In {\em {Proceedings of the International AAAI Conference on Web and Social Media}}, volume~14, pages 24--35.

\bibitem[Athey et~al., 2024]{athey2024using}
Athey, S., Imbens, G.~W., Metzger, J., and Munro, E. (2024).
\newblock {Using Wasserstein Generative Adversarial Networks for the Design of Monte Carlo Simulations}.
\newblock {\em Journal of Econometrics}, 240(2):105076.

\bibitem[Athey et~al., 2019]{athey2019generalized}
Athey, S., Tibshirani, J., and Wager, S. (2019).
\newblock {Generalized Random Forests}.
\newblock {\em {The Annals of Statistics}}, 47(2):1148--1178.

\bibitem[Berk et~al., 2017]{berk2017convex}
Berk, R., Heidari, H., Jabbari, S., Joseph, M., Kearns, M., Morgenstern, J., Neel, S., and Roth, A. (2017).
\newblock {A Convex Framework for Fair Regression}.
\newblock {\em {arXiv preprint arXiv:1706.02409}}.

\bibitem[Bertrand and Mullainathan, 2004]{bertrand2004emily}
Bertrand, M. and Mullainathan, S. (2004).
\newblock {Are Emily and Greg More Employable Than Lakisha and Jamal? A Field Experiment on Labor Market Discrimination}.
\newblock {\em {American Economic Review}}, 94(4):991--1013.

\bibitem[Bohren et~al., 2023]{bohren2023inaccurate}
Bohren, J.~A., Haggag, K., Imas, A., and Pope, D.~G. (2023).
\newblock Inaccurate statistical discrimination: An identification problem.
\newblock {\em Review of Economics and Statistics}, 2023:1--45.

\bibitem[Bolukbasi et~al., 2016]{bolukbasi2016man}
Bolukbasi, T., Chang, K.-W., Zou, J.~Y., Saligrama, V., and Kalai, A.~T. (2016).
\newblock {Man is to Computer Programmer as Woman is to Homemaker? Debiasing Word Embeddings}.
\newblock {\em {Advances in Neural Information Processing Systems}}, 29.

\bibitem[Brock and De~Haas, 2021]{brock2021discriminatory}
Brock, J.~M. and De~Haas, R. (2021).
\newblock {Discriminatory Lending: Evidence from Bankers in the Lab}.
\newblock {\em {CentER Discussion Paper}}.

\bibitem[Burke et~al., 2022]{burke2022performance}
Burke, R., Ragothaman, P., Mattei, N., Kimmig, B., Voida, A., Sonboli, N., Kathait, A., and Fabros, M. (2022).
\newblock {A Performance-Preserving Fairness Intervention for Adaptive Microfinance Recommendation}.
\newblock In {\em {KDD Workshop on Online and Adaptive Recommender Systems at the 28th SIGKDD Conference on Knowledge Discovery and Data Mining}}.

\bibitem[Carroll et~al., 2006]{carroll2006measurement}
Carroll, R.~J., Ruppert, D., Stefanski, L.~A., and Crainiceanu, C.~M. (2006).
\newblock {\em Measurement error in nonlinear models: a modern perspective}.
\newblock Chapman and Hall/CRC.

\bibitem[Chen et~al., 2023]{chen2023bias}
Chen, J., Dong, H., Wang, X., Feng, F., Wang, M., and He, X. (2023).
\newblock {Bias and Debias in Recommender System: A Survey and Future Directions}.
\newblock {\em {ACM Transactions on Information Systems}}, 41(3):1--39.

\bibitem[Cook et~al., 2021]{cook2021gender}
Cook, C., Diamond, R., Hall, J.~V., List, J.~A., and Oyer, P. (2021).
\newblock {The Gender Earnings Gap in the Gig Economy: Evidence from Over a Million Rideshare Drivers}.
\newblock {\em {The Review of Economic Studies}}, 88(5):2210--2238.

\bibitem[Cook and Stefanski, 1994]{cook1994simulation}
Cook, J.~R. and Stefanski, L.~A. (1994).
\newblock Simulation-extrapolation estimation in parametric measurement error models.
\newblock {\em Journal of the American Statistical association}, 89(428):1314--1328.

\bibitem[Dash et~al., 2023]{dash2023review}
Dash, A., Ye, J., and Wang, G. (2023).
\newblock {A Review of Generative Adversarial Networks (GANs) and Its Applications in a Wide Variety of Disciplines: From Medical to Remote Sensing}.
\newblock {\em {IEEE Access}}.

\bibitem[Davidson and Gleim, 2023]{davidson2023gender}
Davidson, A. and Gleim, M.~R. (2023).
\newblock {The Gender Earnings Gap in Sharing Economy Services: The Role of Price, Number of Stays, and Guests Accommodated on Airbnb}.
\newblock {\em {Journal of Marketing Theory and Practice}}, 31(4):490--501.

\bibitem[Dupas et~al., 2024]{dupas2024keeping}
Dupas, P., Fafchamps, M., and Hernandez-Nunez, L. (2024).
\newblock {Keeping Up Appearances: An Experimental Investigation of Relative Rank Signaling}.
\newblock Technical report, National Bureau of Economic Research.

\bibitem[Dwork et~al., 2012]{dwork2012fairness}
Dwork, C., Hardt, M., Pitassi, T., Reingold, O., and Zemel, R. (2012).
\newblock Fairness through awareness.
\newblock In {\em Proceedings of the 3rd innovations in theoretical computer science conference}, pages 214--226.

\bibitem[D’Espallier et~al., 2011]{DESPALLIER2011758}
D’Espallier, B., Guérin, I., and Mersland, R. (2011).
\newblock {Women and Repayment in Microfinance: A Global Analysis}.
\newblock {\em {World Development}}, 39(5):758--772.

\bibitem[Eckel and Grossman, 1996]{eckel1996altruism}
Eckel, C.~C. and Grossman, P.~J. (1996).
\newblock Altruism in anonymous dictator games.
\newblock {\em Games and economic behavior}, 16(2):181--191.

\bibitem[Edelman et~al., 2017]{edelman2017racial}
Edelman, B., Luca, M., and Svirsky, D. (2017).
\newblock {Racial Discrimination in the Sharing Economy: Evidence from a Field Experiment}.
\newblock {\em {American Economic Journal: Applied Economics}}, 9(2):1--22.

\bibitem[Ert et~al., 2016]{ert2016trust}
Ert, E., Fleischer, A., and Magen, N. (2016).
\newblock {Trust and Reputation in the Sharing Economy: The Role of Personal Photos in Airbnb}.
\newblock {\em {Tourism Management}}, 55:62--73.

\bibitem[Esponda and Pouzo, 2016]{esponda2016berk}
Esponda, I. and Pouzo, D. (2016).
\newblock Berk--nash equilibrium: A framework for modeling agents with misspecified models.
\newblock {\em Econometrica}, 84(3):1093--1130.

\bibitem[Friedman, 2001]{friedman2001greedy}
Friedman, J.~H. (2001).
\newblock {Greedy Function Approximation: A Gradient Boosting Machine}.
\newblock {\em {Annals of Statistics}}, pages 1189--1232.

\bibitem[Ge et~al., 2020]{ge2016racial}
Ge, Y., Knittel, C.~R., MacKenzie, D., and Zoepf, S. (2020).
\newblock {Racial Discrimination in Transportation Network Companies}.
\newblock {\em {Journal of Public Economics}}, 190:104205.

\bibitem[Gelbach, 2016]{gelbach2016covariates}
Gelbach, J.~B. (2016).
\newblock {When Do Covariates Matter? And Which Ones, and How Much?}
\newblock {\em {Journal of Labor Economics}}, 34(2):509--543.

\bibitem[Glynn and Quinn, 2010]{glynn2010introduction}
Glynn, A.~N. and Quinn, K.~M. (2010).
\newblock {An Introduction to the Augmented Inverse Propensity Weighted Estimator}.
\newblock {\em {Political Analysis}}, 18(1):36--56.

\bibitem[Goodfellow et~al., 2014]{goodfellow2014generative}
Goodfellow, I., Pouget-Abadie, J., Mirza, M., Xu, B., Warde-Farley, D., Ozair, S., Courville, A., and Bengio, Y. (2014).
\newblock {Generative Adversarial Nets}.
\newblock In {\em {Advances in Neural Information Processing Systems 27 (NIPS 2014)}}, pages 2672--2680.

\bibitem[Haferkamp et~al., 2012]{haferkamp2012men}
Haferkamp, N., Eimler, S.~C., Papadakis, A.-M., and Kruck, J.~V. (2012).
\newblock {Men Are from Mars, Women Are from Venus? Examining Gender Differences in Self-Presentation on Social Networking Sites}.
\newblock {\em {Cyberpsychology, Behavior, and Social Networking}}, 15(2):91--98.

\bibitem[Hainmueller et~al., 2014]{hainmueller_hopkins_yamamoto_2014}
Hainmueller, J., Hopkins, D.~J., and Yamamoto, T. (2014).
\newblock {Causal Inference in Conjoint Analysis: Understanding Multidimensional Choices via Stated Preference Experiments}.
\newblock {\em {Political Analysis}}, 22(1):1--30.

\bibitem[Han et~al., 2024]{han2024unveiling}
Han, L., Fang, J., Zheng, Q., George, B.~T., Liao, M., and Hossin, M.~A. (2024).
\newblock {Unveiling the Effects of Livestream Studio Environment Design on Sales Performance: A Machine Learning Exploration}.
\newblock {\em {Industrial Marketing Management}}, 117:161--172.

\bibitem[Hapek, 2021]{hapek2021fairness}
Hapek, K. (2021).
\newblock {\em {A Fairness-Based Recommender System for Charitable Lending Platform Kiva Using Classification and \ensuremath{\epsilon}-Greedy Policy}}.
\newblock PhD thesis, Dublin, National College of Ireland.

\bibitem[He et~al., 2017]{he2017maskrcnn}
He, K., Gkioxari, G., Doll{\'a}r, P., and Girshick, R. (2017).
\newblock Mask {R-CNN}.
\newblock In {\em Proceedings of the IEEE International Conference on Computer Vision (ICCV)}, pages 2961--2969.

\bibitem[Imai et~al., 2010a]{imai2010general}
Imai, K., Keele, L., and Tingley, D. (2010a).
\newblock A general approach to causal mediation analysis.
\newblock {\em Psychological methods}, 15(4):309.

\bibitem[Imai et~al., 2010b]{imai2010identification}
Imai, K., Keele, L., and Yamamoto, T. (2010b).
\newblock Identification, inference, and sensitivity analysis for causal mediation effects.
\newblock {\em Statistical Science}, 25(1):51--71.

\bibitem[Jabbar et~al., 2021]{jabbar2021survey}
Jabbar, A., Li, X., and Omar, B. (2021).
\newblock {A Survey on Generative Adversarial Networks: Variants, Applications, and Training}.
\newblock {\em {ACM Computing Surveys (CSUR)}}, 54(8):1--49.

\bibitem[Johannemann et~al., 2019]{johannemann2019sufficient}
Johannemann, J., Hadad, V., Athey, S., and Wager, S. (2019).
\newblock {Sufficient Representations for Categorical Variables}.
\newblock {\em {arXiv preprint arXiv:1908.09874}}.

\bibitem[Kajackaite and Gneezy, 2017]{kajackaite2017incentives}
Kajackaite, A. and Gneezy, U. (2017).
\newblock {Incentives and Cheating}.
\newblock {\em {Games and Economic Behavior}}, 102:433--444.

\bibitem[Kamiran and Calders, 2012]{kamiran2012data}
Kamiran, F. and Calders, T. (2012).
\newblock {Data Preprocessing Techniques for Classification Without Discrimination}.
\newblock {\em {Knowledge and Information Systems}}, 33(1):1--33.

\bibitem[Karras et~al., 2019]{karras2019stylebased}
Karras, T., Laine, S., and Aila, T. (2019).
\newblock {A Style-Based Generator Architecture for Generative Adversarial Networks}.
\newblock In {\em {Proceedings of the IEEE/CVF Conference on Computer Vision and Pattern Recognition}}, pages 4401--4410.

\bibitem[Kleinberg et~al., 2018]{kleinberg2018algorithmic}
Kleinberg, J., Ludwig, J., Mullainathan, S., and Rambachan, A. (2018).
\newblock {Algorithmic Fairness}.
\newblock In {\em {AEA Papers and Proceedings}}, volume 108, pages 22--27.

\bibitem[Kleinberg et~al., 2017]{kleinberg2016inherent}
Kleinberg, J., Mullainathan, S., and Raghavan, M. (2017).
\newblock {Inherent Trade-offs in the Fair Determination of Risk Scores}.
\newblock In {\em {8th Innovations in Theoretical Computer Science Conference (ITCS 2017)}}, volume~67 of {\em {LIPIcs}}, pages 43:1--43:23.

\bibitem[Kristof and WuDunn, 2010]{kristof2010half}
Kristof, N.~D. and WuDunn, S. (2010).
\newblock {\em {Half the Sky: Turning Oppression into Opportunity for Women Worldwide}}.
\newblock Vintage.

\bibitem[Kung et~al., 2018]{kung2018attention}
Kung, F.~Y., Kwok, N., and Brown, D.~J. (2018).
\newblock {Are Attention Check Questions a Threat to Scale Validity?}
\newblock {\em {Applied Psychology}}, 67(2):264--283.

\bibitem[Lambin and Palikot, 2022]{lambin2022impact}
Lambin, X. and Palikot, E. (2022).
\newblock {The Impact of Online Reputation on Ethnic Discrimination}.
\newblock Technical report, {Working Paper}.

\bibitem[Lambrecht and Tucker, 2019]{lambrecht2019algorithmic}
Lambrecht, A. and Tucker, C. (2019).
\newblock Algorithmic bias? an empirical study of apparent gender-based discrimination in the display of stem career ads.
\newblock {\em Management science}, 65(7):2966--2981.

\bibitem[Lee et~al., 2014]{lee2014fairness}
Lee, E.~L., Lou, J.-K., Chen, W.-M., Chen, Y.-C., Lin, S.-D., Chiang, Y.-S., and Chen, K.-T. (2014).
\newblock {Fairness-Aware Loan Recommendation for Microfinance Services}.
\newblock In {\em {Proceedings of the 2014 International Conference on Social Computing}}, pages 1--4.

\bibitem[Li et~al., 2018]{li2018balancing}
Li, F., Morgan, K.~L., and Zaslavsky, A.~M. (2018).
\newblock {Balancing Covariates via Propensity Score Weighting}.
\newblock {\em {Journal of the American Statistical Association}}, 113(521):390--400.

\bibitem[Lin et~al., 2023]{lin2306can}
Lin, J., Dai, X., Xi, Y., Liu, W., Chen, B., Li, X., Zhu, C., Guo, H., Yu, Y., Tang, R., et~al. (2023).
\newblock {How Can Recommender Systems Benefit from Large Language Models: A Survey (2023)}.
\newblock {\em {arXiv preprint arXiv:2306.05817}}.

\bibitem[Lin et~al., 2021]{lin2021happiness}
Lin, Y., Yao, D., and Chen, X. (2021).
\newblock {Happiness Begets Money: Emotion and Engagement in Live Streaming}.
\newblock {\em {Journal of Marketing Research}}, 58(3):417--438.

\bibitem[Luca et~al., 2026]{luca2024evolution}
Luca, M., Pronkina, E., and Rossi, M. (2026).
\newblock {The Evolution of Discrimination in Online Markets: How the Rise in Anti-Asian Bias Affected Airbnb During the Pandemic}.
\newblock {\em {Marketing Science}}, 45(1):108--122.

\bibitem[Ludwig and Mullainathan, 2024]{ludwig2024machine}
Ludwig, J. and Mullainathan, S. (2024).
\newblock {Machine Learning as a Tool for Hypothesis Generation}.
\newblock {\em {The Quarterly Journal of Economics}}, 139(2):751--827.

\bibitem[Luo and Toubia, 2024]{luo2024using}
Luo, L.~E. and Toubia, O. (2024).
\newblock {Using AI for Controllable Stimuli Generation: An Application to Gender Discrimination with Faces}.
\newblock {\em {Available at SSRN 4865798}}.

\bibitem[Marchenko, 2019]{marchenko2019impact}
Marchenko, A. (2019).
\newblock {The Impact of Host Race and Gender on Prices on Airbnb}.
\newblock {\em {Journal of Housing Economics}}, 46:101635.

\bibitem[Mehrabi et~al., 2021]{mehrabi2021survey}
Mehrabi, N., Morstatter, F., Saxena, N., Lerman, K., and Galstyan, A. (2021).
\newblock {A Survey on Bias and Fairness in Machine Learning}.
\newblock {\em {ACM Computing Surveys (CSUR)}}, 54(6):1--35.

\bibitem[Mendes and Koslov, 2013]{mendes2013brittle}
Mendes, W.~B. and Koslov, K. (2013).
\newblock {Brittle Smiles: Positive Biases Toward Stigmatized and Outgroup Targets}.
\newblock {\em {Journal of Experimental Psychology: General}}, 142(3):923.

\bibitem[Naghiaei et~al., 2022]{naghiaei2022cpfair}
Naghiaei, M., Rahmani, H.~A., and Deldjoo, Y. (2022).
\newblock {Cpfair: Personalized Consumer and Producer Fairness Re-Ranking for Recommender Systems}.
\newblock In {\em {Proceedings of the 45th International ACM SIGIR Conference on Research and Development in Information Retrieval}}, pages 770--779.

\bibitem[Nosek et~al., 2007]{nosek2007pervasiveness}
Nosek, B.~A., Smyth, F.~L., Hansen, J.~J., Devos, T., Lindner, N.~M., Ranganath, K.~A., Smith, C.~T., Olson, K.~R., Chugh, D., Greenwald, A.~G., et~al. (2007).
\newblock {Pervasiveness and Correlates of Implicit Attitudes and Stereotypes}.
\newblock {\em {European Review of Social Psychology}}, 18(1):36--88.

\bibitem[Ozer et~al., 2023]{ozer2023digital}
Ozer, G.~T., Greenwood, B.~N., and Gopal, A. (2023).
\newblock {Digital Multisided Platforms and Women’s Health: An Empirical Analysis of Peer-to-Peer Lending and Abortion Rates}.
\newblock {\em {Information Systems Research}}, 34(1):223--252.

\bibitem[Park et~al., 2019]{park2019beauty}
Park, J., Kim, K., and Hong, Y.-Y. (2019).
\newblock {Beauty, Gender, and Online Charitable Giving}.
\newblock {\em {Available at SSRN 3405823}}.

\bibitem[Peterson et~al., 2022]{peterson2022deep}
Peterson, J.~C., Uddenberg, S., Griffiths, T.~L., Todorov, A., and Suchow, J.~W. (2022).
\newblock {Deep Models of Superficial Face Judgments}.
\newblock {\em {Proceedings of the National Academy of Sciences}}, 119(17):e2115228119.

\bibitem[Pope and Sydnor, 2011]{pope2011s}
Pope, D.~G. and Sydnor, J.~R. (2011).
\newblock {What’s in a Picture? Evidence of Discrimination from Prosper.com}.
\newblock {\em {Journal of Human Resources}}, 46(1):53--92.

\bibitem[Ravina, 2019]{ravina2008love}
Ravina, E. (2019).
\newblock {Love \& Loans: The Effect of Beauty and Personal Characteristics in Credit Markets}.
\newblock {\em {Available at SSRN 1107307}}.

\bibitem[Reuben et~al., 2014]{reuben2014stereotypes}
Reuben, E., Sapienza, P., and Zingales, L. (2014).
\newblock {How Stereotypes Impair Women’s Careers in Science}.
\newblock {\em {Proceedings of the National Academy of Sciences}}, 111(12):4403--4408.

\bibitem[Robins et~al., 1994]{robins1994estimation}
Robins, J.~M., Rotnitzky, A., and Zhao, L.~P. (1994).
\newblock {Estimation of Regression Coefficients When Some Regressors Are Not Always Observed}.
\newblock {\em {Journal of the American Statistical Association}}, 89(427):846--866.

\bibitem[Shen et~al., 2021]{shen2021study}
Shen, B., RichardWebster, B., O'Toole, A., Bowyer, K., and Scheirer, W.~J. (2021).
\newblock {A Study of the Human Perception of Synthetic Faces}.
\newblock In {\em {2021 16th IEEE International Conference on Automatic Face and Gesture Recognition (FG 2021)}}, pages 1--8. IEEE.

\bibitem[Shishido et~al., 2016]{shishido2016tell}
Shishido, J., Narasimhan, J., and Haller, M. (2016).
\newblock {Tell Me Something I Don't Know: Analyzing OkCupid Profiles}.
\newblock In {\em {SciPy}}, pages 75--81.

\bibitem[Sisodia et~al., 2025]{sisodia2024generative}
Sisodia, A., Burnap, A., and Kumar, V. (2025).
\newblock {Generative Interpretable Visual Design: Using Disentanglement for Visual Conjoint Analysis}.
\newblock {\em {Journal of Marketing Research}}, 62(3):405--428.

\bibitem[Stigler, 2018]{gelbachimplementation}
Stigler, M. (2018).
\newblock {dec\_covar: R Implementation of Gelbach Covariate Decomposition}.
\newblock {\em https://github.com/MatthieuStigler/Misconometrics/tree/master/Gelbach\_decompo}.

\bibitem[Theseira, 2009]{theseira2009competition}
Theseira, W. (2009).
\newblock {\em {Competition to Default: Racial Discrimination in the Market for Online Peer-to-Peer Lending}}.
\newblock PhD thesis, Dissertation, Wharton.

\bibitem[VanderWeele and Vansteelandt, 2014]{vanderweele2014mediation}
VanderWeele, T. and Vansteelandt, S. (2014).
\newblock Mediation analysis with multiple mediators.
\newblock {\em Epidemiologic methods}, 2(1):95--115.

\bibitem[Wang et~al., 2025]{wang2024recommending}
Wang, Y., Tao, L., and Zhang, X.~X. (2025).
\newblock {Recommending for a Multi-Sided Marketplace: A Multi-Objective Hierarchical Approach}.
\newblock {\em {Marketing Science}}, 44(1):1--29.

\bibitem[Wang et~al., 2021]{wang2021understanding}
Wang, Y., Wang, X., Beutel, A., Prost, F., Chen, J., and Chi, E.~H. (2021).
\newblock {Understanding and Improving Fairness-Accuracy Trade-Offs in Multi-Task Learning}.
\newblock In {\em {Proceedings of the 27th ACM SIGKDD Conference on Knowledge Discovery \& Data Mining}}, pages 1748--1757.

\bibitem[Williams et~al., 2018]{williams2018algorithms}
Williams, B.~A., Brooks, C.~F., and Shmargad, Y. (2018).
\newblock {How Algorithms Discriminate Based on Data They Lack: Challenges, Solutions, and Policy Implications}.
\newblock {\em {Journal of Information Policy}}, 8(1):78--115.

\bibitem[Wu et~al., 2019]{wu2019detectron2}
Wu, Y., Kirillov, A., Massa, F., Lo, W.-Y., and Girshick, R. (2019).
\newblock Detectron2.
\newblock \url{https://github.com/facebookresearch/detectron2}.

\bibitem[Yang et~al., 2023]{yang2023styleganex}
Yang, S., Jiang, L., Liu, Z., and Loy, C.~C. (2023).
\newblock {StyleGANEX: StyleGAN-Based Manipulation Beyond Cropped Aligned Faces}.
\newblock In {\em {Proceedings of the IEEE/CVF International Conference on Computer Vision}}, pages 21000--21010.

\bibitem[Younkin and Kuppuswamy, 2018]{younkin2018colorblind}
Younkin, P. and Kuppuswamy, V. (2018).
\newblock {The Colorblind Crowd? Founder Race and Performance in Crowdfunding}.
\newblock {\em {Management Science}}, 64(7):3269--3287.

\bibitem[Yuan et~al., 2024]{yuan2024gender}
Yuan, Y., Liu, X., Zhang, S., and Srinivasan, K. (2024).
\newblock Gender and racial price disparities in the nft marketplace.
\newblock {\em International Journal of Research in Marketing}.

\bibitem[Zhang et~al., 2022a]{zhang2022reducing}
Zhang, L., Xiong, S., Zhang, L., Bai, L., and Yan, Q. (2022a).
\newblock {Reducing Racial Discrimination in the Sharing Economy: Empirical Results from Airbnb}.
\newblock {\em {International Journal of Hospitality Management}}, 102:103151.

\bibitem[Zhang et~al., 2025]{zhang2025serving}
Zhang, S., Friedman, E.~M., Srinivasan, K., Dhar, R., and Zhang, X. (2025).
\newblock Serving with a smile on airbnb: Analyzing the economic returns and behavioral underpinnings of the host’s smile.
\newblock {\em Journal of Consumer Research}, 51(6):1073--1097.

\bibitem[Zhang et~al., 2022b]{zhang2022makes}
Zhang, S., Lee, D., Singh, P.~V., and Srinivasan, K. (2022b).
\newblock {What Makes a Good Image? Airbnb Demand Analytics Leveraging Interpretable Image Features}.
\newblock {\em {Management Science}}, 68(8):5644--5666.

\bibitem[Zhang et~al., 2021]{zhang2021can}
Zhang, S., Mehta, N., Singh, P.~V., and Srinivasan, K. (2021).
\newblock Frontiers: Can an artificial intelligence algorithm mitigate racial economic inequality? an analysis in the context of airbnb.
\newblock {\em Marketing Science}, 40(5):813--820.

\end{thebibliography}
\end{document}